\documentclass[namedreferences]{SolarPhysics}
\usepackage[optionalrh]{spr-sola-addons} 
\usepackage{graphicx}                    
\usepackage{courier}                    
\usepackage{amssymb}                     

\usepackage{color}                       
\usepackage{url}                         

\begin{document}
\begin{article}
\begin{opening}

\title{Solar Magnetic Carpet II: Coronal Interactions of Small-Scale Magnetic Fields}
\author{K. A. \surname{Meyer}$^{1}$\sep D. H. \surname{Mackay}$^{1}$}
\author{A. A. \surname{van Ballegooijen}$^{2}$}

\runningauthor{Meyer, Mackay \& van Ballegooijen}
\runningtitle{Coronal Interactions of Small-Scale Solar Magnetic Fields}

\institute{ $^{1}$ School of Mathematics and Statistics, University of 
                  St Andrews, North Haugh, St Andrews, Fife, KY16 9SS, Scotland, U.K. \\
                  \url{karen@mcs.st-and.ac.uk}\\
                  $^{2}$ Harvard-Smithsonian Center for Astrophysics, 60 Garden Street, Cambridge, MA 02138}

\date{Received ; accepted }
\begin{abstract}\\
This paper is the second in a series of studies working towards constructing a realistic, evolving, non-potential coronal model for the solar magnetic carpet.
In the present study, the interaction of two magnetic elements is considered. Our objectives are to study magnetic energy build up, storage and dissipation as a result of emergence, cancellation, and flyby of these magnetic elements. In the future these interactions will be the basic building blocks of more complicated simulations involving hundreds of elements. Each interaction is simulated in the presence of an overlying uniform magnetic field, which lies at various orientations with respect to the evolving magnetic elements.
For these three small-scale interactions, the free energy stored in the field at the end of the simulation ranges from $0.2-2.1\times 10^{26}$ ergs, while the total energy dissipated ranges from $1.3-6.3\times 10^{26}$ ergs. For all cases, a stronger overlying field results in higher energy storage and dissipation. For the cancellation and emergence simulations, motion perpendicular to the overlying field results in the highest values. For the flyby simulations, motion parallel to the overlying field gives the highest values. In all cases, the free energy built up is sufficient to explain small-scale phenomena such as X-ray bright points or nanoflares. In addition, if scaled for the correct number of magnetic elements for the volume considered, the energy continually dissipated provides a significant fraction of the quiet Sun coronal heating budget.

\end{abstract}
\end{opening}

\section{Introduction}\label{intro} 

The \emph{magnetic carpet} is the term used to describe the small-scale photospheric magnetic field of the quiet Sun \cite{title1998}.
Magnetic features in the magnetic carpet are generally categorised into three main classifications: ephemeral regions, network features and internetwork features. The nature of these small-scale features has been extensively studied by many authors \cite{harvey1973,martin1988,martin1990,wang1995,schrijver1997,chae2001,hagenaar2001,parnell2002,hagenaar2003,dewijn2008,parnell2009,thornton2011}. The main properties of the magnetic features are summarised in Table~\ref{tab:features}.

Supergranular motions \cite{simon1964,hagenaar1997,paniveni2004} provide the dominant flow pattern within the magnetic carpet. In addition to this, magnetic carpet features may evolve via the processes of \emph{emergence}, \emph{cancellation}, \emph{coalescence} and \emph{fragmentation}.
These evolution processes, coupled with the photospheric flow pattern, mean that the quiet Sun photosphere is highly dynamic. \inlinecite{hagenaar2008} determined the photospheric flux replacement time to be just 1 - 2 hours. Since magnetic fields from the magnetic carpet extend up through the chromosphere and into the lower corona, it is expected that the quiet Sun corona is also highly dynamic.

Within the quiet Sun corona, energy may be released in a variety of ways. \inlinecite{withbroe1977} determined that the amount of heating required to maintain the quiet Sun corona is $3\times 10^5$ ergs cm$^{-2}$ s$^{-1}$. There are a number of small-scale, transient phenomena associated with sporadic energy release, such as X-ray bright points (XBP) or nanoflares. XBPs are localised brightenings within the quiet sun corona, observed as point-like features or small loops \cite{golub1974}. They are typically associated with opposite polarity magnetic features on the photosphere, and in most cases with cancellation events \cite{webb1993}. While XBP are not believed to be the primary source of quiet Sun heating, as they radiate only around $5\times 10^4$ ergs cm$^{-2}$ s$^{-1}$ \cite{habbal1991}, they do provide a contribution \cite{longcope1999}. Many authors have considered the interaction between pairs of small-scale magnetic elements in association with XBP (e.g. \inlinecite{priest1994}, \inlinecite{longcope1998}, \inlinecite{longcope1999}, \inlinecite{vonrekowski2006}).
Nanoflares are localised, impulsive bursts of energy, releasing energy on the order of $10^{24}$ ergs \cite{parker1988}. It is believed that nanoflares provide a contribution to coronal heating, by dissipating energy through magnetic reconnection. several authors have considered nanoflares in this context, including \inlinecite{cargill1993}, \inlinecite{browning2004}, \inlinecite{browning2008} and \inlinecite{sakamoto2009}.

In the past, several models for the solar magnetic carpet have been produced. These fall roughly into two categories, although there is a natural overlap between them. The first are models that aim to reproduce the evolution of photospheric magnetic features as seen in magnetogram observations \cite{schrijver1997,vanballegooijen1998,simon2001,parnell2001,cranmer2010,meyer2011}. The second are those that simulate the small-scale coronal field, usually using static photospheric boundary conditions \cite{schrijver2002,close2003,close2004,regnier2008,cranmer2010}. These models tend to use the potential field approximation to study the coronal field.
Previous modelling has shown that coronal heating inferred from the number of null points deduced from potential field models is insufficient to explain the hot corona \cite{schrijver2002,regnier2008}. As a result, more complex models that follow the history and distribution of electric currents and non-potentiality within the coronal field are required.

With this in mind, our long term goal is to produce a non-potential model for the continuous evolution of the magnetic carpet, that models both the photospheric and coronal field. To construct this, we have formulated a two component model: a 2D photospheric model acts as the lower boundary for the full 3D coronal simulation.
The photospheric component includes the effects of supergranular flows in addition to the four main flux evolution processes of emergence, cancellation, coalescence and fragmentation. In \inlinecite{meyer2011}, hereafter Paper I, we show that our photospheric model reproduces many observed properties of the magnetic carpet. The purpose of the current paper is to present and test a new modelling technique that follows the continuous evolution of the 3D magnetic carpet coronal field through non-linear force-free states. This is in contrast to the models described above, which use static potential field extrapolations. A key difference between our non-potential model and the above potential field models is that our model retains a memory of flux connectivity and magnetic interactions over relevant timescales. We apply this technique to commonly occurring magnetic carpet interactions. These interactions are emergence, cancellation, and flyby. All three are simulated under the presence of an overlying magnetic field. Studying these interactions with the overlying field is useful, since small-scale phenomena on the solar photosphere can be influenced by stronger overlying fields.
In a following paper, we will carry out full 3D magnetic carpet simulations of hundreds of magnetic elements, using the synthetic magnetograms described in Paper I as photospheric boundary data. Therefore the present paper is designed to consider three basic building blocks of the more complicated simulations.

\begin{table}
\begin{center}
\begin{tabular}{cccc}
\hline
Name & Origin & Flux (Mx) & Size (km) \\
\hline
 Ephemeral  & Emerge within  & $10^{18}-10^{20}$ & $3\,000 - 5\,000$ \\
 Regions (ERs) & supergranules &  &  \\
& & & \\
 Internetwork & Emerge within & $10^{16}-2\times 10^{18}$ & $2\,000$ (mean) \\
(IN) Features & supergranules &  &  \\
& & & \\
 Network  & Formed from residuals of & $10^{18}-10^{19}$ & $1\,000 - 10\,000$ \\
    Features  & ERs (90\%) and IN features (10\%) & & \\
\hline
\end{tabular}
\caption{Magnetic flux features within the solar magnetic carpet.}\label{tab:features}
\end{center}
\end{table}

Previous studies have also considered such interactions, these include full MHD simulations of magnetic flux emergence (e.g. \inlinecite{archontis2004}, \inlinecite{mactaggart2009}) and of a magnetic flyby (e.g. \inlinecite{galsgaard2000}, \inlinecite{parnell2010}). In the paper of \inlinecite{galsgaard2005}, the authors study the heating associated with a flyby using full MHD simulations, and they show that the amount of energy stored or dissipated within the corona depends on several factors. They conclude that it is not sufficient to know the evolution of the magnetic field at the photospheric boundary to be able to predict the evolution of the coronal field. Knowledge of the strength and direction of the overlying field is also important.

Within the literature, there are no previous studies that compare all three interactions of emergence, cancellation and flyby to one another under the same modelling approximations and using the same parameters. Therefore within the present study, we will consider the flux connectivity of the magnetic elements, the build up and storage of free magnetic energy, and the energy dissipated during these three basic magnetic carpet interactions. In particular, we wish to identify the factors involved in determining the amount of magnetic energy that is stored or dissipated.
We use this study as a preliminary analysis to quantify results expected in more complex simulations involving more magnetic elements (see Paper I).

The paper is outlined as follows. Section~\ref{model} discusses our treatment of the photospheric boundary condition and how it couples to the coronal magnetic field model. The set-up and analysis of the three basic interactions are described in Section~\ref{basic}. We present the discussion and conclusions in Section~\ref{conclusion}.

\section{Model}\label{model}

A non-linear force-free magnetic field is a useful approximation to the coronal field, as it allows for the existence of electric currents and free magnetic energy. For a magnetic field to be force-free, it must satisfy:
\begin{equation}
 \nabla \times \mathbf{B} = \alpha(\mathbf{r})\mathbf{B},
\end{equation}
and
\begin{equation}\label{eqn:divb}
 \nabla \cdot \mathbf{B} = 0.
\end{equation}
The physical conditions required for such an approximation to be valid are described by \inlinecite{regnier2007}. The parameter $\alpha(\mathbf{r})$ describes the twist of the magnetic field. It is a scalar function of position, but must be constant along a given magnetic field line. An important property of a non-linear force-free field is that it allows for regions of both high and low $\alpha$, along with varying sign of $\alpha$, so may therefore model a wide variety of coronal structures.

There are several methods for constructing non-linear force-free fields from fixed photospheric boundary conditions; a summary is given by \inlinecite{schrijver2006}. In contrast to these methods, which produce single independent extrapolations, we choose to model a continuous evolution of the coronal field through a series of quasi-static, non-linear force-free equilibria, that are driven by an evolving photospheric boundary condition. The magnetofrictional technique employed to carry this out is described next.

\subsection{Magnetofrictional Method}

Our model for the 3D coronal magnetic field is based upon that of \inlinecite{vanballegooijen2000}; the method is described fully in \inlinecite{mackay2011}. It has been used in the past to study the global coronal magnetic field \cite{yeates2008}, the evolving magnetic structure of solar filaments \cite{mackay2009}, and the decay of an active region \cite{mackay2011}. In the model, we evolve the coronal field by the induction equation,
\begin{equation}\label{eqn:induction}
 \frac{\partial \mathbf{A}}{\partial t} = \mathbf{v} \times \mathbf{B} + \textrm{\boldmath$\epsilon$},
\end{equation}
where $\mathbf{A}$ is the vector potential and $\mathbf{B}=\nabla\times\mathbf{A}$. The plasma velocity $\mathbf{v}$ is specified by:
\begin{equation}\label{eqn:v}
 \mathbf{v} = \frac{1}{\nu}\frac{\mathbf{j}\times\mathbf{B}}{B^2},
\end{equation}
following the magnetofrictional method of \inlinecite{yang1986}, where $\nu$ is the coefficient of friction and $\mathbf{j}=\nabla\times\mathbf{B}$. Through using this formalisation, the coronal field evolves through a series of approximate, quasi-static, non-linear force-free equilibria. The second term on the right hand side of Equation~\ref{eqn:induction} is a non-ideal term that represents hyperdiffusion. The form is chosen to be:
\begin{equation}\label{eqn:hd}
 \textrm{\boldmath$\epsilon$} = \frac{\mathbf{B}}{B^{2}}\nabla\cdot(\eta_{4}B^{2}\nabla\alpha),
\end{equation}
\cite{boozer1986,bhattacharjee1986,strauss1988}, where $\eta_4$, the hyperdiffusivity constant, is chosen to be $4.7 \times 10^{5}$ km$^4$ s$^{-1}$. The key effect of hyperdiffusion is that it conserves magnetic helicity whilst smoothing gradients in $\alpha$. For any non-linear force-free field, hyperdiffusion acts to reduce the field towards a linear force-free state containing the same magnetic helicity. It should be noted that the timescales of the present simulations are far too short for a linear force-free field to be reached.

\subsection{Magnetic Energy Storage and Dissipation}\label{energy}

To consider the effect of the magnetic carpet on the corona, one aspect of the simulations that we are interested in is the build-up and release of energy. At any instant in time in our numerical box of volume $V$, the total magnetic energy is
\begin{equation}\label{eqn:energy}
 W = \int_V \frac{B^2}{8\pi} dV.
\end{equation}
Following this, the rate of change of the total magnetic energy is
\begin{displaymath}
 \frac{dW}{dt}  =  \frac{1}{4\pi}\int_V \frac{d}{dt}\bigg( \frac{B^2}{2} \bigg) dV.
\end{displaymath}
Substituting in Equation~\ref{eqn:induction}:
\begin{eqnarray}\nonumber
\frac{d}{dt}\bigg( \frac{B^2}{2} \bigg) & = &  \mathbf{B} \cdot \frac{\partial \mathbf{B}}{\partial t} \\
        & = & \mathbf{B} \cdot \nabla \times (\mathbf{v} \times \mathbf{B} + \textrm{\boldmath$\epsilon$}) \nonumber \\
\end{eqnarray}
\begin{equation} \label{eqn:dbdt}
 \therefore \frac{d W}{dt}  =  \int \int_S \mathbf{I}\: d\mathbf{S} -\int \int \int_V Q\: dV,
\end{equation}
where
\begin{equation}
\mathbf{I} \equiv \frac{1}{4\pi}\bigg[(\mathbf{v} \times \mathbf{B} + \textrm{\boldmath$\epsilon$})\times\mathbf{B} +\eta_4 B^2 \alpha\nabla\alpha\bigg]
\end{equation}
and
\begin{equation}\label{eqn:q}
Q \equiv \frac{B^2}{4\pi}(\nu|\mathbf{v}|^2+\eta_4 |\nabla\alpha|^2).
\end{equation}
The first term on the right hand side of Equation~\ref{eqn:dbdt} represents the energy injected due to surface motions, along with that injected or removed due to flux emergence or cancellation respectively. In addition, there is a contribution from hyperdiffusion. The second term is the rate at which energy is dissipated per unit time, due to the coronal evolution. This dissipated energy, which is released in the coronal volume, may be considered as energy that is available to be converted into heat or plasma motions. While it may be regarded as this, for simplicity in the present paper we only consider the size of $Q$, and do not follow the corresponding plasma processes.

For the coronal dissipation term (Equation~\ref{eqn:q}), the first term represents energy dissipation due to magnetofriction, which is released as the field relaxes to a force-free state. The second term represents energy dissipation due to hyperdiffusion, which is described by \inlinecite{vanballegooijen2008}. In the present simulations, it is found that $\nabla \alpha$ and the relaxation velocity are both largest near the magnetic elements and at locations where the magnetic field lines reconnect. Hence the dissipative term has its largest contribution at the sources, where there is strong magnetic field, $\mathbf{B}$, and at locations of changing magnetic topology.

Another quantity that we study is the free magnetic energy stored within the magnetic field. This is defined to be the difference between the energy of the non-linear force-free field and the corresponding potential field. The free energy at any instant is
\begin{equation}\label{eqn:free}
E_\textrm{f}(t)=W_{\textrm{\small nl}}(t) - W_\textrm{\small p}(t) = \int_V \frac{B_{\textrm{\small nl}}^2-B_\textrm{\small p}^2}{8\pi} dV,
\end{equation}
where $B_{\textrm{\small nl}}$ is the non-linear force-free field and $B_\textrm{\small p}$ is the corresponding potential field.
Within the simulations we will study first the energy that is continually released and may contribute to the heating of the corona (Equation~\ref{eqn:q}). Second, we will consider the energy stored in the magnetic field that may be related to more sporadic events, such as XBPs or nanoflares (Equation~\ref{eqn:free}). The relative importance of these two forms of energy will be discussed for each of the three interactions between magnetic elements.

\subsection{Photospheric Boundary Condition}\label{lower}

To simulate the processes of emergence, cancellation and flyby, each magnetic element at the photosphere is assumed to have a simple Gaussian form. Therefore the normal field component ($B_{z}$) of an element is given by:
\begin{equation}
 B_{z}=B_{0} e^{-r^{2}/r_{0}^2},
\end{equation}
where $B_0$ is the peak strength, $r_0$ the Gaussian half-width and $r$ the distance from the centre of the element. Each magnetic element within our simulations has a peak strength of $B_0=88$ G, a Gaussian half-width of $r_0=0.6$ Mm and an absolute flux of $10^{18}$ Mx. The total photospheric magnetic field is made up of the sum of a number of these elements.

A unique feature of the simulations is that $B_{z}$ during the evolution of the photospheric field is specified analytically at discrete time intervals, $T_i$ (= 200 s = 3.3 min) apart. Movement of the sources between these time intervals is obtained by changing their central positions, $(x,y)$, rather than advecting them numerically. Through doing so, we avoid undesirable numerical effects such as overshoot due to numerical differentiation, and pile-up at cancellation sites due to forcing.

\begin{figure}
 \begin{center}
  \includegraphics[width=1.0\textwidth]{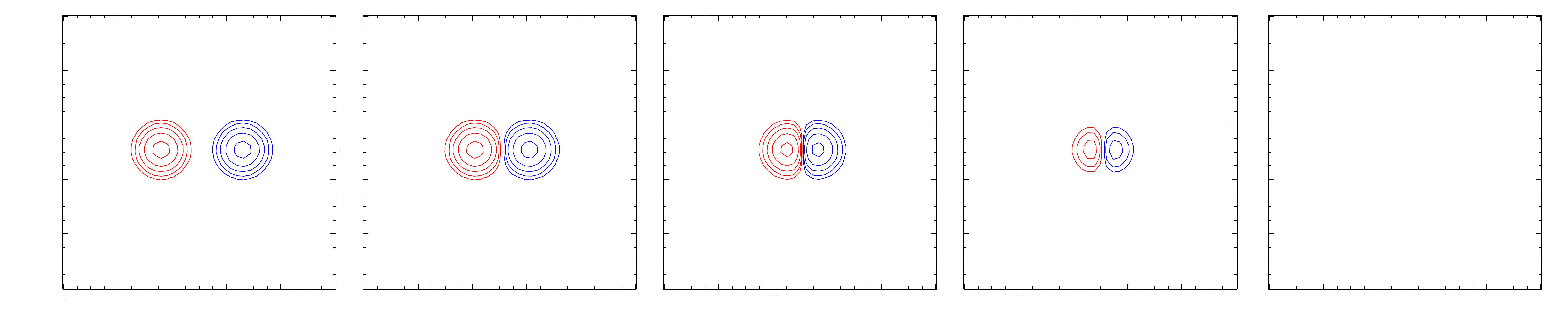}

     \vspace{-0.01\textwidth}  
     \centerline{ \bf     
      \hspace{0.045 \textwidth}  \color{black}{\small{(i) $t=50$}}
      \hspace{0.035\textwidth}  \color{black}{\small{(ii) $t=66.7$}}
      \hspace{0.025\textwidth}  \color{black}{\small{(iii) $t=83.3$}}
      \hspace{0.025\textwidth}  \color{black}{\small{(iv) $t=96.7$}}
      \hspace{0.025\textwidth}  \color{black}{\small{(v) $t=100$}}
         \hfill}

     \vspace{-0.23\textwidth}
     \centerline{ \bf     
      \hspace{-0.04 \textwidth}  \color{black}{\small{(a)}}
         \hfill}
     \vspace{0.23\textwidth}

  \includegraphics[width=1.0\textwidth]{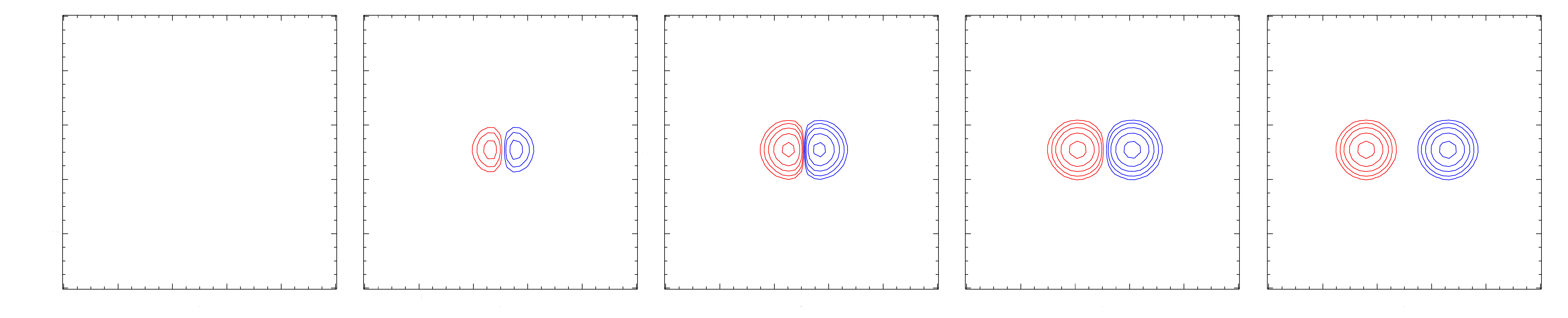}

     \vspace{-0.01\textwidth}  
     \centerline{ \bf     
      \hspace{0.045 \textwidth}  \color{black}{\small{(i) $t=0$}}
      \hspace{0.055\textwidth}  \color{black}{\small{(ii) $t=3.3$}}
      \hspace{0.035\textwidth}  \color{black}{\small{(iii) $t=16.7$}}
      \hspace{0.025\textwidth}  \color{black}{\small{(iv) $t=33.3$}}
      \hspace{0.025\textwidth}  \color{black}{\small{(v) $t=50$}}
         \hfill}

     \vspace{-0.23\textwidth}
     \centerline{ \bf     
      \hspace{-0.04 \textwidth}  \color{black}{\small{(b)}}
         \hfill}
     \vspace{0.23\textwidth}

  \includegraphics[width=1.0\textwidth]{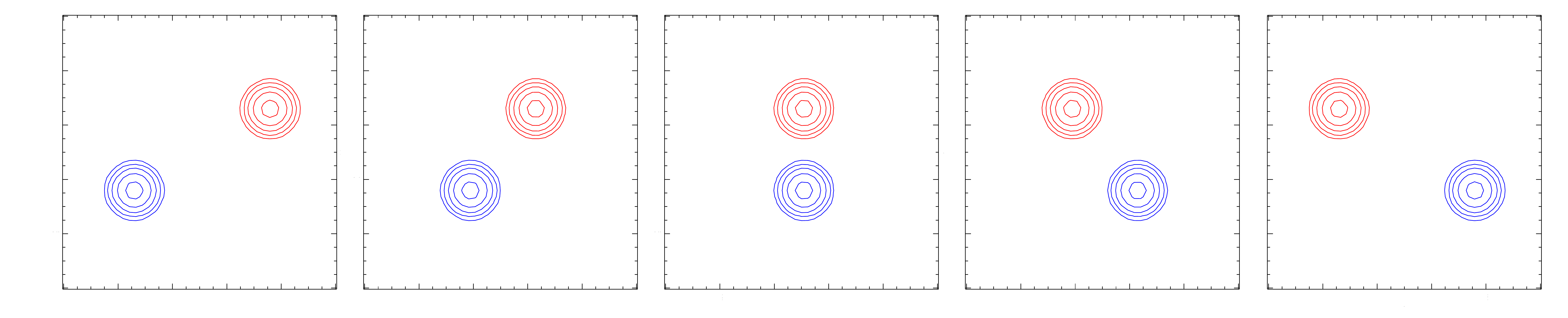}

     \vspace{-0.01\textwidth}  
     \centerline{ \bf     
      \hspace{0.045 \textwidth}  \color{black}{\small{(i) $t=0$}}
      \hspace{0.055\textwidth}  \color{black}{\small{(ii) $t=43.3$}}
      \hspace{0.025\textwidth}  \color{black}{\small{(iii) $t=83.3$}}
      \hspace{0.015\textwidth}  \color{black}{\small{(iv) $t=123.3$}}
      \hspace{0.01\textwidth}  \color{black}{\small{(v) $t=166.7$}}
         \hfill}

     \vspace{-0.23\textwidth}
     \centerline{ \bf     
      \hspace{-0.04 \textwidth}  \color{black}{\small{(c)}}
         \hfill}
     \vspace{0.19\textwidth}

 \end{center}
\caption{Three magnetic flux interactions modelled at the photospheric level. The interactions occur between a positive and a negative magnetic element of equal strength: (a) cancellation, (b) emergence and (c) flyby. In each case, red and blue contours represent the positive and negative polarities respectively, at levels of $\pm$[4,7,14,28,57] G. The area shown is of size 10 Mm $\times$ 10 Mm. For each image, the time is given in minutes.}\label{fig:int}
\end{figure}

Using this description for magnetic elements, we may model a wide range of magnetic flux interactions, such as:
\begin{description}
 \item[(a) Cancellation:] Two magnetic elements, one positive and one negative, move together until their centres coincide. If they are of the same strength, they completely cancel. A cancellation occurring over a time period of 100 min (30 $T_i$) can be seen in Figure~\ref{fig:int}(a).

 \item[(b) Emergence:] Two magnetic elements, one positive and one negative, have the same strength and initially the same central position. Moving the sources apart then simulates emergence as is seen in observations. An emergence occurring over a time period of 100 min (30 $T_i$) can be seen in Figure~\ref{fig:int}(b).

 \item[(c) Flyby:] Two magnetic elements that move relative to one another, but never interact at the photospheric level. A flyby occurring over a time period of 166.7 min (50 $T_i$) can be seen in Figure~\ref{fig:int}(c).

\end{description}
At any instant in time, the flux through the photosphere is given by
\begin{equation}
 \textrm{Flux}=\int_{S} B_z dS,
\end{equation}
where $S$ is the photospheric boundary surface.
\begin{figure}
   \begin{center}
     \includegraphics[width=0.7\textwidth,clip=]{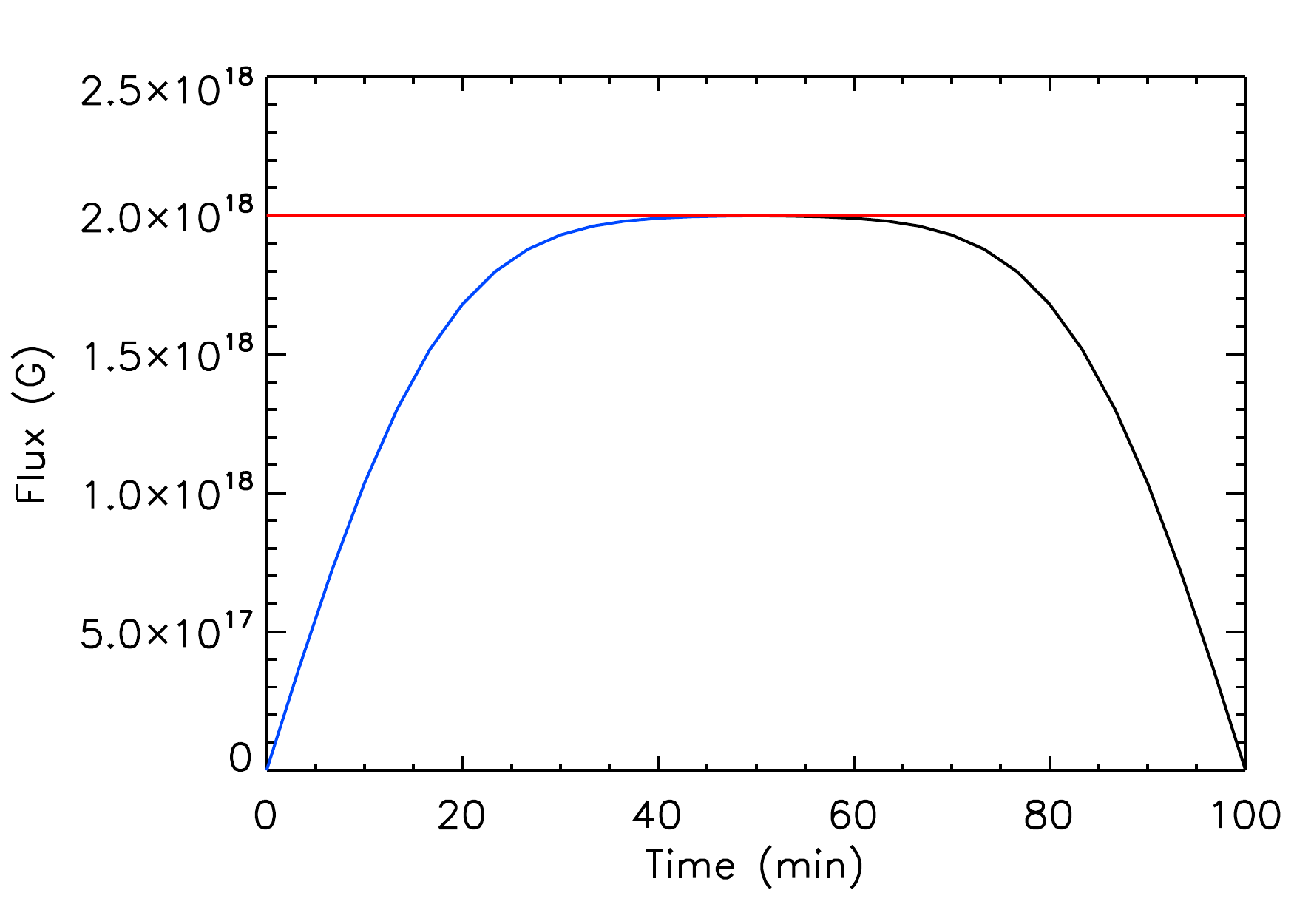}
   \end{center}
\caption{Total absolute flux through the photosphere as a function of time for a cancellation (black), emergence (blue) and flyby (red) event.}\label{fig:flux}
\end{figure}
Figure~\ref{fig:flux} shows a plot of the total absolute flux through the photosphere as a function of time for each of the events in Figure~\ref{fig:int}. It can be seen that the curves are all completely smooth. The cancellation curve (black line) is level until the magnetic elements encounter one another (around $t=60$ min); the curve then decreases smoothly to zero at which point the elements completely cancel. The emergence curve (blue line) shows the opposite behaviour, while the flyby curve (red line) remains completely level as no flux emergence or cancellation occurs\footnote{Note that for the flyby, only $t$=0-100 min is plotted, however the curve remains completely level throughout the entire simulation ($t$=0-166.7 min).}.

To ensure that $\nabla\cdot\mathbf{B}=0$ within the simulations, the coronal field induction equation is specified in terms of $\mathbf{A}$. Therefore to drive the evolution of the coronal field we require as photospheric boundary data the $A_{x}$ and $A_{y}$ corresponding to the analytically specified $B_{z}$ at each time step $T_i$.
Without loss of generality, we can write $A_{x}$ and $A_{y}$ at $z=0$ in terms of a scalar potential $\Phi(x,y)$, where
\begin{equation}
(A_x,A_y)_{(z=0)} = \nabla\times(\Phi \mathbf{\hat{z}}) = \bigg( \frac{\partial \Phi}{\partial y}, -\frac{\partial \Phi}{\partial x} \bigg).
\end{equation}
Then from the $z$-component of $\mathbf{B}=\nabla\times\mathbf{A}$ we have:
\begin{equation}
  B_{z(z=0)} = -\frac{\partial^{2} \Phi}{\partial x^{2}} - \frac{\partial^{2} \Phi}{\partial y^{2}}.
\end{equation}
This may be solved using a fast Fourier transform to find $\Phi$, and hence the $A_x$ and $A_y$ corresponding to $B_z$ at the level of the photosphere.

To produce a continuous time sequence between each interval $T_i$, where $B_z$ and subsequently $A_x$ and $A_y$ are analytically specified, a linear interpolation of $A_x$ and $A_y$ is carried out between each $T_i$ and $T_{i+1}$, where 500 interpolation steps are taken between each analytical specification. This means that at each time $T_i$, the normal field at the photospheric boundary matches the exact analytically specified $B_z$ given by the sum of the Gaussian profiles of the discrete magnetic elements. In response to the photospheric evolution of $A_x$ and $A_y$ at z=0, the vector potential within the coronal volume, and therefore the coronal field, evolves through a series of quasi-static equilibria as described by Equation~\ref{eqn:induction}.
This treatment of the magnetic field at the photospheric boundary ensures that we still have freedom for $A_{z}$, which sits half a grid point up from the photosphere (due to a staggered grid). Any non-potentiality in the coronal field near the photosphere arises from the $z$ component of the vector potential.
The initial condition for the coronal field in each simulation is a potential field extrapolated from $A_x$ and $A_y$ on z=0, at $t=T_0$.

Since the evolution of our field is continuous, connectivity within the coronal magnetic field is maintained throughout the simulation. The resultant series of non-potential equilibria retain a memory of flux interactions from one step to the next, and subsequently the build up of non-potential effects. This is a significantly different approach compared to independent potential field extrapolations, and provides a new insight into the energy budget of the quiet Sun corona.

\section{Basic Interactions of Magnetic Elements} \label{basic}

The three basic interactions studied - cancellation, emergence and flyby - are processes that commonly occur between magnetic elements of equal but opposite flux. Each simulation uses the photospheric boundary treatment described in Section~\ref{lower}, which is then applied to the 3D magnetofrictional model in order to drive the coronal field evolution. Section~\ref{setup} describes the features of the set-up that are common to all three cases; Sections~\ref{canc} -~\ref{fly} then give the results for individual cases. Section~\ref{compare} compares the three basic interactions to one another.

\subsection{Set-up}\label{setup}

A numerical box of size 30 $\times$ 30 $\times$ 17.58 Mm is chosen, composed of 256 $\times$ 256 $\times$ 150 grid cells. In each case, the interaction between magnetic elements is centred around the midpoint of the box. The box is periodic in the $x$-direction, and closed in the $y$-direction and at the top.

For each interaction, the simulation is run under the presence of an overlying uniform magnetic field of strength 1 G, 5 G or 10 G, which points in the $x$-direction. Each interaction is also simulated with three different orientations of the bipole's axis with respect to the overlying field (parallel, anti-parallel or perpendicular). The effects of varying the strength of the overlying field and direction of motion of the bipole with respect to the overlying field are investigated. For each simulation, the magnetic elements are advected at a constant velocity of 0.5 km s$^{-1}$.

\subsection{Cancellation}\label{canc}

\begin{figure}
 \begin{center}
  \includegraphics[width=0.7\textwidth]{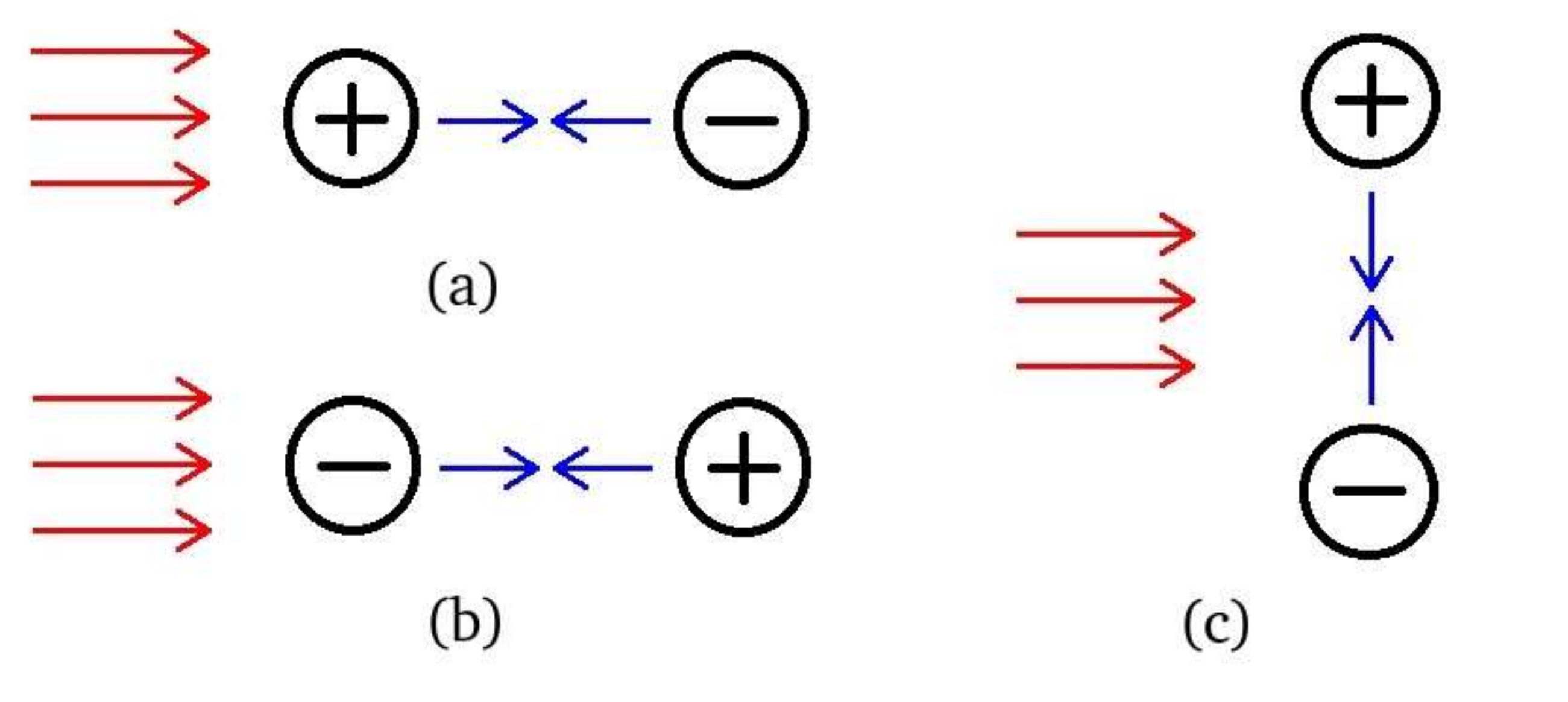}
 \end{center}
\caption{Cancellation: The red arrows represent the direction of the overlying field, the blue arrows represent the direction of motion of the magnetic elements. The bipole's axis is oriented (a) parallel to, (b) anti-parallel to, and (c) perpendicular to the overlying field.}\label{fig:cancel}
\end{figure}

The initial set-up for each of the cancellation simulations is illustrated in Figure~\ref{fig:cancel}. Each magnetic element is initially positioned 3 Mm from box centre. It then takes 100 min to reach the midpoint, where the opposite polarity elements cancel. As seen in Figure~\ref{fig:flux}, the total absolute flux through the photosphere (black line) remains constant until the magnetic elements come into contact, at which point the flux rapidly decreases to zero.

\subsubsection{Field Lines}

  \begin{figure}
   \centerline{\hspace*{-0.01\textwidth}
               \includegraphics[width=0.45\textwidth,clip=]{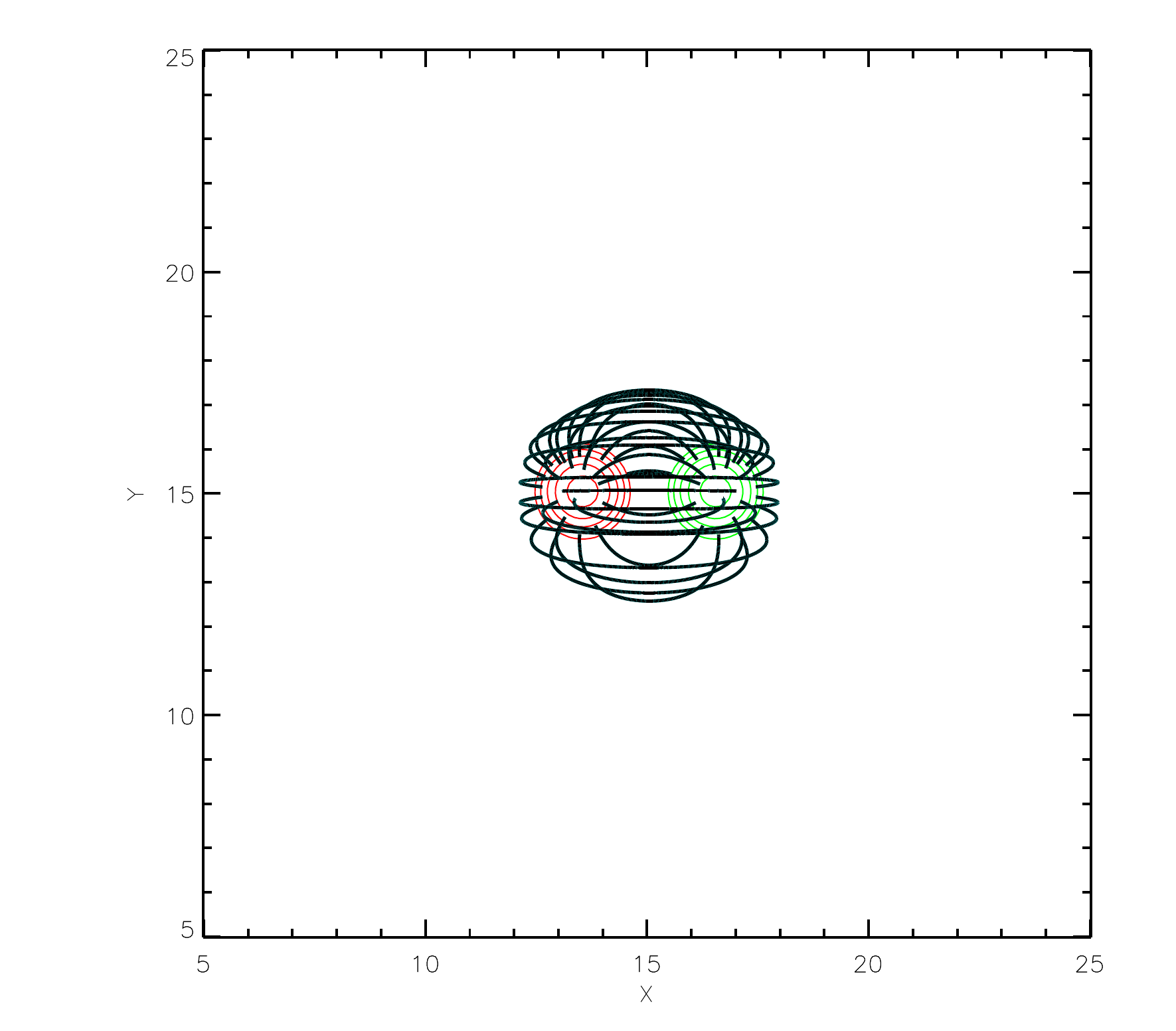}
               \hspace*{0.\textwidth}
               \includegraphics[width=0.45\textwidth,clip=]{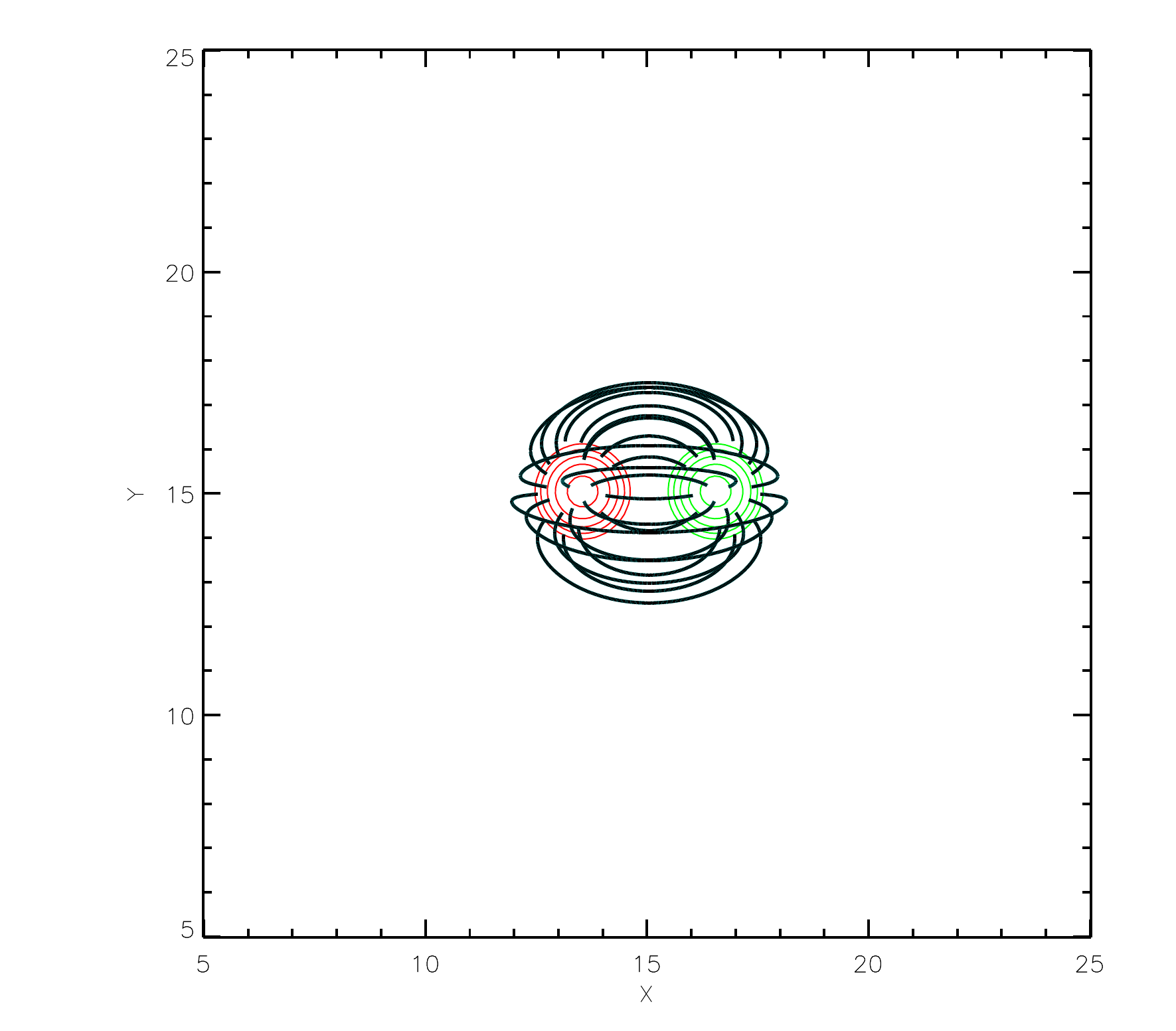}
              }
     \vspace{0.00\textwidth}   
     \centerline{ \bf     
      \hspace{0.245 \textwidth}  \color{black}{(a)}
      \hspace{0.405\textwidth}  \color{black}{(b)}       \hfill}
     \vspace{-0.1\textwidth}    

   \centerline{\hspace*{-0.01\textwidth}
               \includegraphics[width=0.45\textwidth,clip=]{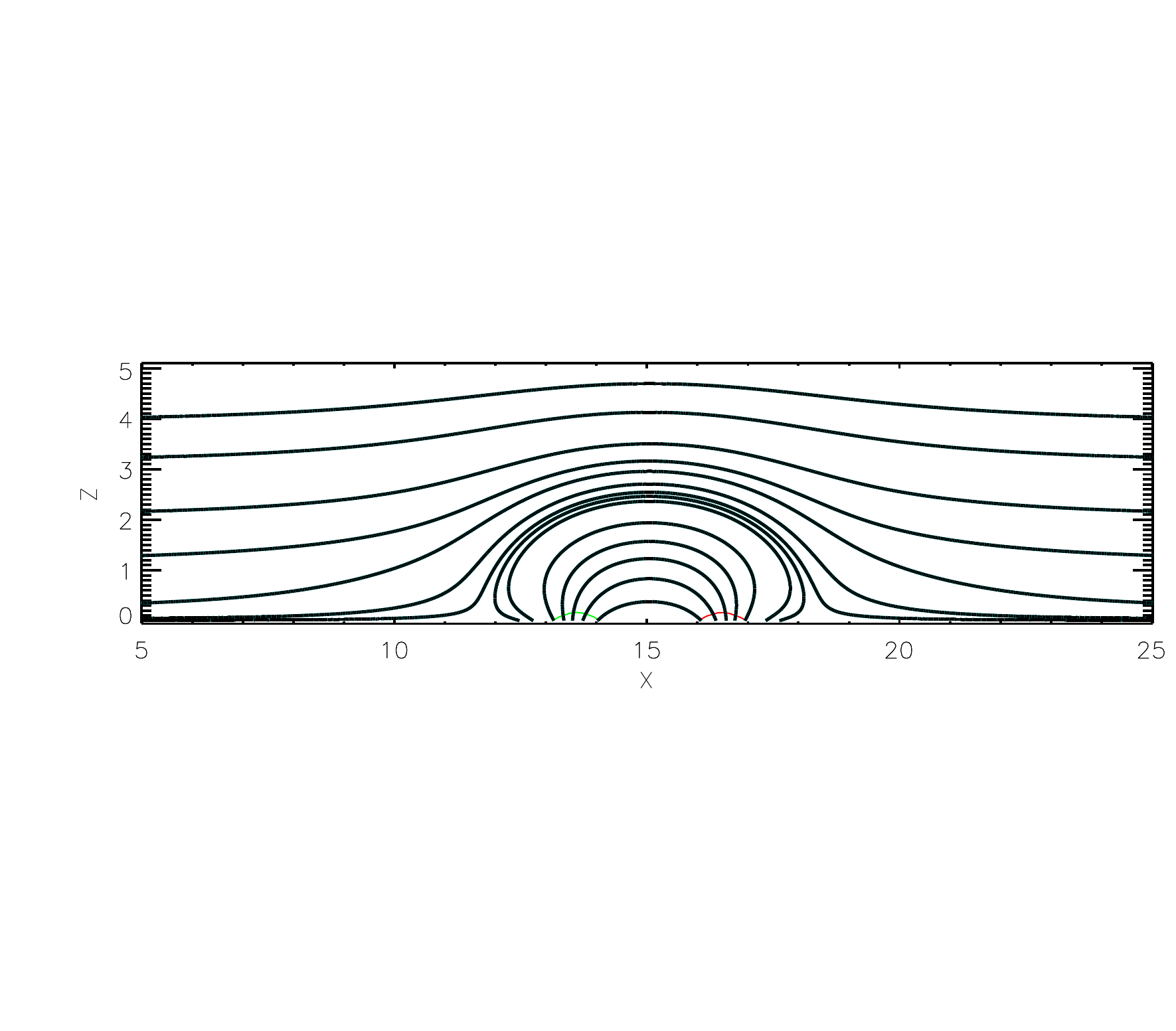}
               \hspace*{0.\textwidth}
               \includegraphics[width=0.45\textwidth,clip=]{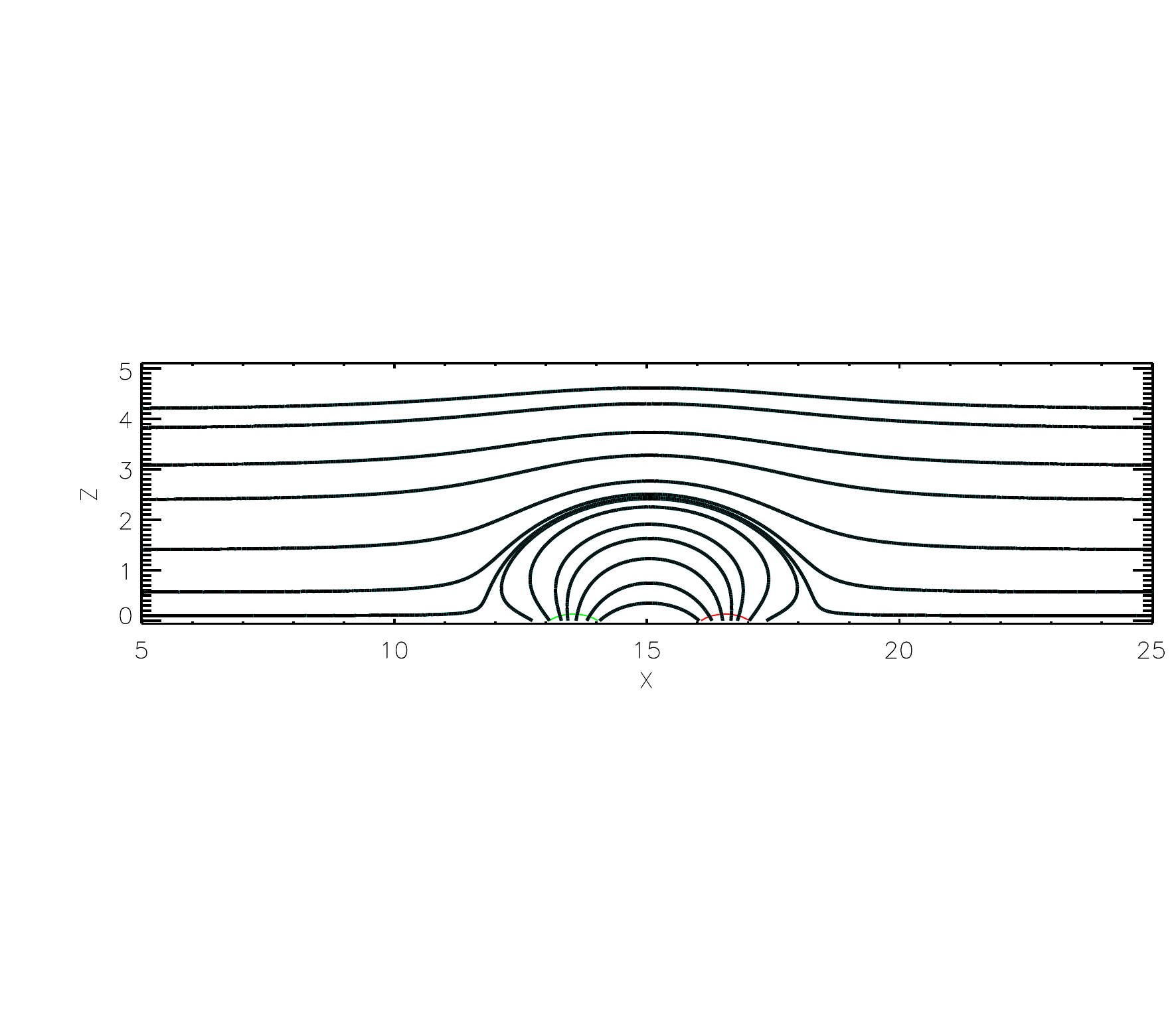}
              }
     \vspace{-0.1\textwidth}   
     \centerline{ \bf     
      \hspace{0.245 \textwidth}  \color{black}{(c)}
      \hspace{0.405\textwidth}  \color{black}{(d)}         \hfill}
     \vspace{0.02\textwidth}    

\caption{Cancellation of a bipole that is aligned parallel to a 5 G overlying field. (a) Non-linear force-free field and (b) potential field as seen in the x-y plane at z=0. (c) Non-linear force-free field and (d) potential field as seen in the x-z plane at y=15. A selection of magnetic field lines originating from the bipole at the photospheric level is plotted in each image. In images (c) and (d), some of the overlying field lines have also been plotted. The images are taken at $t=50$ min. Red and green contours represent positive and negative magnetic field.}\label{fig:cancelfield1}
   \end{figure}

  \begin{figure}
   \centerline{\hspace*{-0.01\textwidth}
               \includegraphics[width=0.45\textwidth,clip=]{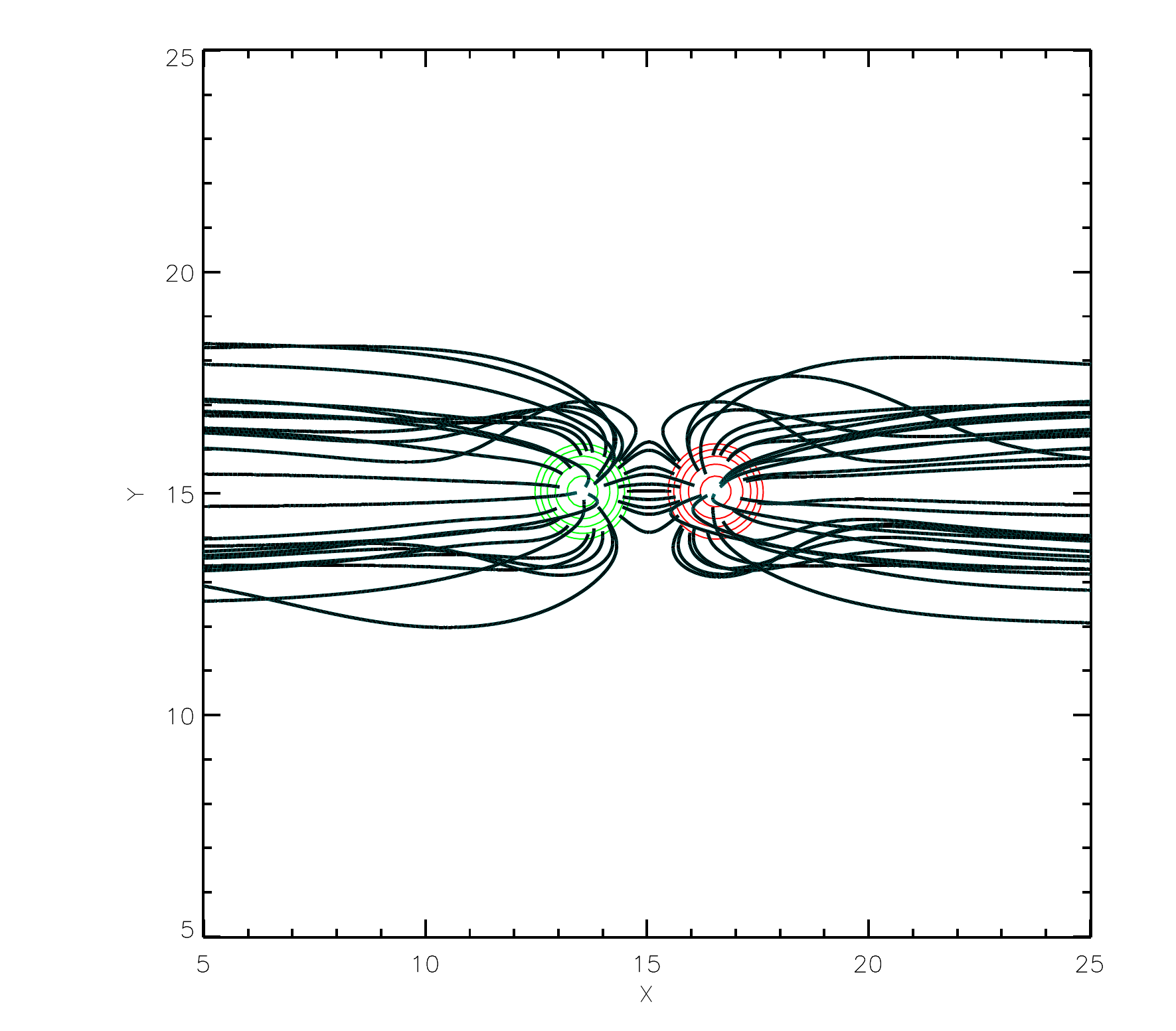}
               \hspace*{0.\textwidth}
               \includegraphics[width=0.45\textwidth,clip=]{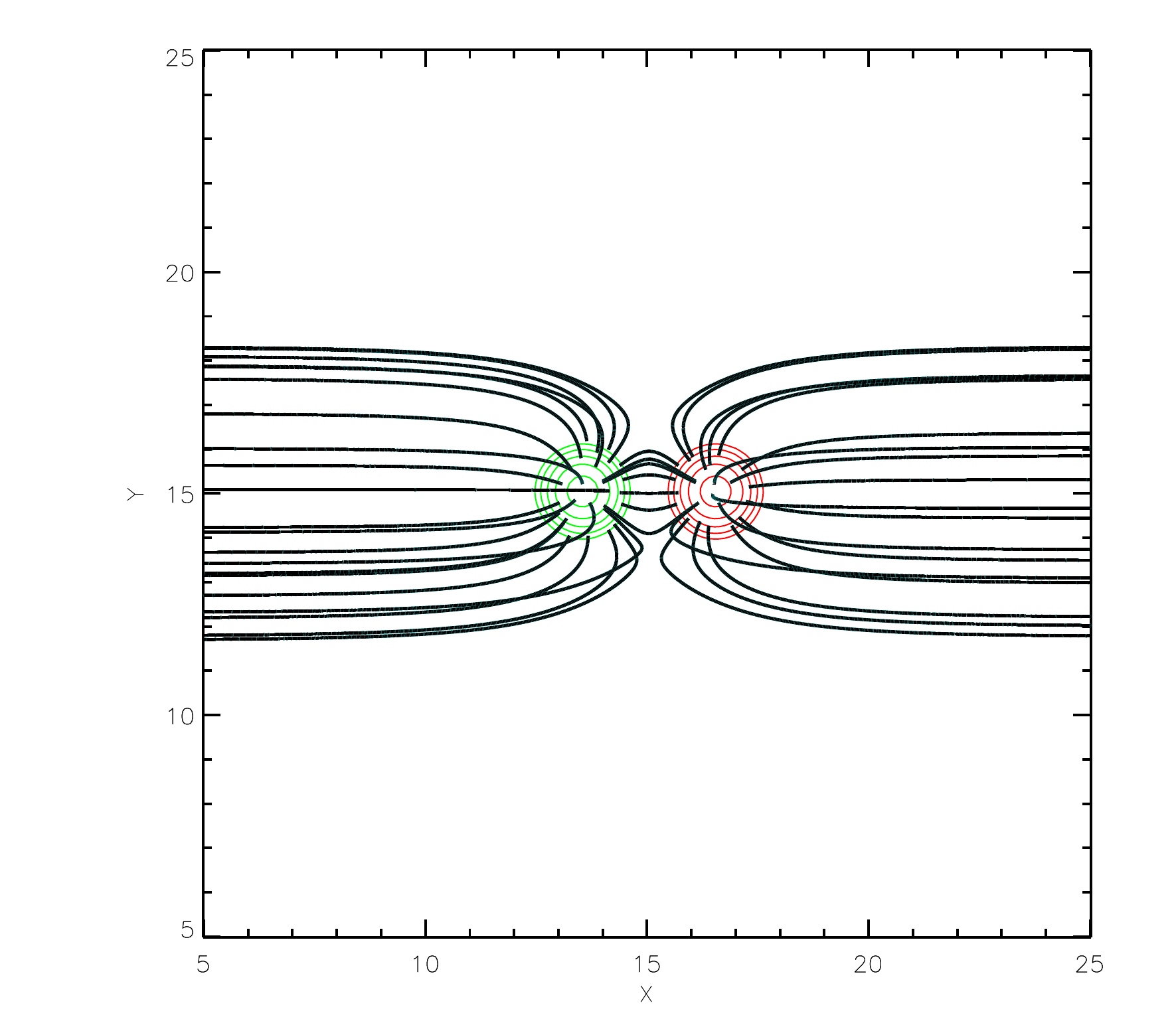}
              }
     \vspace{0.00\textwidth}   
     \centerline{ \bf     
      \hspace{0.245 \textwidth}  \color{black}{(a)}
      \hspace{0.405\textwidth}  \color{black}{(b)}       \hfill}
     \vspace{-0.1\textwidth}    

   \centerline{\hspace*{-0.01\textwidth}
               \includegraphics[width=0.45\textwidth,clip=]{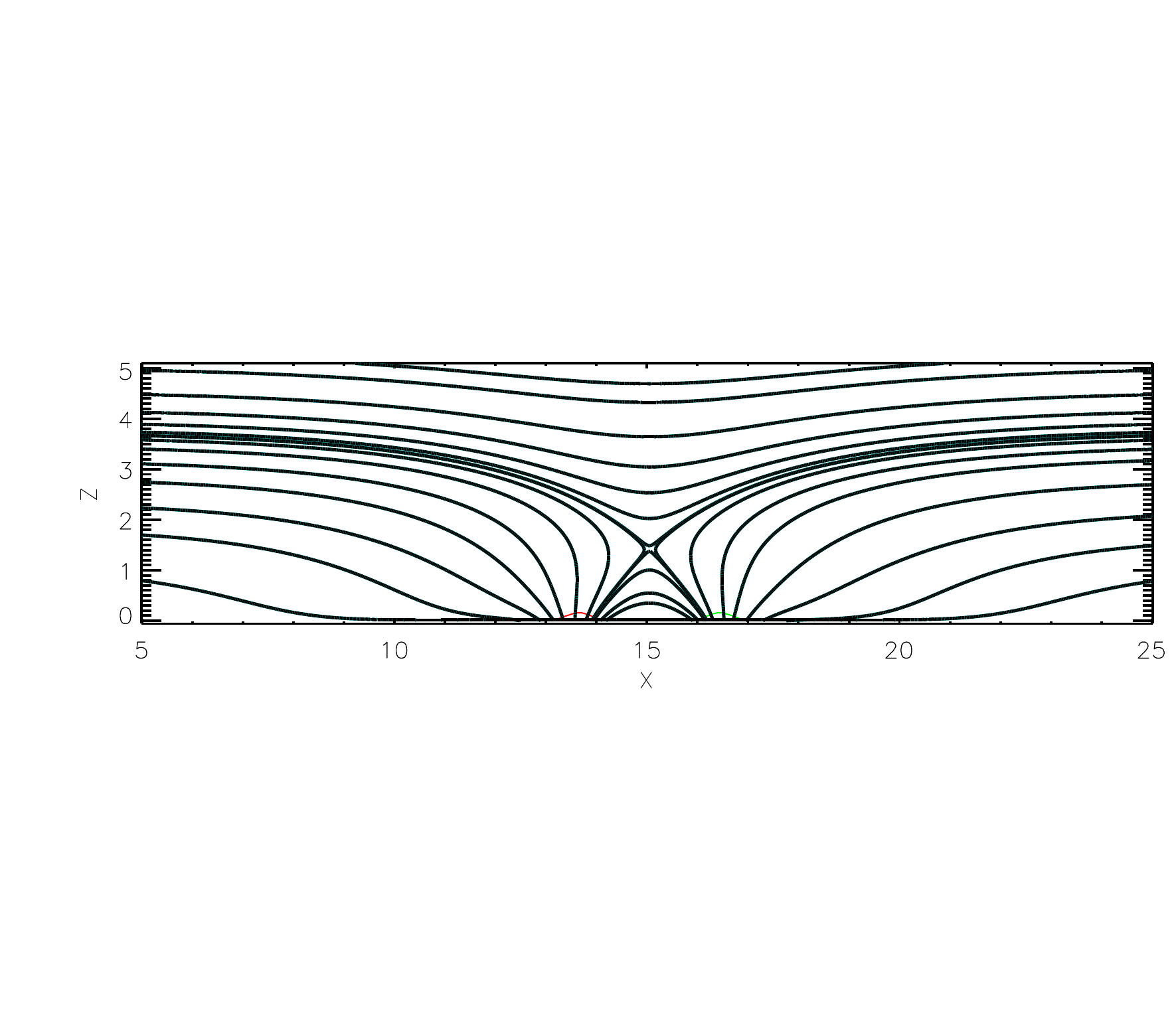}
               \hspace*{0.\textwidth}
               \includegraphics[width=0.45\textwidth,clip=]{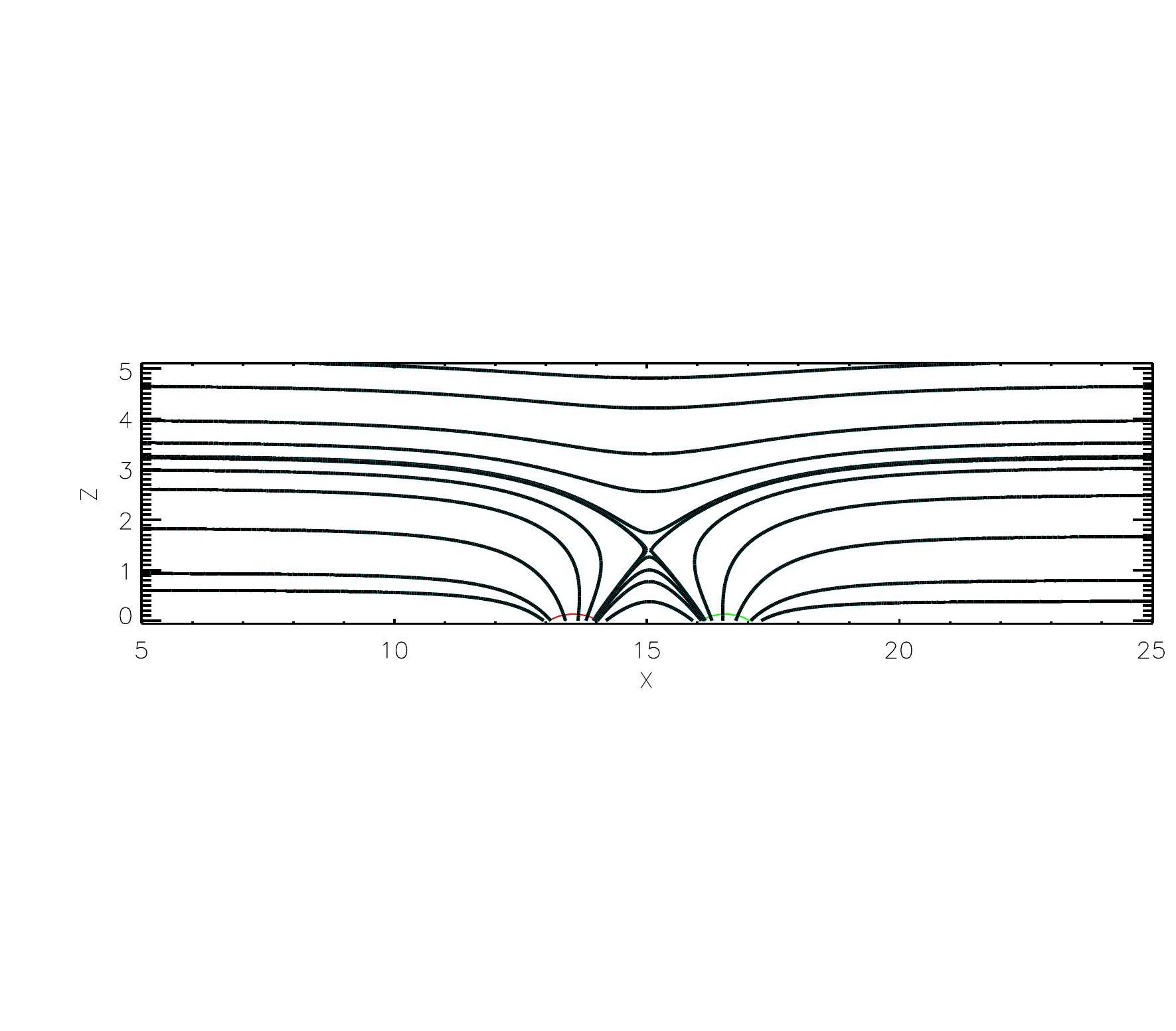}
              }
     \vspace{-0.1\textwidth}   
     \centerline{ \bf     
      \hspace{0.245 \textwidth}  \color{black}{(c)}
      \hspace{0.405\textwidth}  \color{black}{(d)}         \hfill}
     \vspace{0.02\textwidth}    

\caption{Cancellation of a bipole that is aligned anti-parallel to a 5 G overlying field. (a) Non-linear force-free field and (b) potential field as seen in the x-y plane  at z=0. (c) Non-linear force-free field and (d) potential field as seen in the x-z plane at y=15. A selection of magnetic field lines originating from the bipole at the photospheric level is plotted in each image. In images (c) and (d), some of the overlying field lines have also been plotted. The images are taken at $t=50$ min. Red and green contours represent positive and negative magnetic field.}\label{fig:cancelfield2}
   \end{figure}

  \begin{figure}
   \centerline{\hspace*{-0.01\textwidth}
               \includegraphics[width=0.45\textwidth,clip=]{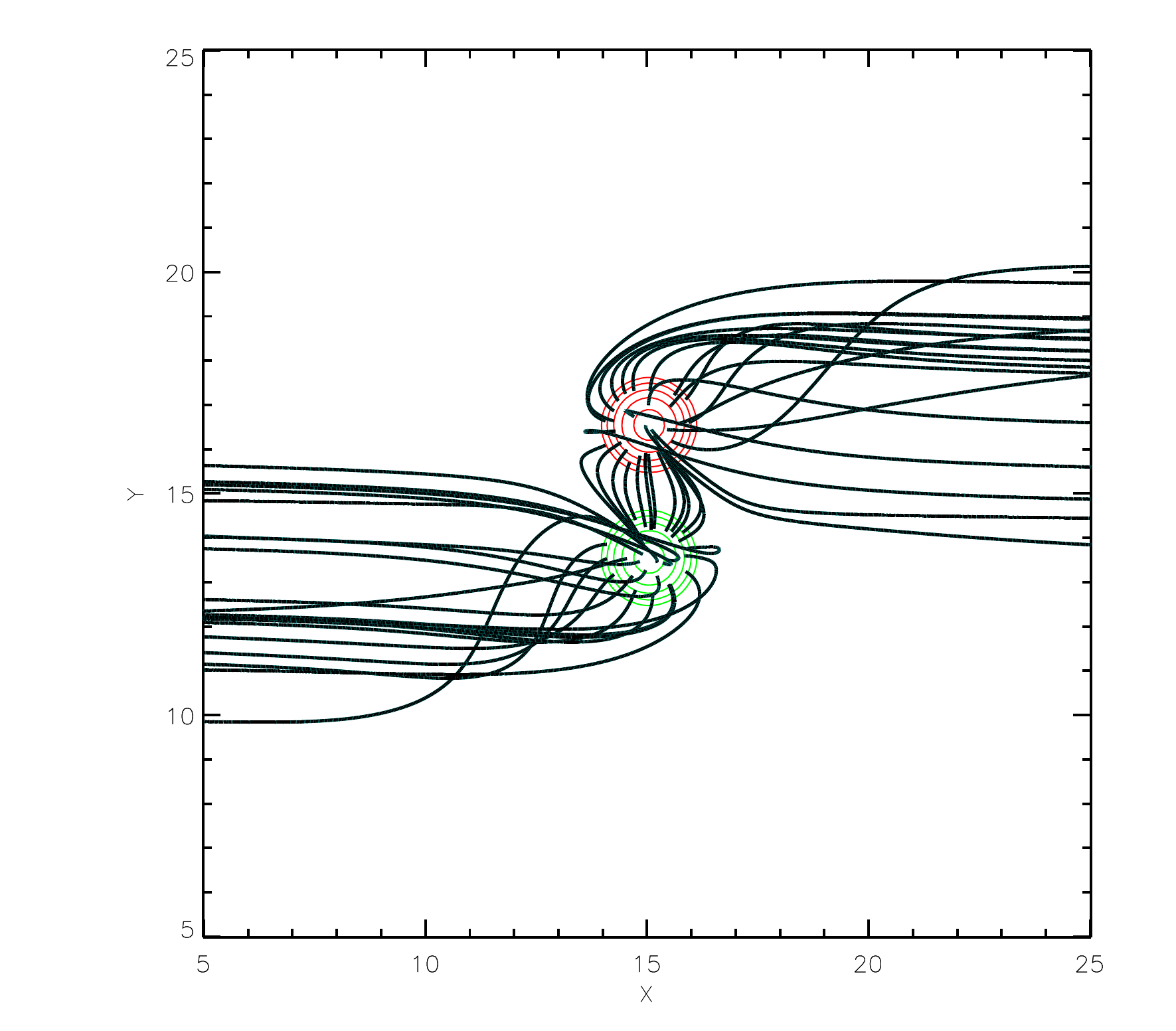}
               \hspace*{0.\textwidth}
               \includegraphics[width=0.45\textwidth,clip=]{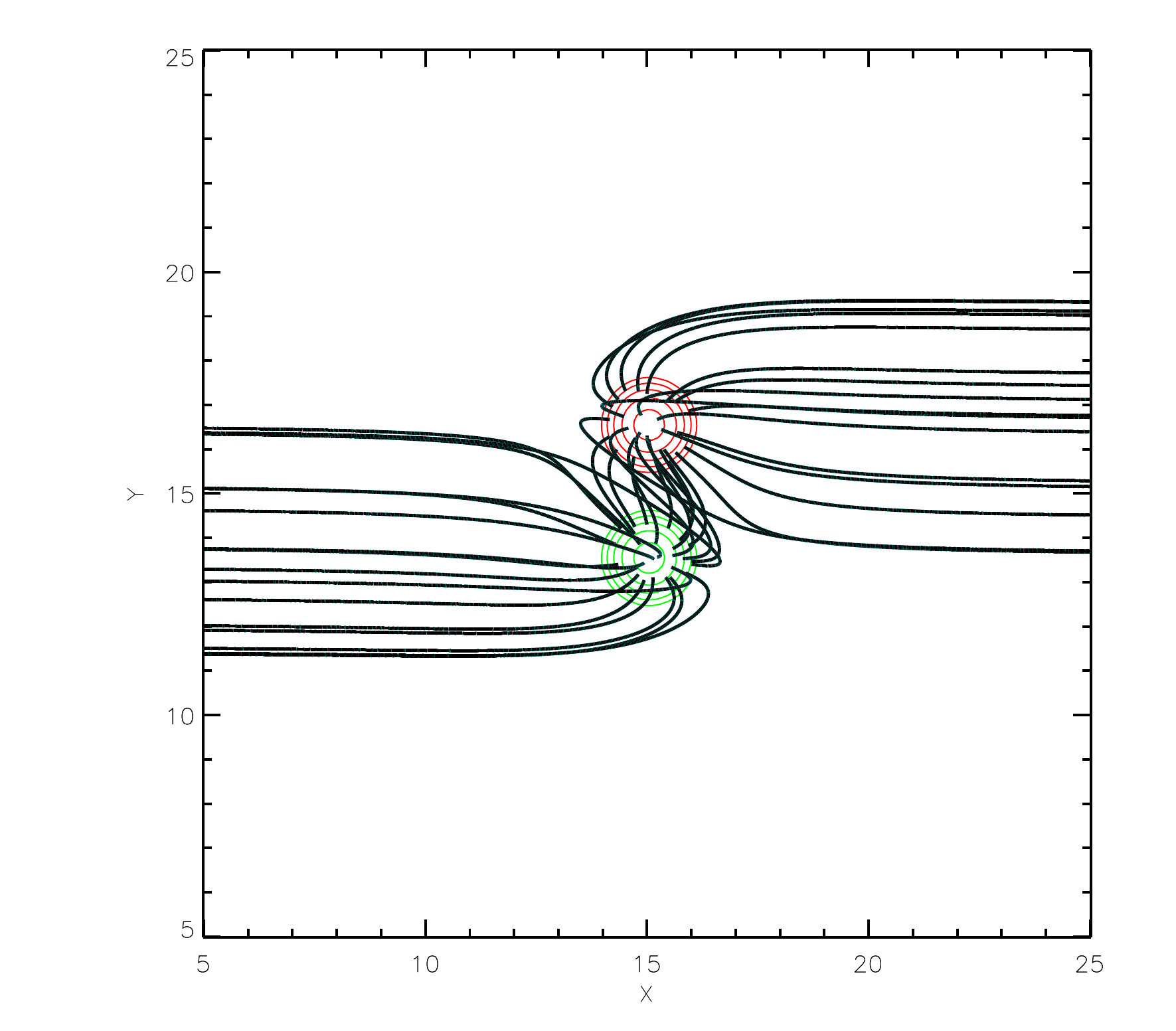}
              }
     \vspace{0.00\textwidth}   
     \centerline{ \bf     
      \hspace{0.245 \textwidth}  \color{black}{(a)}
      \hspace{0.405\textwidth}  \color{black}{(b)}       \hfill}
     \vspace{-0.1\textwidth}    

   \centerline{\hspace*{-0.01\textwidth}
               \includegraphics[width=0.45\textwidth,clip=]{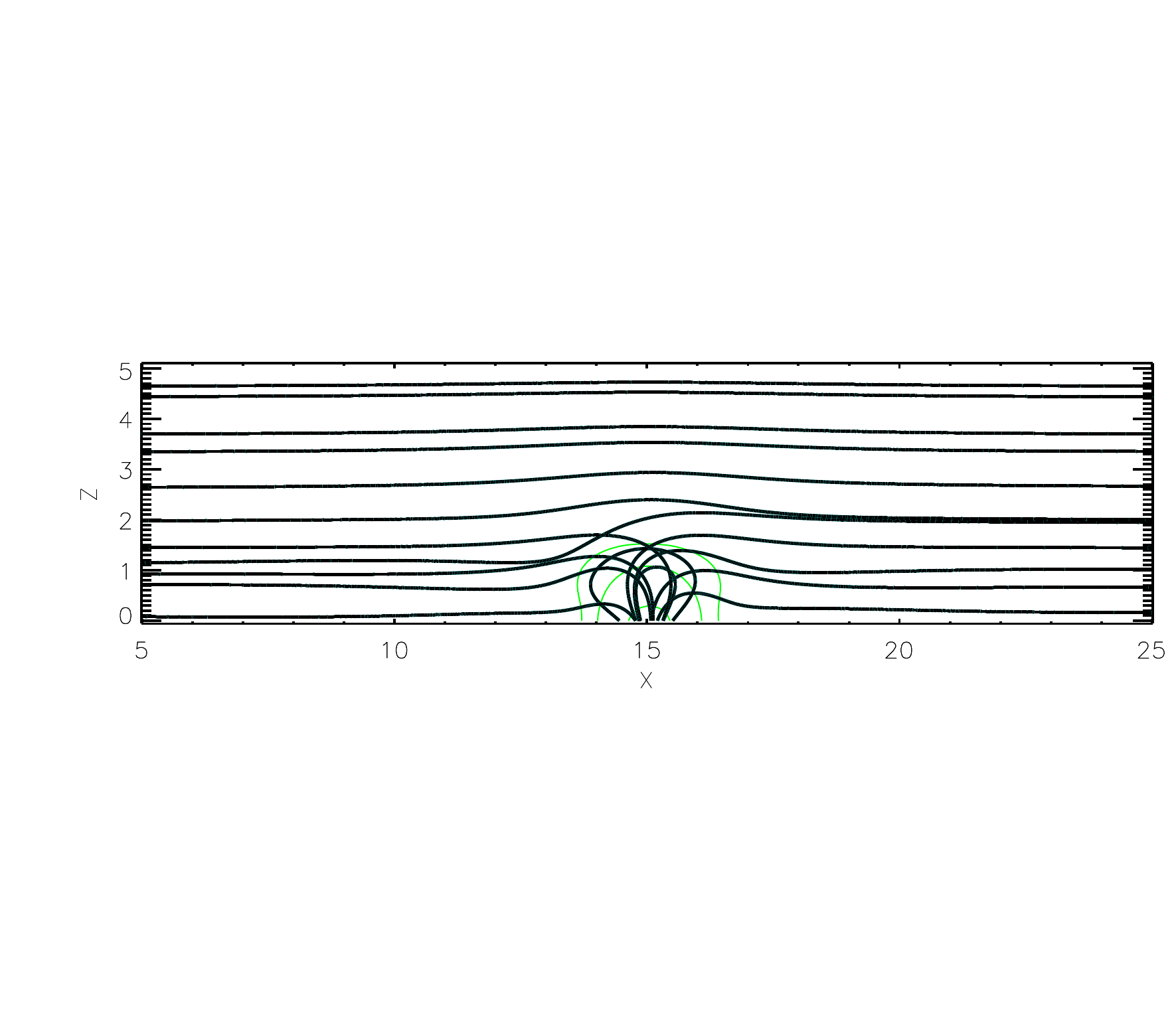}
               \hspace*{0.\textwidth}
               \includegraphics[width=0.45\textwidth,clip=]{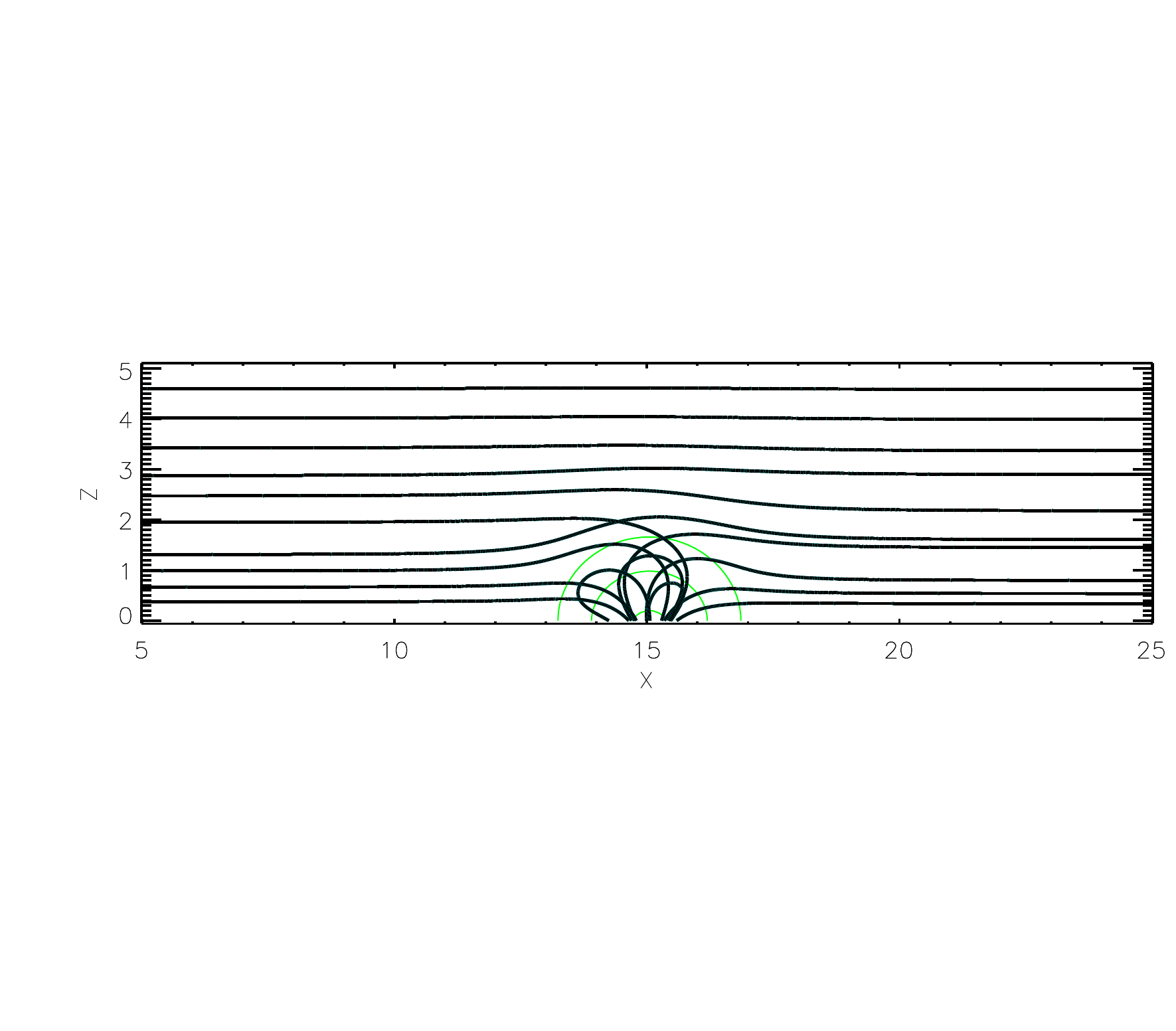}
              }
     \vspace{-0.1\textwidth}   
     \centerline{ \bf     
      \hspace{0.245 \textwidth}  \color{black}{(c)}
      \hspace{0.405\textwidth}  \color{black}{(d)}         \hfill}
     \vspace{0.02\textwidth}    

\caption{Cancellation of a bipole that is aligned perpendicular to a 5 G overlying field. (a) Non-linear force-free field and (b) potential field as seen in the x-y plane at z=0. (c) Non-linear force-free field and (d) potential field as seen in the x-z plane at y=15. A selection of magnetic field lines originating from the bipole at the photospheric level is plotted in each image. In images (c) and (d), some of the overlying field lines have also been plotted. The images are taken at $t=50$ min. Red and green contours represent positive and negative magnetic field.}\label{fig:cancelfield3}
   \end{figure}

Typical examples of the magnetic connectivity during cancellation, for a 5 G overlying field and each orientation of the bipole can be seen in Figures~\ref{fig:cancelfield1} (parallel), \ref{fig:cancelfield2} (anti-parallel) and \ref{fig:cancelfield3} (perpendicular), at $t=50$ min. Images (a) and (c) show the non-linear force-free field. Although the field configurations for each case are only shown at one time, similar configurations occur up until the point at which the magnetic elements cancel.
For parallel cancellation, no matter the strength of the overlying field, all flux from the positive polarity connects to the negative polarity. In contrast, for the anti-parallel and perpendicular cases, as the strength of the overlying field is increased, connectivity between the two magnetic elements decreases.

In Figures~\ref{fig:cancelfield1}, \ref{fig:cancelfield2} and \ref{fig:cancelfield3}, images (b) and (d) show potential field extrapolations at $t=50$ min, for the same photospheric field distribution as in images (a) and (c). For each case, magnetic field lines are plotted from the same starting points as in images (a) and (b). For the parallel cancellation (Figure~\ref{fig:cancelfield1}), the non-linear force-free field lines are similar to those of the potential field. However, for the anti-parallel and perpendicular cancellation, more twisting and bending of the magnetic field lines can be seen. This indicates that for the anti-parallel and perpendicular cases, the non-linear force-free field produces significantly different results.

\subsubsection{Flux Connectivity and Energetics}

\begin{figure}
   \centerline{\hspace*{0.015\textwidth}
               \includegraphics[width=0.45\textwidth,clip=]{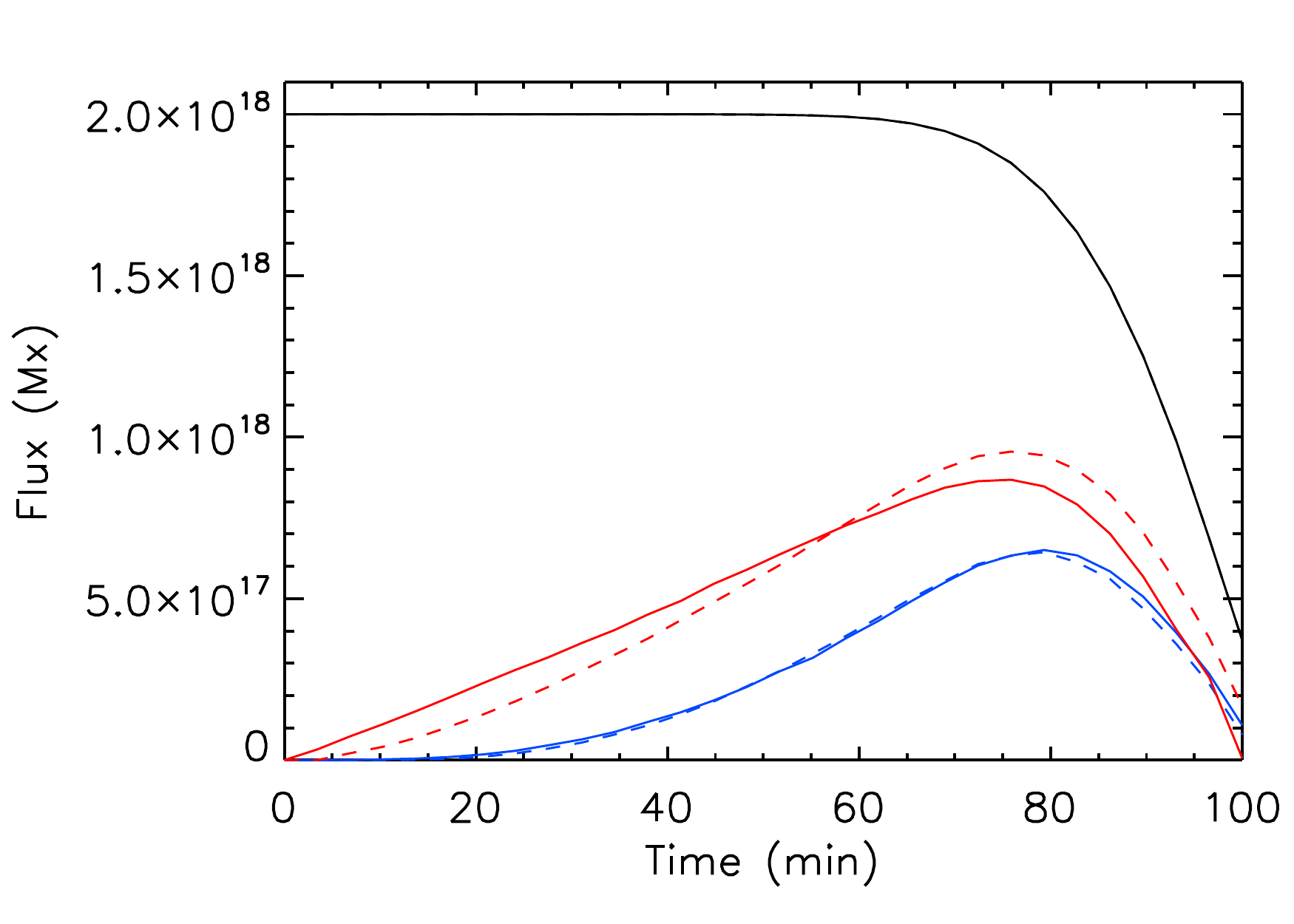}
               \hspace*{-0.0\textwidth}
               \includegraphics[width=0.45\textwidth,clip=]{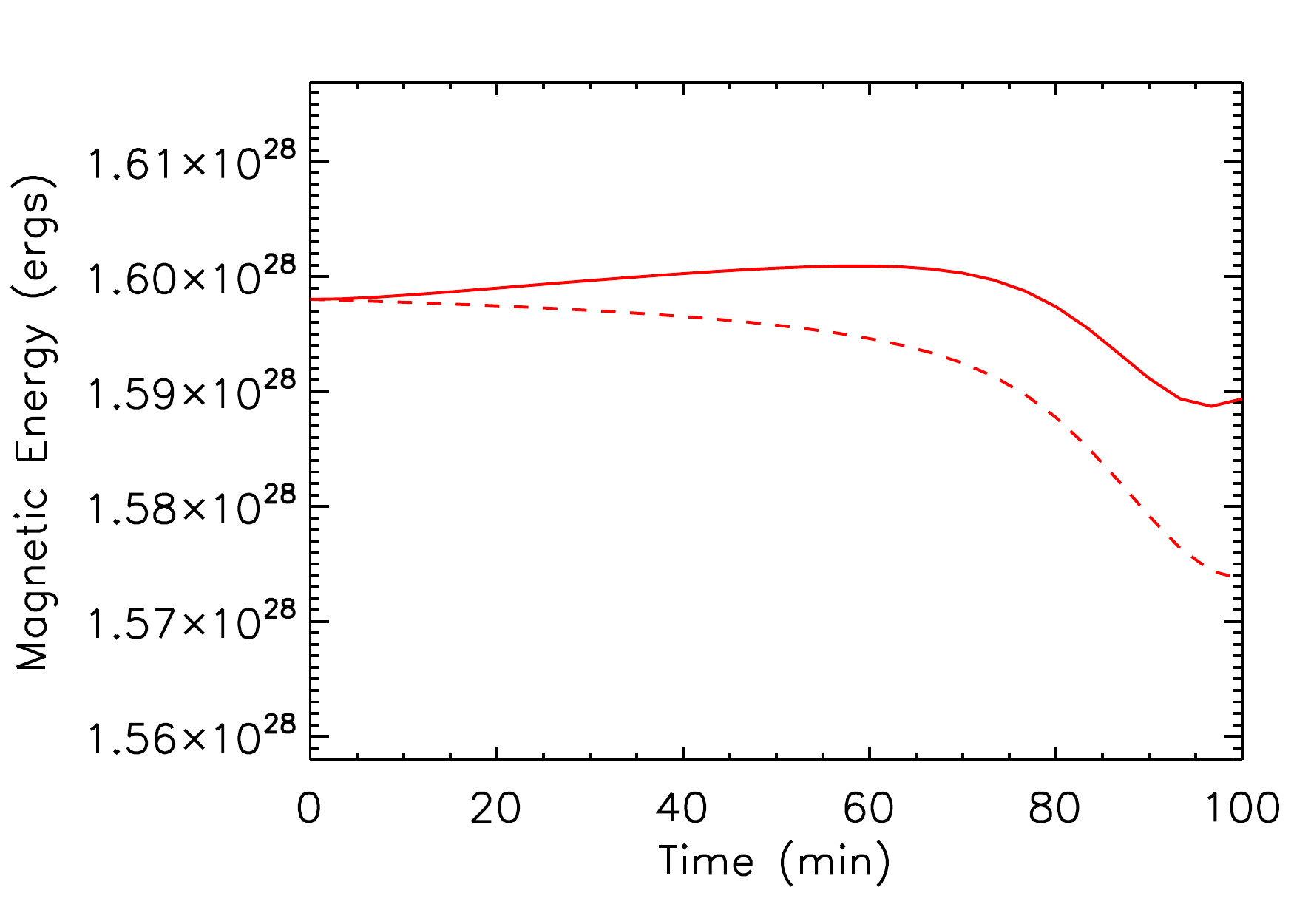}
              }
     \vspace{-0.33\textwidth}   
     \centerline{ \bf     
      \hspace{-0.0 \textwidth}  \color{black}{(a)}
      \hspace{0.42\textwidth}  \color{black}{(b)}
         \hfill}
     \vspace{0.3\textwidth}    
   \centerline{\hspace*{0.015\textwidth}
               \includegraphics[width=0.45\textwidth,clip=]{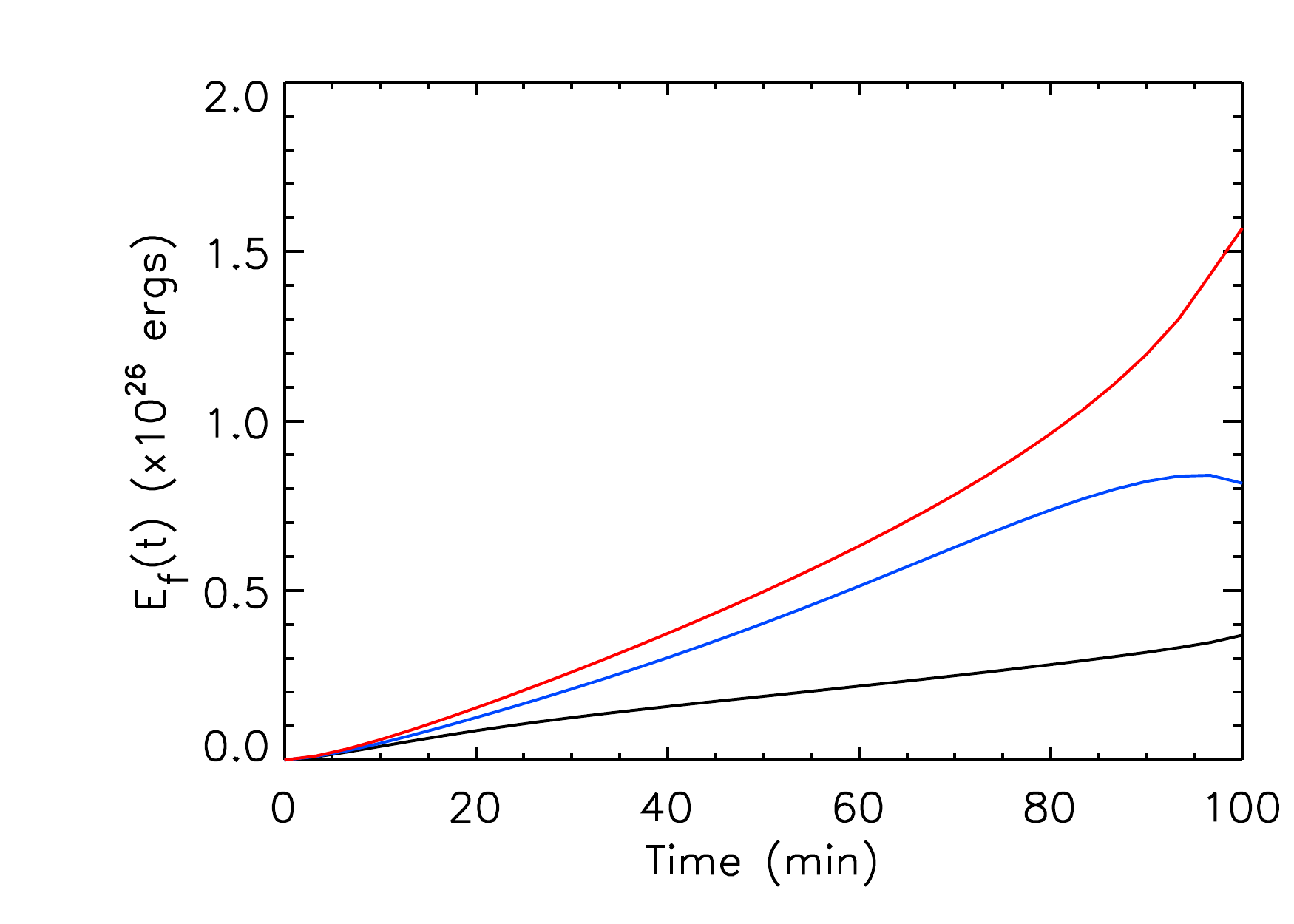}
               \hspace*{-0.0\textwidth}
               \includegraphics[width=0.45\textwidth,clip=]{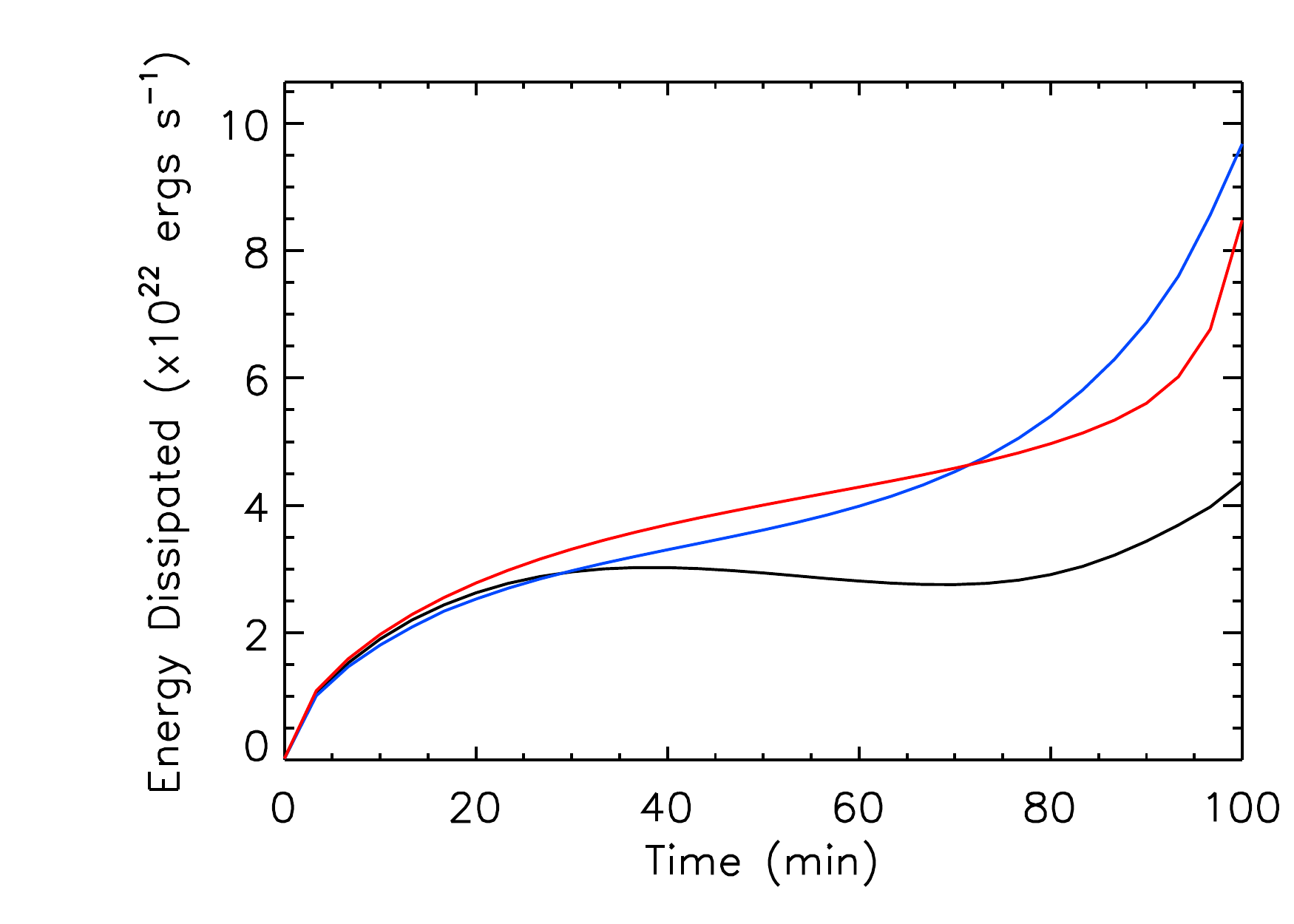}
              }
     \vspace{-0.33\textwidth}   
     \centerline{ \bf     
      \hspace{-0.0 \textwidth}  \color{black}{(c)}
      \hspace{0.42\textwidth}  \color{black}{(d)}
         \hfill}
     \vspace{0.3\textwidth}    
   \centerline{\hspace*{0.015\textwidth}
               \includegraphics[width=0.45\textwidth,clip=]{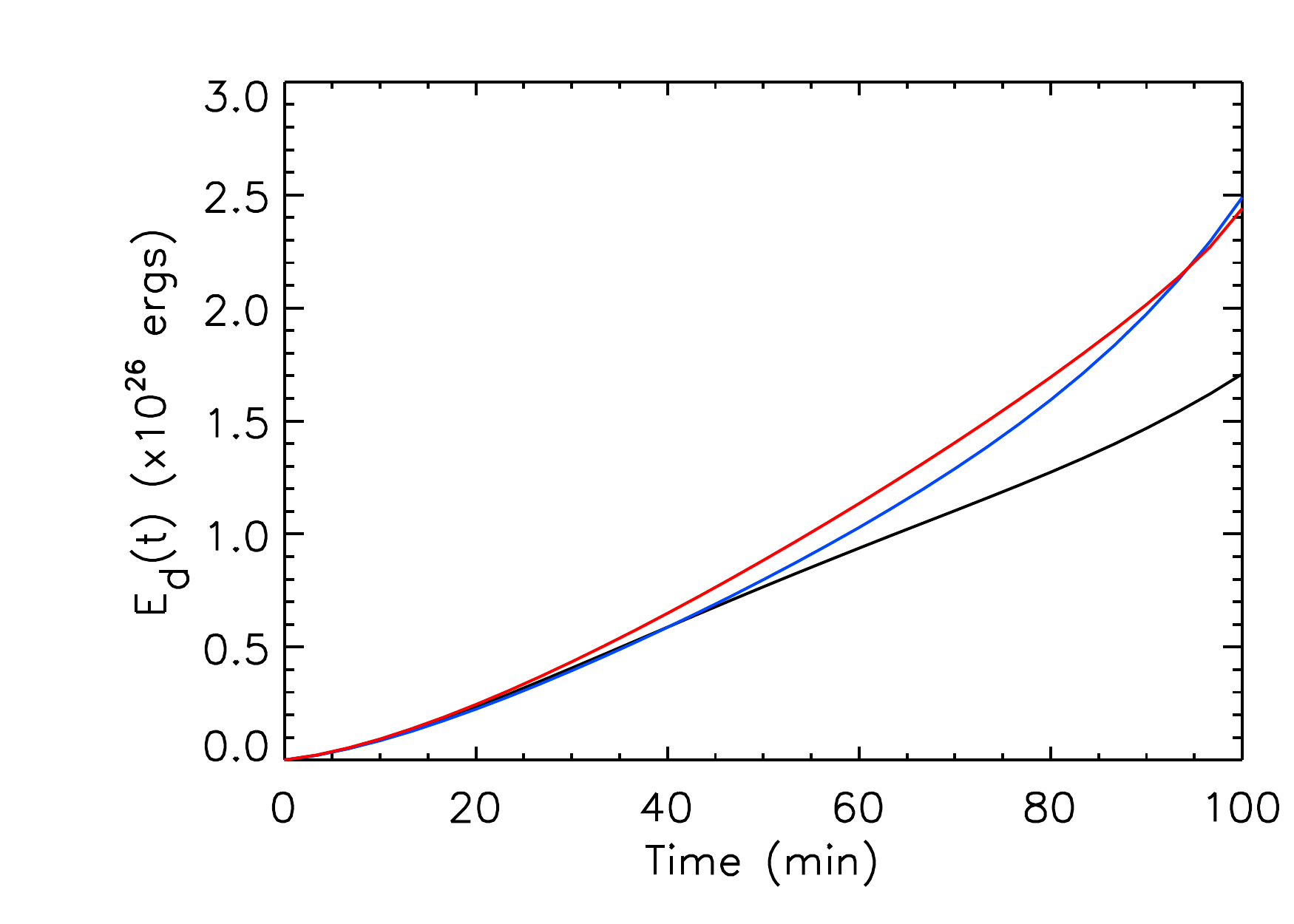}
               \hspace*{-0.0\textwidth}
               \includegraphics[width=0.45\textwidth,clip=]{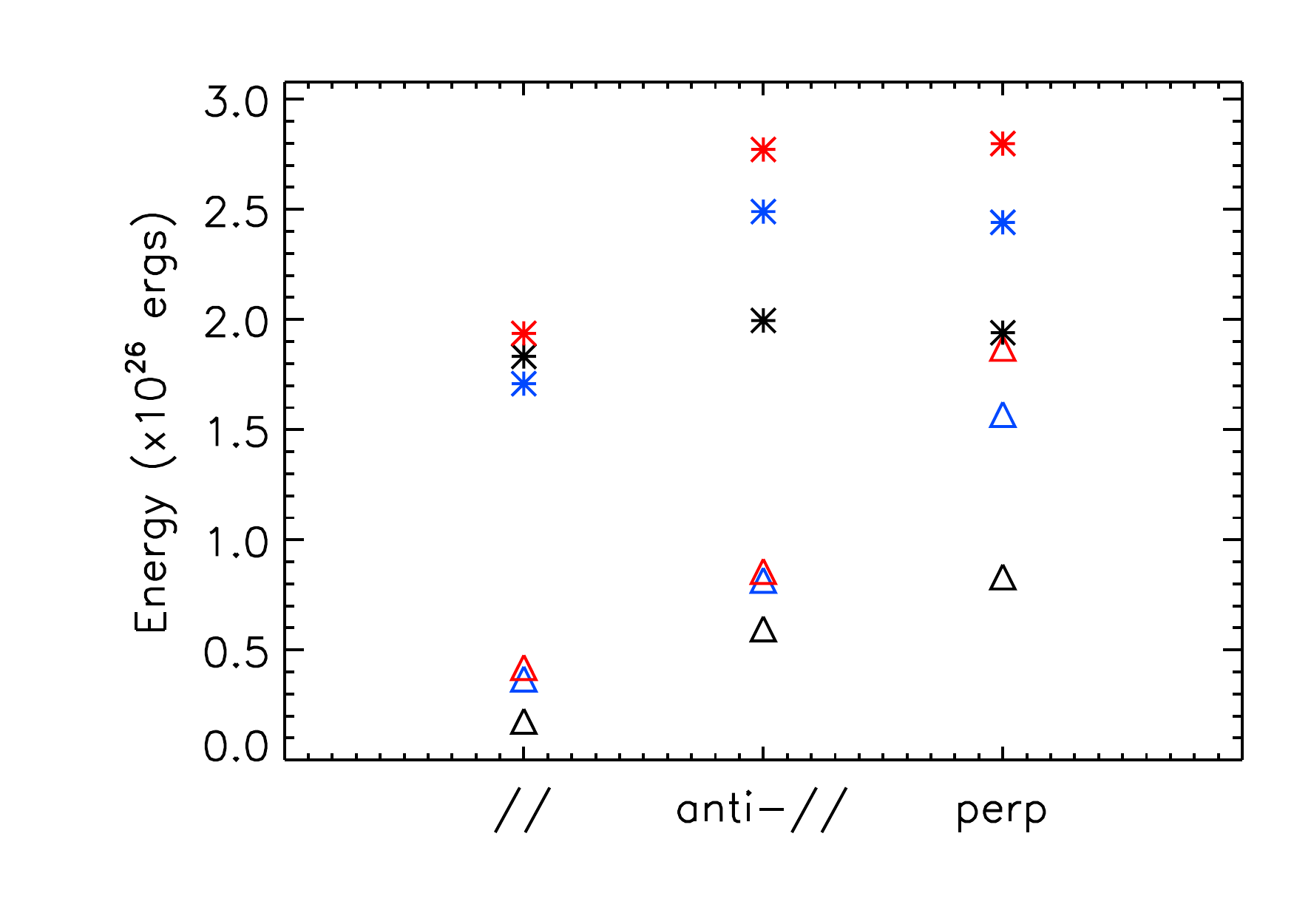}
              }
     \vspace{-0.33\textwidth}   
     \centerline{ \bf     
      \hspace{-0.0 \textwidth} \color{black}{(e)}
      \hspace{0.42\textwidth}  \color{black}{(f)}
         \hfill}
     \vspace{0.33\textwidth}    
              
\caption{Plots as a function of time for a cancelling bipole moving parallel (black), anti-parallel (blue) and perpendicular (red) to a 5 G overlying field, for the non-linear force-free field (solid) and corresponding potential field (dashed). (a) Total flux connecting magnetic elements, (b) total magnetic energy, (c) free magnetic energy, $E_\textrm{f}(t)$, (d) energy dissipated, $\int_V Q dV$, and (e) cumulative energy dissipated, $E_\textrm{d}(t)$. (f) Free magnetic energy (triangles) and cumulative energy dissipated (stars) at the end of each simulation, for a 1 G (black), 5 G (blue) and 10 G (red) overlying field.}\label{fig:cancel1}
   \end{figure}

Figure~\ref{fig:cancel1}(a) shows a plot of the total flux connecting the magnetic elements as a function of time, for parallel (black), anti-parallel (blue) and perpendicular (red) cancellation and a 5 G overlying field. Both the non-linear force-free field (solid lines) and corresponding potential field extrapolation (dashed lines) have been plotted for comparison.
The total flux connecting the magnetic elements does not vary significantly between the non-linear force-free field and potential field cases, because the bipole is simply shrinking from an initially potential state, where many of the properties of the initial potential field are preserved. The most flux connects between the magnetic elements for the parallel case, and the least for the anti-parallel case.
Similar results are found for the 1 G and 10 G overlying field simulations. However, as the overlying field strength increases, the amount of flux connecting the magnetic elements decreases in the anti-parallel and perpendicular cases.

Figure~\ref{fig:cancel1}(b) shows the total magnetic energy as a function of time, for the perpendicular cancellation and a 5 G overlying field. The red solid line shows the non-linear force-free field energy, while the red dashed line shows the energy of the corresponding potential field. The non-linear force-free field energy initially increases as energy is injected due to surface motions, whereas the potential field energy is continually decreasing. However, both curves decrease as the magnetic elements begin to cancel, and there is an outflow of energy through the photospheric boundary. There is a slight increase in the non-linear force-free field energy towards the end of the simulation, as strongly curved overlying field lines form during the final stages of cancellation of the bipole. In contrast, the overlying field is completely straight in the final potential field.

Figure~\ref{fig:cancel1}(c) shows the variation of the free magnetic energy, $E_\textrm{f}(t)$, as a function of time, as computed using Equation~\ref{eqn:free}. Results are shown for the 5 G overlying field, and the line colours show the same orientations as in Figure~\ref{fig:cancel1}(a). The free energy stored ranges from $0.37-1.57\times 10^{26}$ ergs.
The parallel cancellation results in the least free energy, while the perpendicular cancellation results in the most.
On comparing Figures~\ref{fig:cancel1}(b) and (c), one can see that the free magnetic energy built up within each simulation is small compared to the total energy within the box (around 1\% for the 5 G perpendicular case). However, in order to avoid boundary effects, the computational box is large compared to the magnetic elements and their area of evolution. Therefore the evolution of the magnetic elements only perturbs a small volume of the overlying field, and the majority of the overlying field remains in a near potential state.
If we compare the free magnetic energy built up to the potential field energy contribution of the bipole, we see that the build up of free energy is significant. From Figure~\ref{fig:cancel1}(b), we compute the potential energy contribution of the bipole to be $2.43\times10^{26}$ ergs. This is of similar order to the free energy stored in the 5 G perpendicular case.

In addition to free magnetic energy stored within the system, energy is continually being dissipated, as described by the heating term $Q$ (Equation~\ref{eqn:q}). This is illustrated by Figure~\ref{fig:cancel1}(d), which shows $Q$ integrated over the volume at a given instant in time. Again, results are shown for the parallel (black), anti-parallel (blue) and perpendicular (red) cancellation, with a 5 G overlying field. A greater quantity of energy is dissipated towards the end, when full cancellation occurs. More energy is dissipated in the anti-parallel and perpendicular cases than in the parallel case. This is because smaller gradients in $\alpha$ are present in the parallel case as the bipole is simply shrinking and no reconnection occurs with the overlying field.

  \begin{figure}
   \centerline{\hspace*{-0.01\textwidth}
               \includegraphics[width=0.45\textwidth,clip=]{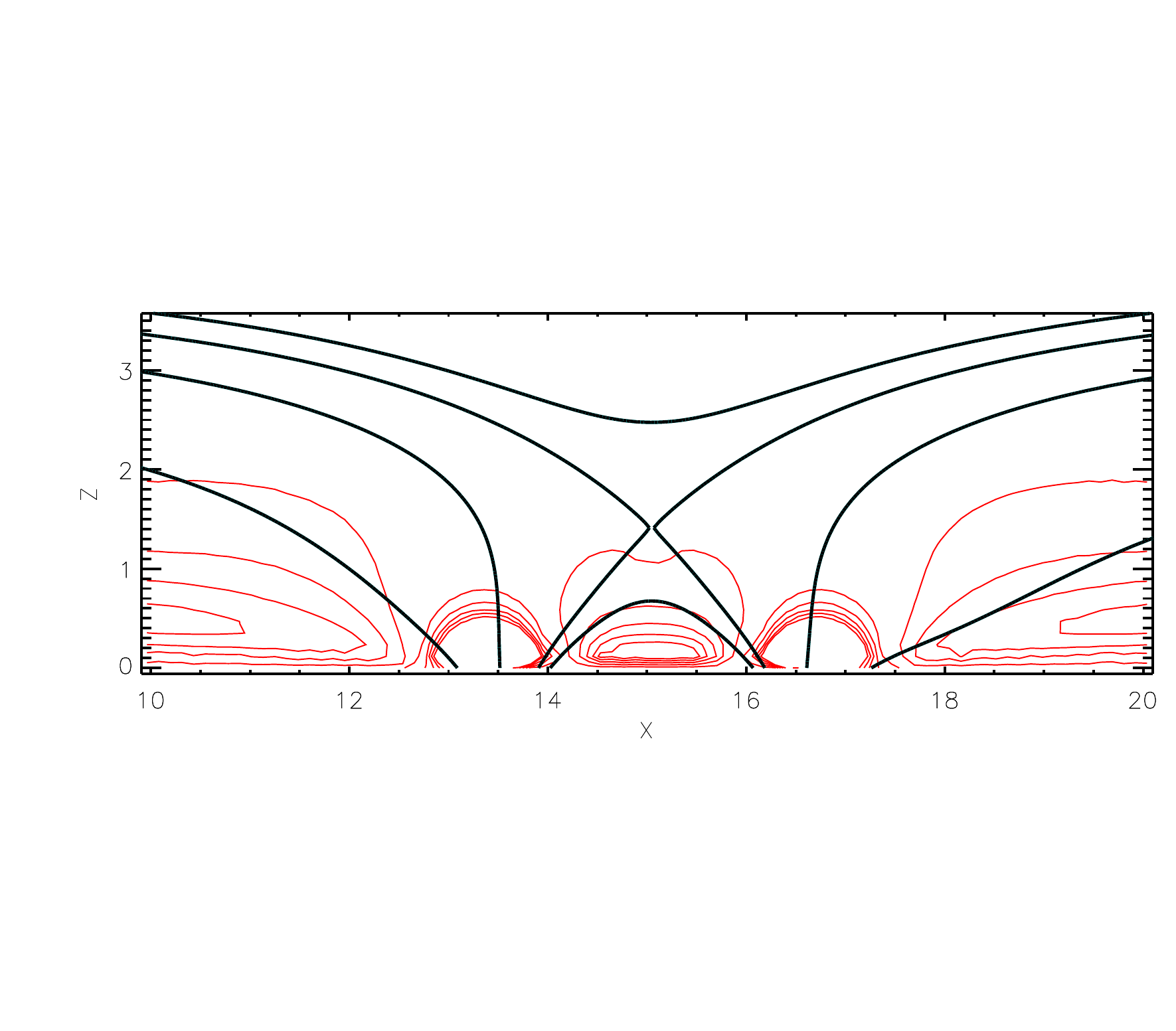}
               \hspace*{0.\textwidth}
               \includegraphics[width=0.45\textwidth,clip=]{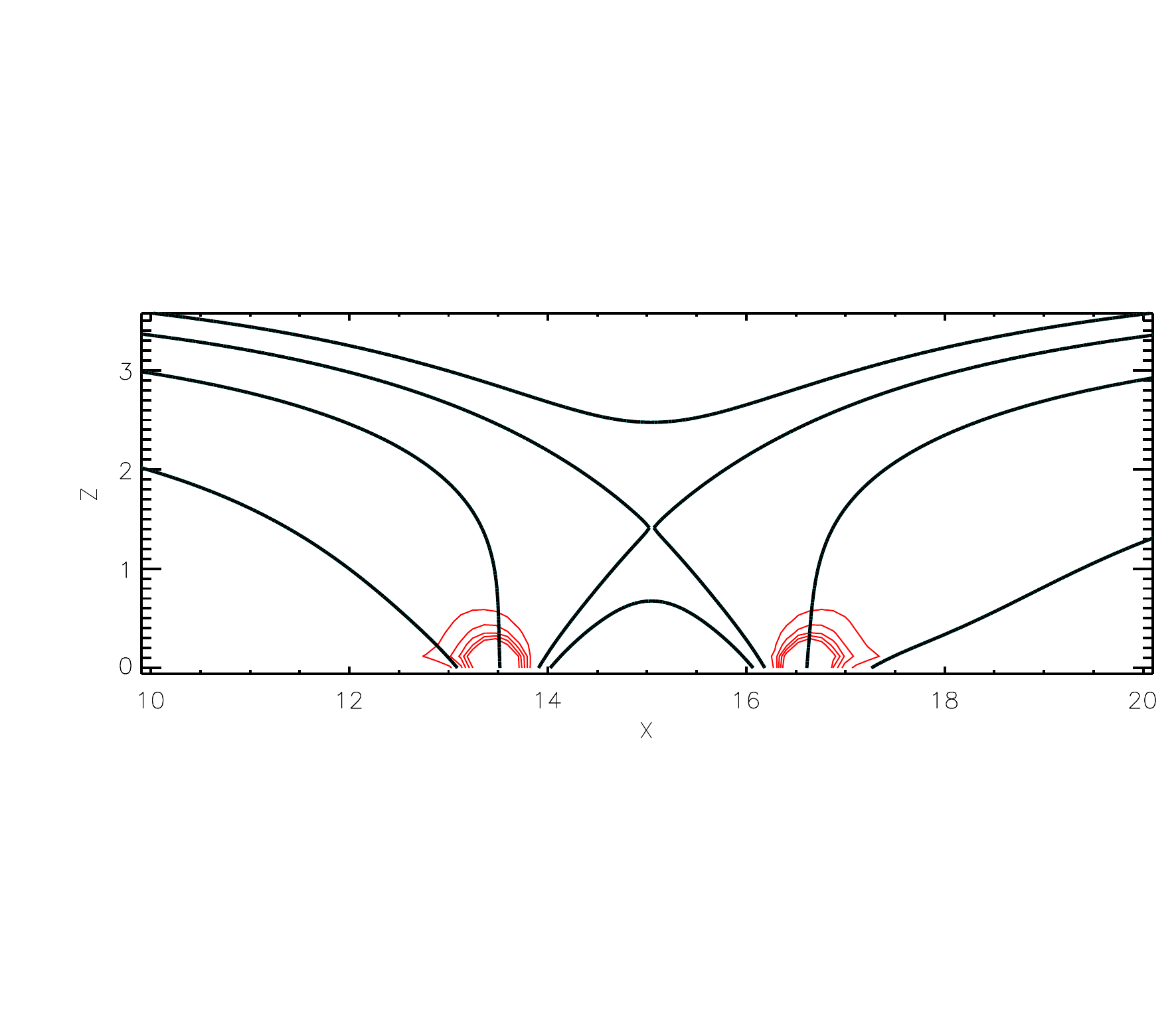}
              }
     \vspace{-0.1\textwidth}   
     \centerline{ \bf     
      \hspace{0.245 \textwidth}  \color{black}{(a)}
      \hspace{0.405\textwidth}  \color{black}{(b)}       \hfill}
     \vspace{-0.1\textwidth}    

   \centerline{\hspace*{-0.01\textwidth}
               \includegraphics[width=0.45\textwidth,clip=]{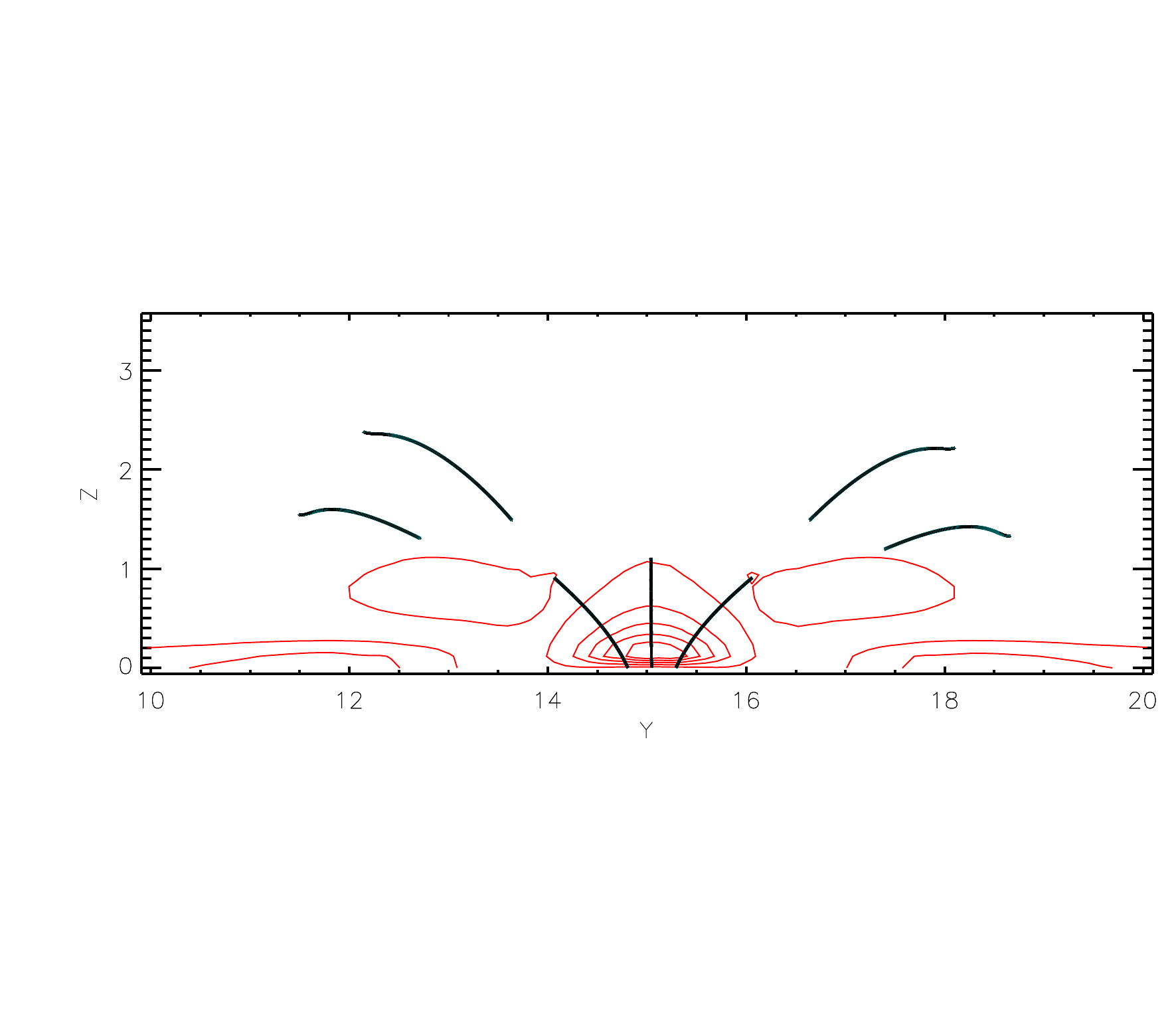}
               \hspace*{0.\textwidth}
               \includegraphics[width=0.45\textwidth,clip=]{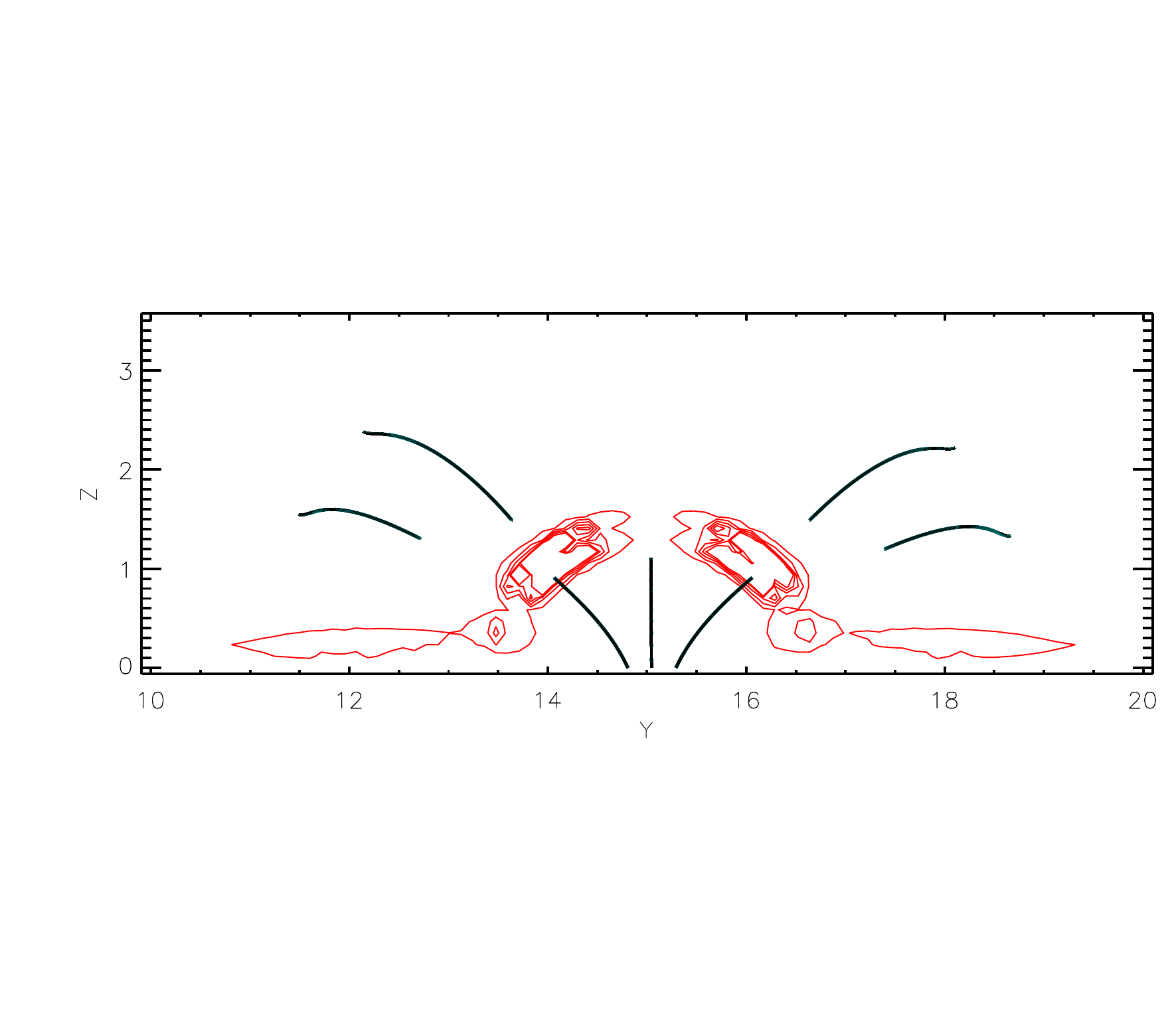}
              }
     \vspace{-0.1\textwidth}   
     \centerline{ \bf     
      \hspace{0.245 \textwidth}  \color{black}{(c)}
      \hspace{0.405\textwidth}  \color{black}{(d)}         \hfill}
     \vspace{0.02\textwidth}    

\caption{Cancellation of a bipole aligned anti-parallel to a 5 G overlying field: normalised contours of (a) $Q_\textrm{\tiny frc}$ and (b) $Q_\textrm{\tiny hd}$ as seen in the $x-z$ plane at $y=15$ Mm, and normalised contours of (c) $Q_\textrm{\tiny frc}$ and (d) $Q_\textrm{\tiny hd}$ as seen in the $y-z$ plane at $x=15$ Mm. In (a), (b) and (c), contour levels are $[0.9, 0.7, 0.5, 0.3, 0.1]$; in (d) contour levels are 200 times less $[0.0045, 0.0035, 0.0025, 0.0015, 0.0005]$. A selection of magnetic field lines originating from the bipole and the overlying field is plotted in each image.}\label{fig:cancel_q}
   \end{figure}

The dissipated energy may be split into two terms representing energy dissipation due to magnetofriction, $Q_\textrm{\tiny frc}$, and hyperdiffusion, $Q_\textrm{\tiny hd}$, where
\begin{displaymath}
 Q_\textrm{\tiny frc}=\frac{B^2}{4 \pi}\nu |\mathbf{v}|^2 \quad \textrm{and} 
    \quad Q_\textrm{\tiny hd}=\frac{B^2}{4 \pi} \eta_4 |\nabla\alpha |^2.
\end{displaymath}
Figure~\ref{fig:cancel_q} shows normalised contours of $Q_\textrm{\tiny frc}$ and $Q_\textrm{\tiny hd}$ in the $x-z$ plane (Figures~\ref{fig:cancel_q}(a) and (b)) and the $y-z$ plane (Figures~\ref{fig:cancel_q}(c) and (d)) taken at $t=50$ min for the anti-parallel cancellation with a 5 G overlying field. In each case, the plots are taken through the centre of the bipole. For (a) - (c), the contours are at the same levels, in (d) they are 200 times lower.
Considering Figures~\ref{fig:cancel_q}(a) and (c), it can be seen that $Q_\textrm{\tiny frc}$ occurs throughout a large part of the coronal volume, both along field lines connecting the magnetic elements, and along those connecting the magnetic elements to the overlying field. In general, the term is largest low down, where the field lines are perturbed by the motion of the magnetic elements. Also, in Figure~\ref{fig:cancel_q}(c), two `wings' of $Q_\textrm{\tiny frc}$ can be seen suspended in the coronal volume. These originate due to strongly curved field lines that reconnect between the bipole and the overlying field.

In contrast, from Figures~\ref{fig:cancel_q}(b) and (d), it can be seen that $Q_\textrm{\tiny hd}$ is more localised than $Q_\textrm{\tiny frc}$; the reason is that $Q_\textrm{\tiny hd}$ only arises where gradients in $\alpha$ occur. This happens mainly at two locations: first, near the footpoints of the field lines connecting the magnetic elements, and secondly around the separatrix surface which separates the overlying field lines from the closed connections between the magnetic elements.
In all of the simulations discussed in this paper, we see a similar trend as to where energy dissipation occurs. In each case, $Q_\textrm{\tiny frc}$ is seen to be space filling, whereas $Q_\textrm{\tiny hd}$ is more localised.

By comparing Figures~\ref{fig:cancel1}(c) and (d), we see that $\int_V Q dV$ (expressed in ergs s$^{-1}$) is around three orders of magnitude smaller than the free energy (expressed in ergs) stored by the end of the simulation.
Although it is three orders of magnitude smaller, the values shown in Figure~\ref{fig:cancel1}(d) are instantaneous values. By integrating $Q$ over both the volume and time, we can see how much energy has been cumulatively dissipated over the whole simulation:
\begin{equation}\label{eqn:edt}
 E_\textrm{d}(t)=\int_0^t \bigg[\int_V Q dV \bigg] dt.
\end{equation}
A plot of the total energy dissipated as a function of time is shown in Figure~\ref{fig:cancel1}(e), for the 5 G case of each orientation (lines are coloured as in \ref{fig:cancel1}(a)). Cumulatively, a significant amount of energy has been dissipated. By comparing Figures~\ref{fig:cancel1}(c) and (e) it can be seen that by the end of the simulation, more energy is cumulatively released ($1.71-2.49\times 10^{26}$ ergs) than is stored as free magnetic energy.

Although Figures~\ref{fig:cancel1}(a)-(e) only show results for the 5 G overlying field cases, Figure~\ref{fig:cancel1}(f) compares the values of free energy (triangles) and total energy dissipated (stars) at the end of the 1 G (black), 5 G (blue) and 10 G (red) simulations. The highest value of free energy is found for a 10 G overlying field, and perpendicular cancellation. In general, higher free energy is found for a stronger overlying field and when a greater volume of the overlying field is disturbed (i. e. perpendicular cancellation). We also see that a stronger overlying field results in more energy dissipation. For all overlying field strengths, parallel cancellation results in the least energy dissipation, however, a similar cumulative amount of energy dissipation occurs for both the anti-parallel and perpendicular cancellation.

\subsection{Emergence}\label{emer}

\begin{figure}
 \begin{center}
  \includegraphics[width=0.7\textwidth]{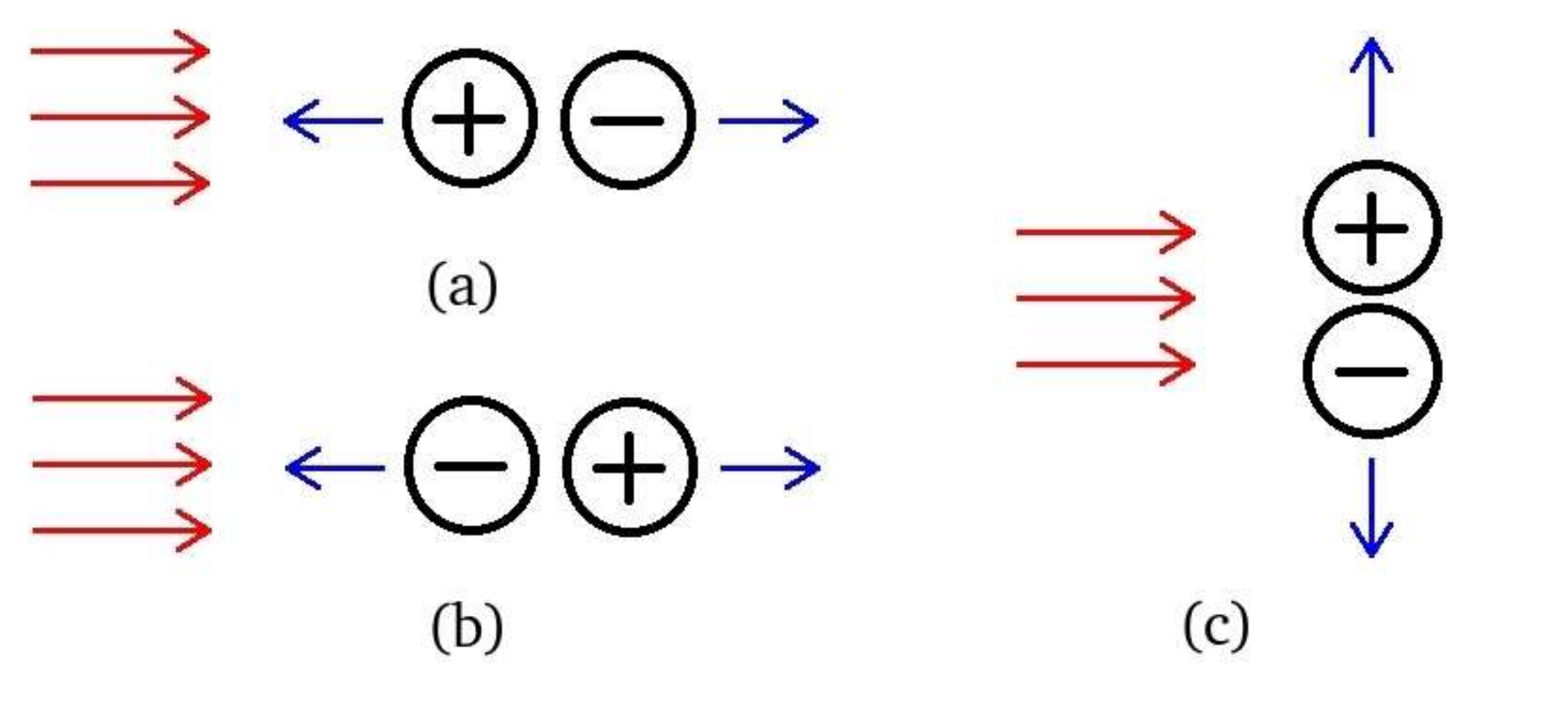}
 \end{center}
\caption{Emergence: The red arrows represent the direction of the overlying field, the blue arrows represent the direction of motion of the magnetic elements. The bipole's axis is oriented (a) parallel to,  (b) anti-parallel to, and (c) perpendicular to the overlying field.}\label{fig:emerge}
\end{figure}

In the emergence simulations, initially, the net flux through the photosphere is zero. The magnetic elements, which coincide, then move apart until they each reach a separation distance of 3 Mm from the box midpoint after 100 min. As a result, a bipole appears in the photospheric distribution, simulating what can be classed in photospheric magnetograms as an emergence. The blue line in Figure~\ref{fig:flux} shows the total absolute flux through the photosphere as a function of time for the emerging bipole. Figure~\ref{fig:emerge} illustrates the initial set-up for each of the simulations.

  \begin{figure}
   \centerline{\hspace*{-0.01\textwidth}
               \includegraphics[width=0.45\textwidth,clip=]{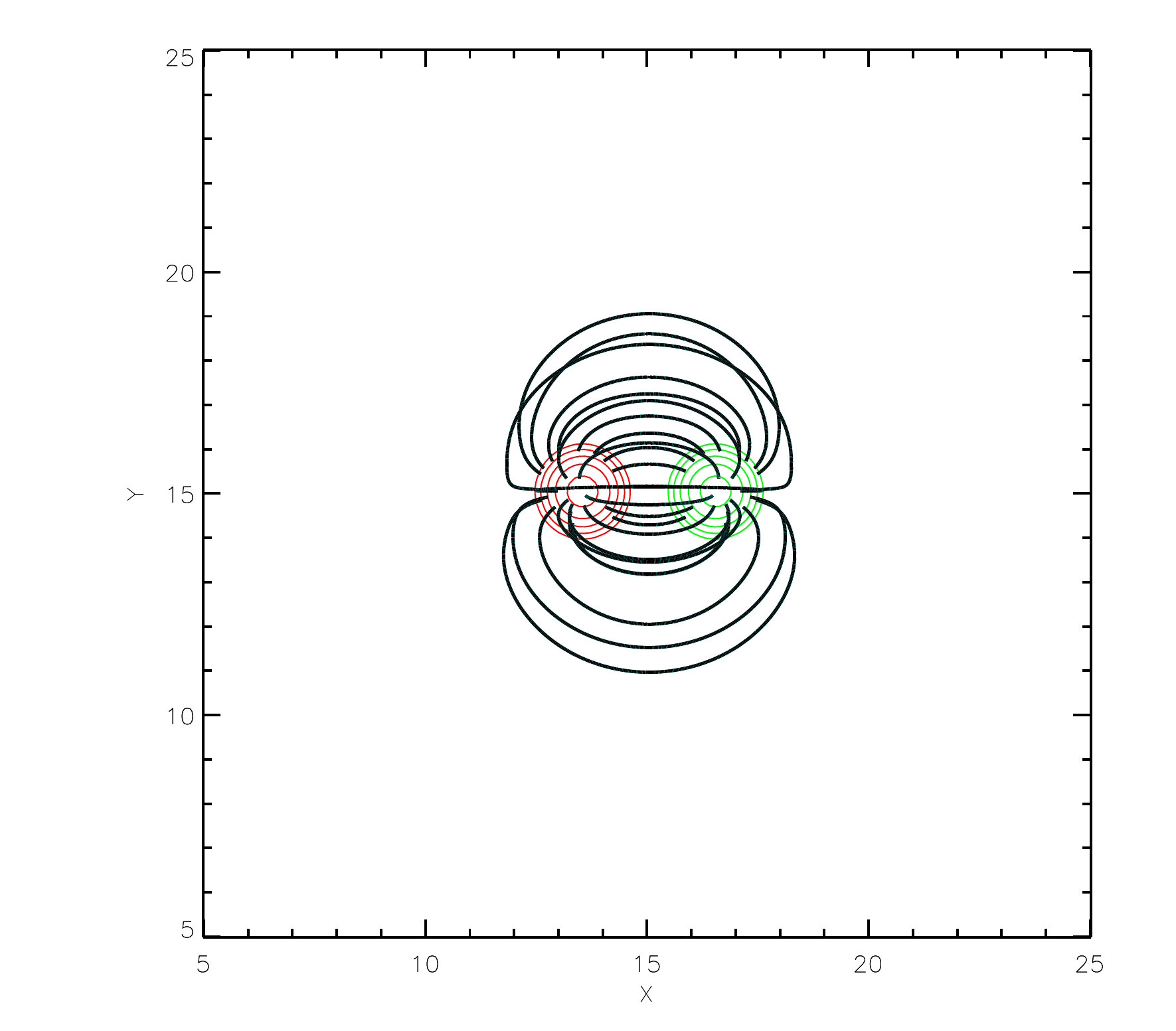}
               \hspace*{0.\textwidth}
               \includegraphics[width=0.45\textwidth,clip=]{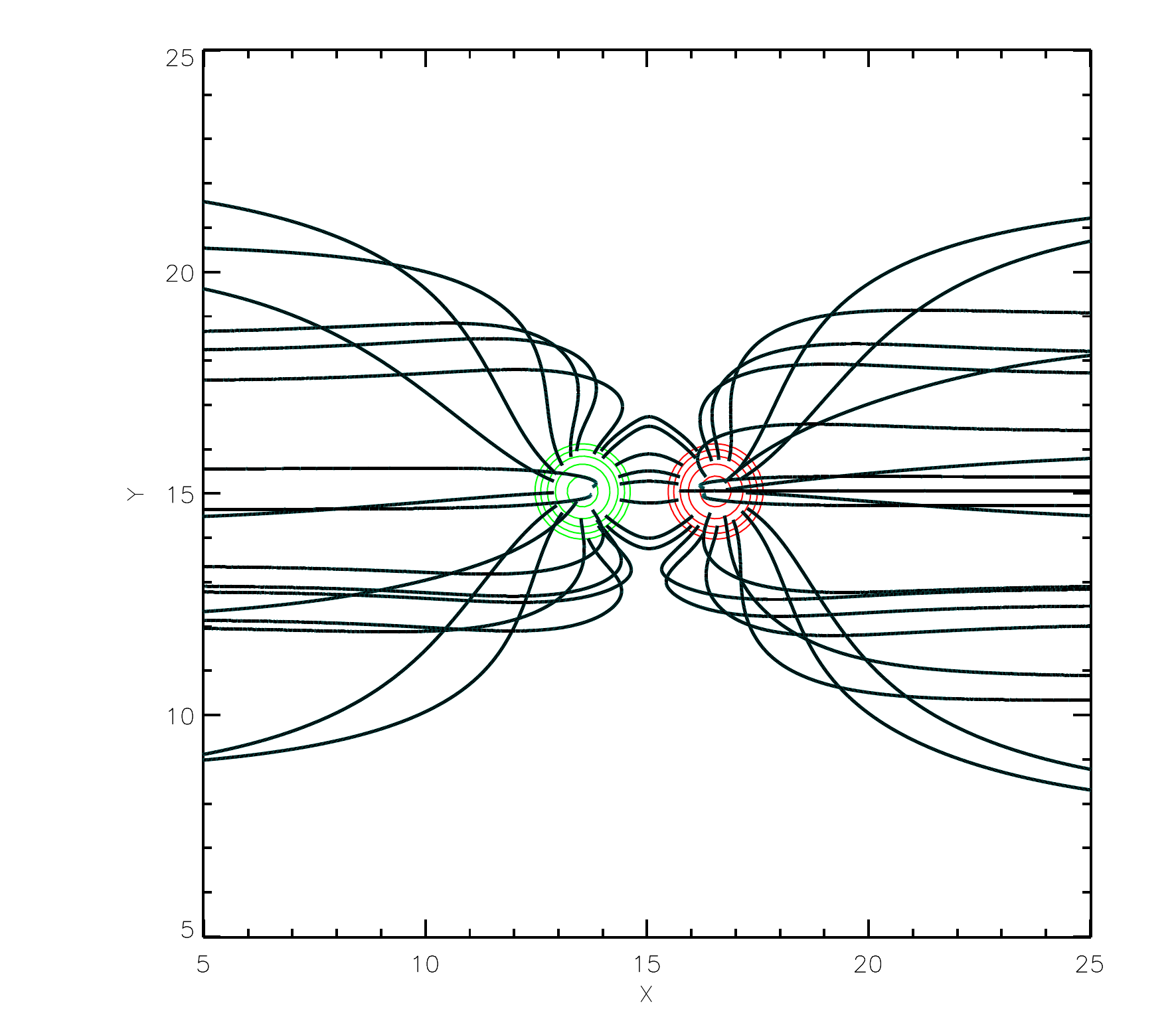}
              }
     \vspace{0.00\textwidth}   
     \centerline{ \bf     
      \hspace{0.245 \textwidth}  \color{black}{(a)}
      \hspace{0.405\textwidth}  \color{black}{(b)}       \hfill}
     \vspace{-0.1\textwidth}    

   \centerline{\hspace*{-0.01\textwidth}
               \includegraphics[width=0.45\textwidth,clip=]{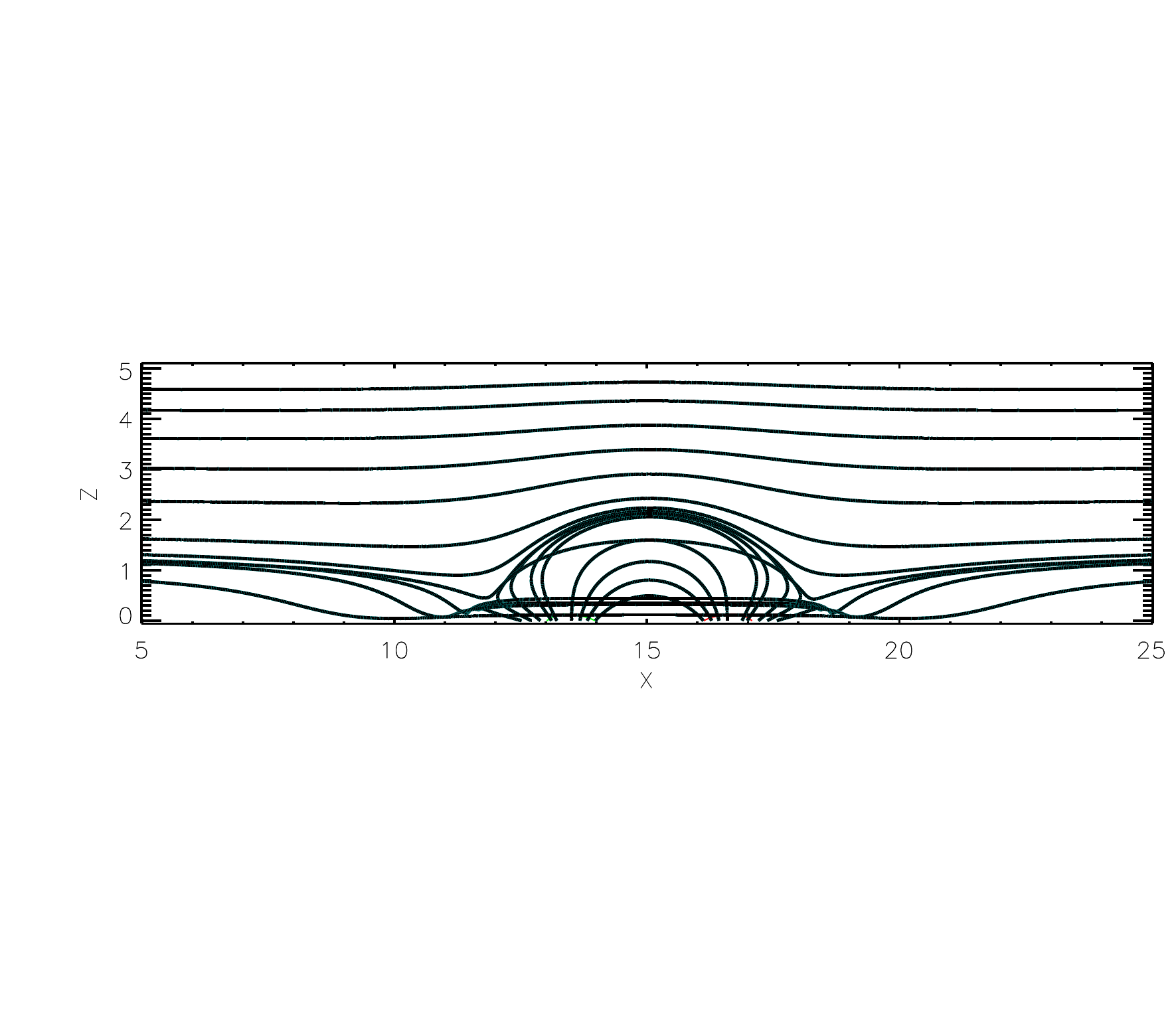}
               \hspace*{0.\textwidth}
               \includegraphics[width=0.45\textwidth,clip=]{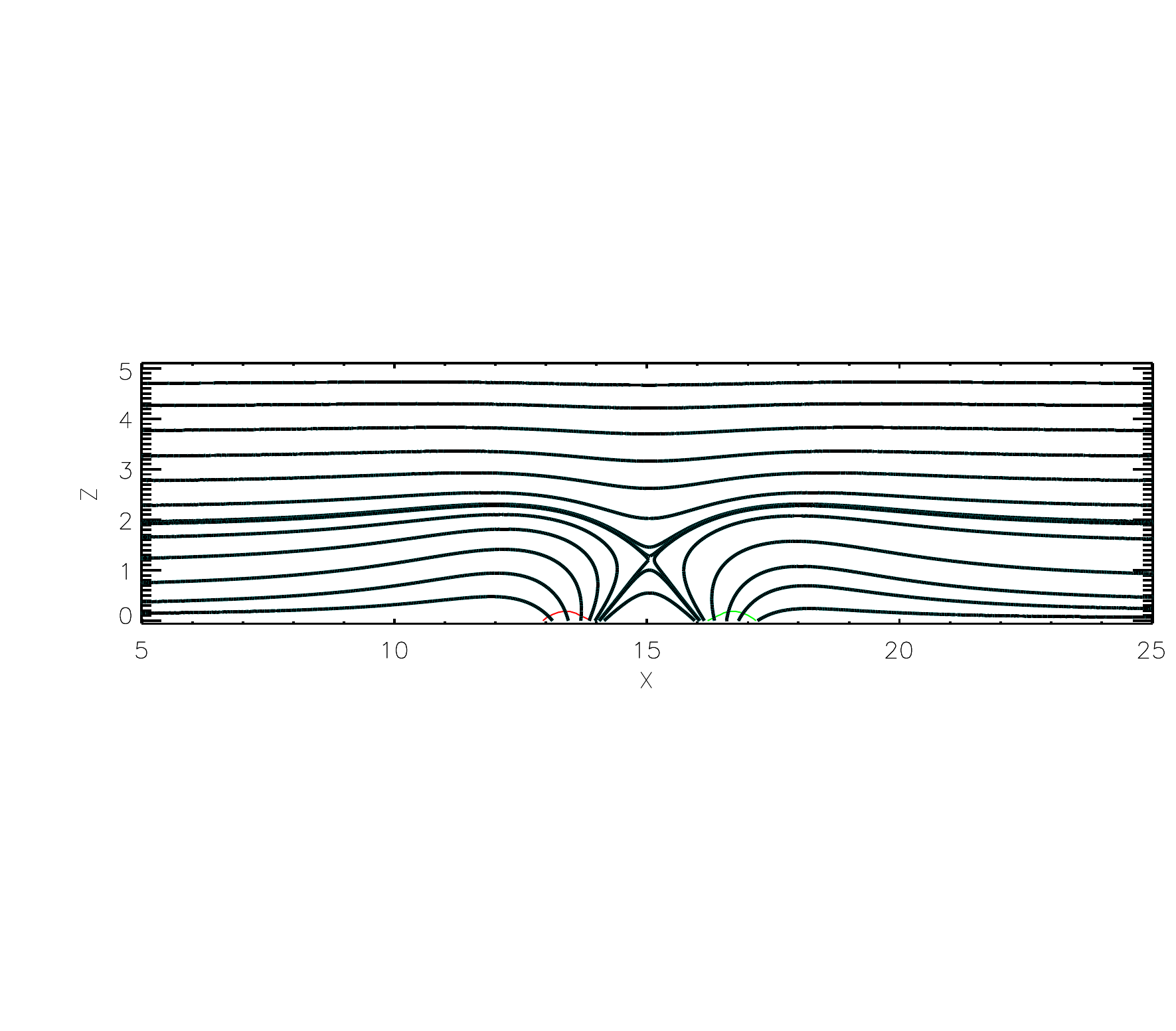}
              }
     \vspace{-0.1\textwidth}   
     \centerline{ \bf     
      \hspace{0.245 \textwidth}  \color{black}{(c)}
      \hspace{0.405\textwidth}  \color{black}{(d)}         \hfill}
     \vspace{0.02\textwidth}    

   \centerline{\hspace*{-0.01\textwidth}
               \includegraphics[width=0.45\textwidth,clip=]{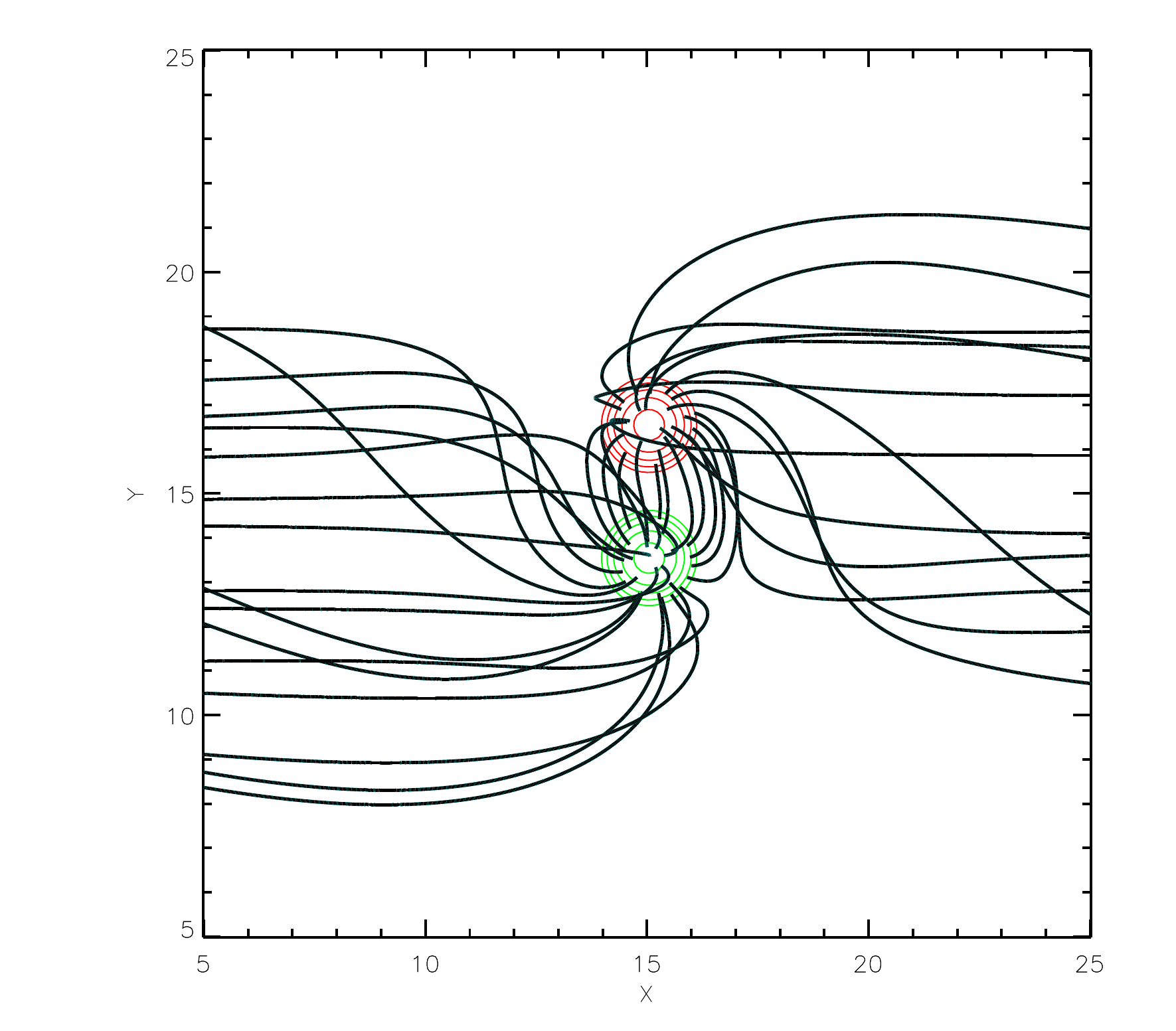}
              }
     \vspace{0.00\textwidth}   
     \centerline{ \bf     
      \hspace{0.48 \textwidth}  \color{black}{(e)}       \hfill}
     \vspace{-0.1\textwidth}    

   \centerline{\hspace*{-0.01\textwidth}
               \includegraphics[width=0.45\textwidth,clip=]{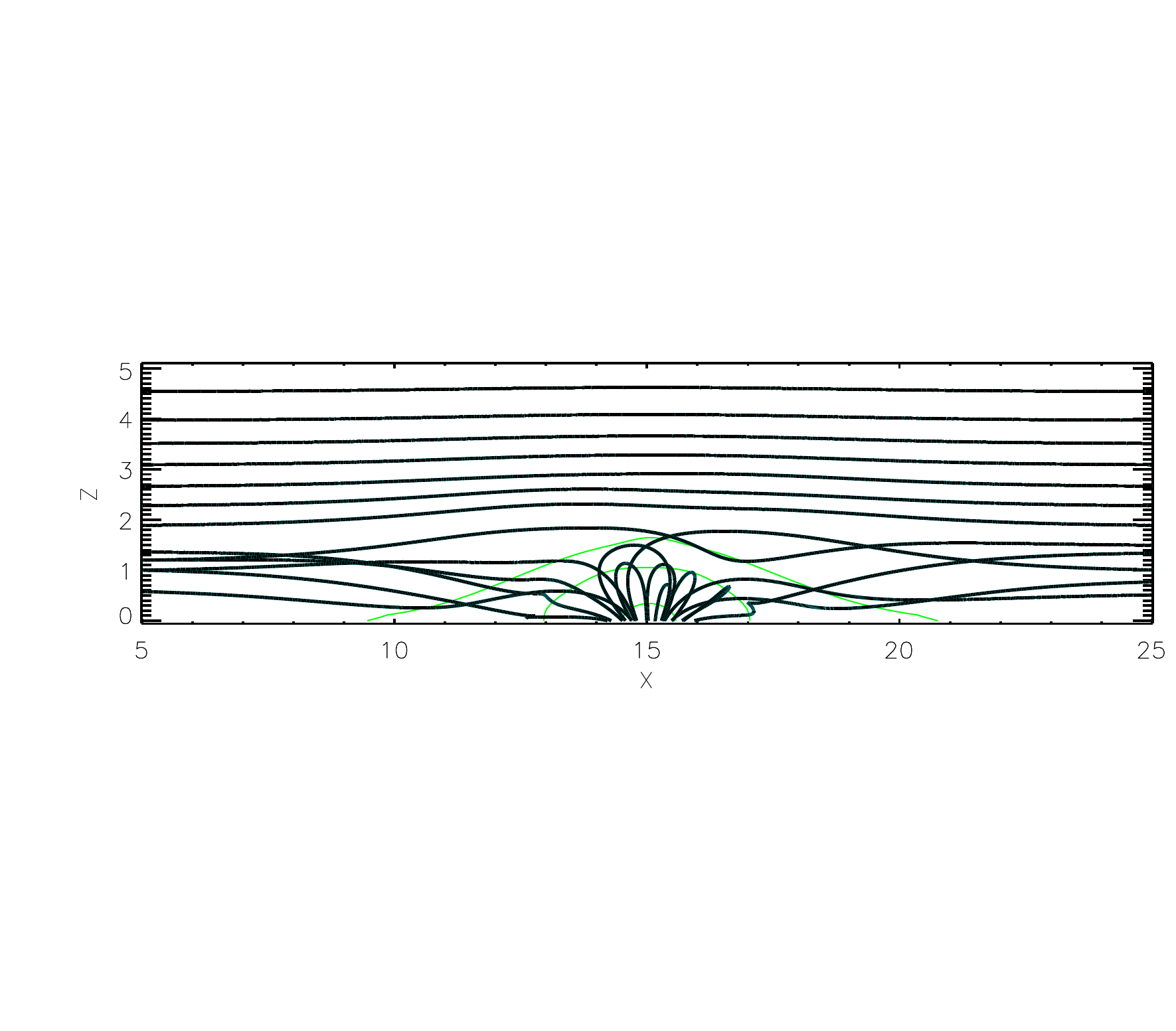}
              }
     \vspace{-0.1\textwidth}   
     \centerline{ \bf     
      \hspace{0.48 \textwidth}  \color{black}{(f)}         \hfill}
     \vspace{0.02\textwidth}    

\caption{Emergence of a bipole in the presence of a 5 G overlying field: (a), (b) and (e) as seen in the x-y plane at z=0, (c), (d) and (f) as seen in the x-z plane at y=15. For each case, the images are taken at $t=50$ min. The bipole's axis is aligned: (a) and (c) parallel to, (b) and (d) anti-parallel to, and (e) and (f) perpendicular to the overlying field. A selection of magnetic field lines originating from the bipole at the photospheric level is plotted on each image. In images (c), (d)  and (f), some of the overlying field lines have also been plotted. Red and green contours represent positive and negative magnetic field.}\label{fig:emergefield}
   \end{figure}

\subsubsection{Field Lines}

Figure~\ref{fig:emergefield} shows images of the 5 G emergence case, for each of the three orientations of the bipole's axis, at $t=50$ min. At this time, the two magnetic elements have separated and no longer overlap. As in the cancellation simulations, when the emerging bipole's axis is parallel to the overlying field, all flux from the positive polarity connects to the negative polarity. In the anti-parallel and perpendicular cases, connections between the two magnetic elements exist in the early stages of emergence. However in both these cases, for strong overlying fields, connections between the magnetic elements can be completely severed by the end of the simulation.

The photospheric boundary conditions for the cancellation and emergence simulations are exactly the reverse of one another. This means that a potential field extrapolation at $t=n$ min of each emergence case is identical to that at $t=(100-n)$ min of the corresponding cancellation case. Thus, the same photospheric field distribution exists for both emergence and cancellation at $t=50$ min. Therefore at this time, the field line plots of the non-linear force-free fields and potential fields in Figures~\ref{fig:cancelfield1}, \ref{fig:cancelfield2} and \ref{fig:cancelfield3} (cancellation) may be compared to the non-linear force-free fields in Figure~\ref{fig:emergefield} (emergence).
A comparison of these images shows that the emerging bipole's field is significantly different from that of the cancelling bipole or the corresponding potential field.
In the parallel emergence case (\ref{fig:emergefield}(c)), the low-lying field lines of the overlying field have been bent and pushed around either side of the bipole as the magnetic elements emerge.  This shows that the special boundary treatment that we use allows the elements to emerge as a single flux system into the overlying coronal field.

\subsubsection{Flux Connectivity and Energetics}

   \begin{figure}
   \centerline{\hspace*{0.015\textwidth}
               \includegraphics[width=0.45\textwidth,clip=]{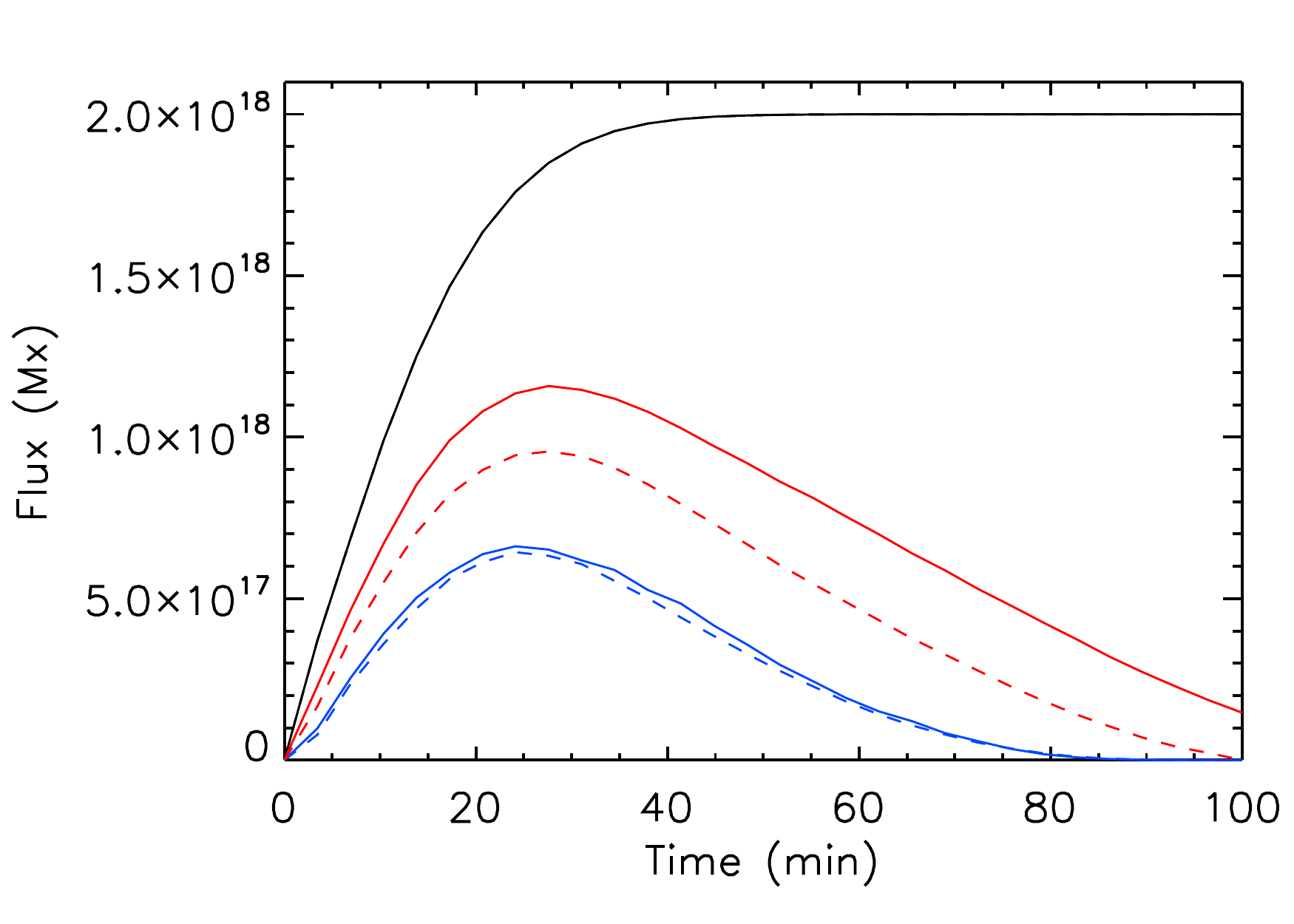}
               \hspace*{-0.0\textwidth}
               \includegraphics[width=0.45\textwidth,clip=]{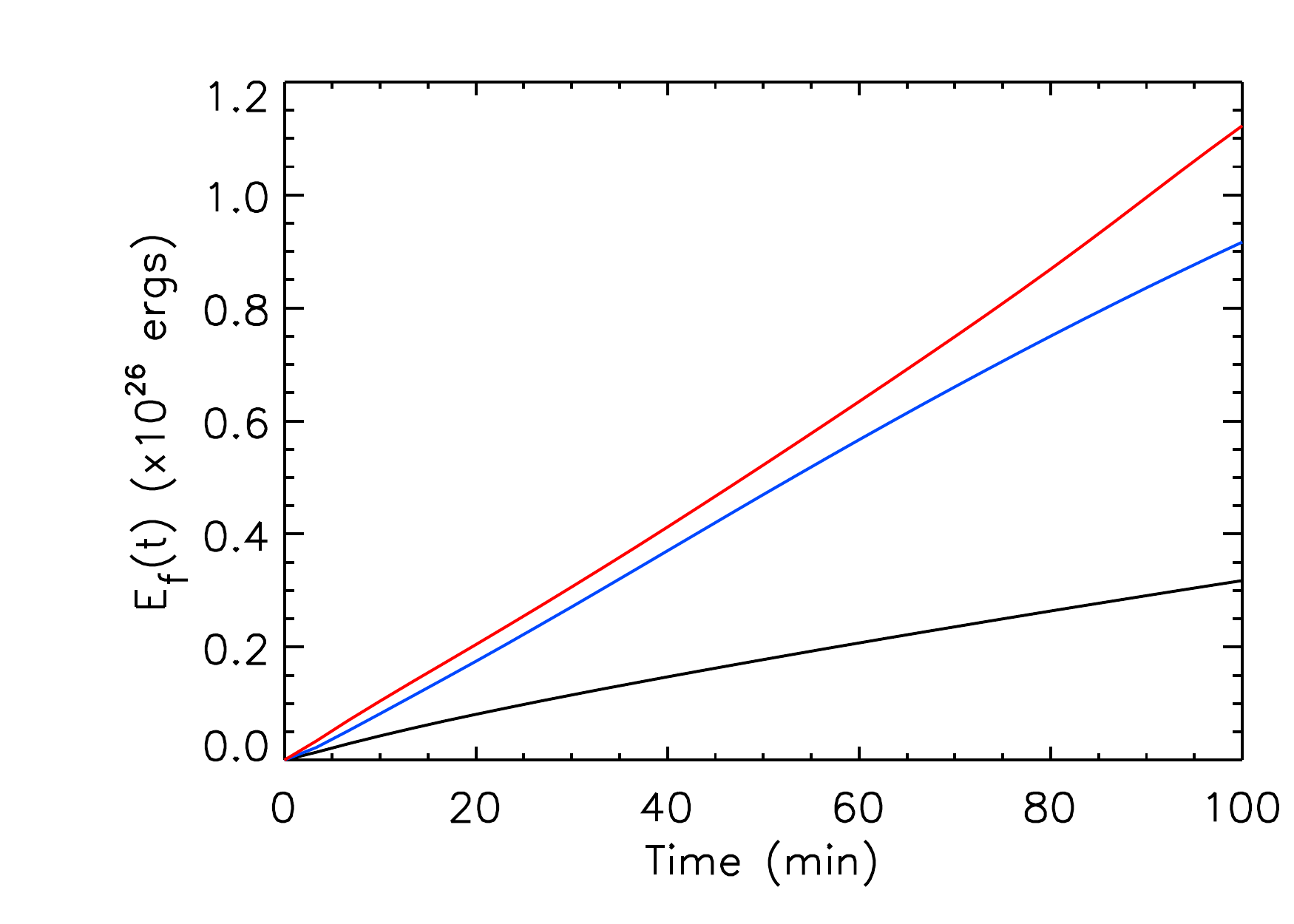}
              }
     \vspace{-0.33\textwidth}   
     \centerline{ \bf     
      \hspace{-0.0 \textwidth}  \color{black}{(a)}
      \hspace{0.42\textwidth}  \color{black}{(b)}
         \hfill}
     \vspace{0.3\textwidth}    
   \centerline{\hspace*{0.015\textwidth}
               \includegraphics[width=0.45\textwidth,clip=]{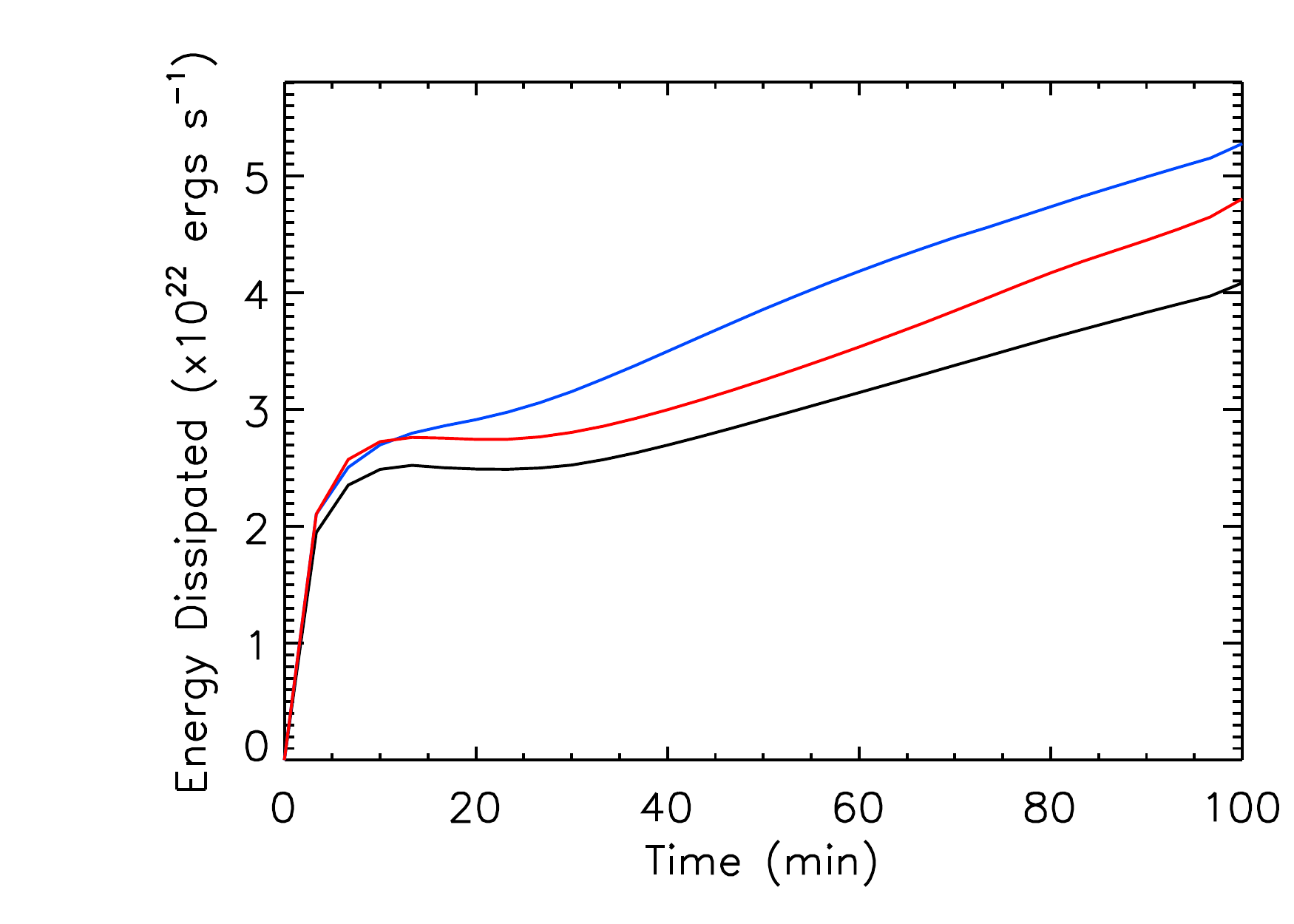}
               \hspace*{-0.0\textwidth}
               \includegraphics[width=0.45\textwidth,clip=]{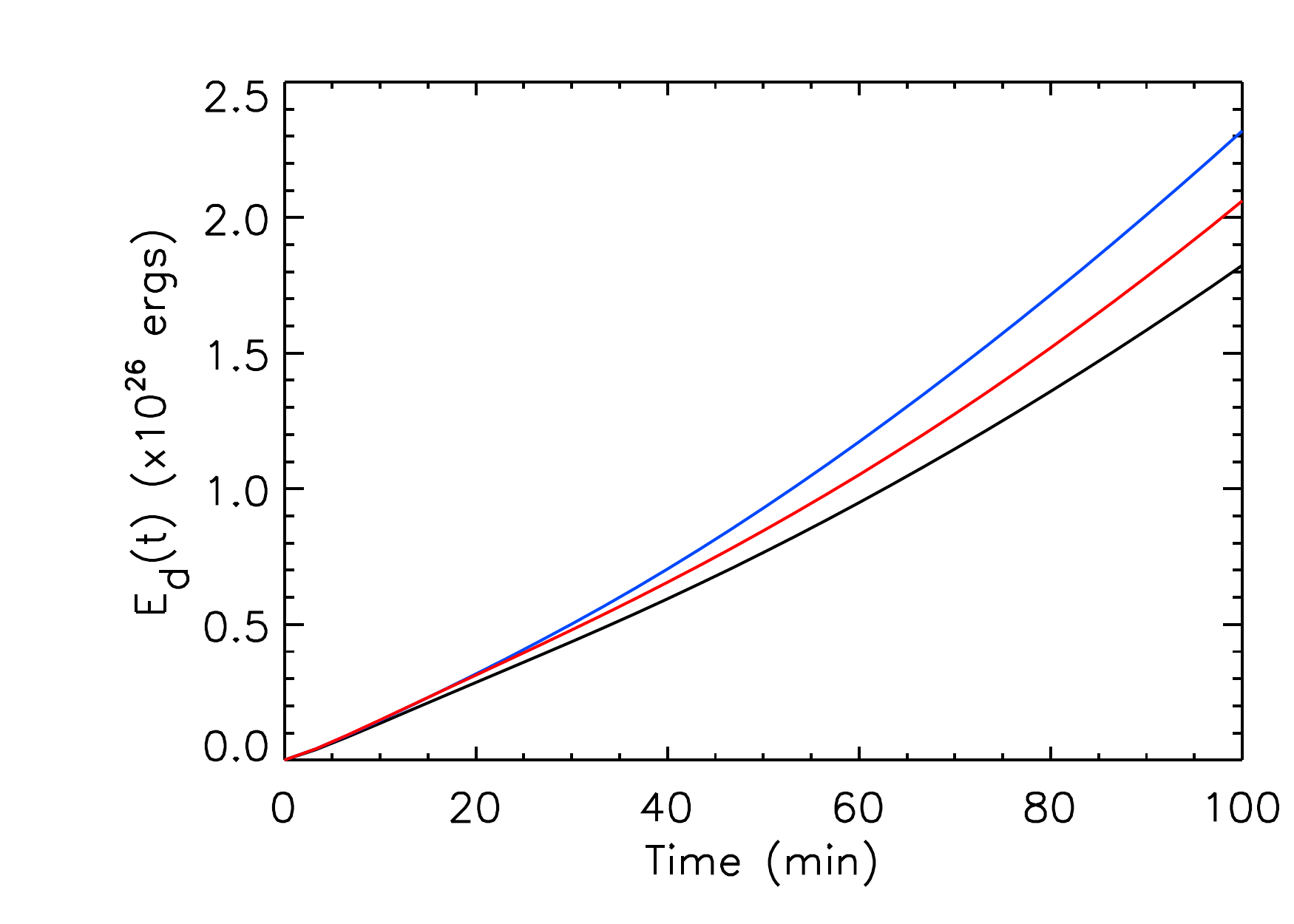}
              }
     \vspace{-0.34\textwidth}   
     \centerline{ \bf     
      \hspace{-0.0 \textwidth} \color{black}{(c)}
      \hspace{0.42\textwidth}  \color{black}{(d)}
         \hfill}
     \vspace{0.34\textwidth}    

\caption{Plots for an emerging bipole moving parallel (black), anti-parallel (blue) and perpendicular (red) to a 5 G overlying field, as a function of time: (a) total flux connecting magnetic elements, (b) free magnetic energy, $E_\textrm{f}(t)$, (c) energy dissipated, $\int_V Q dV$, and (d) cumulative energy dissipated, $E_\textrm{d}(t)$.}\label{fig:emerge1}
   \end{figure}

Figure~\ref{fig:emerge1}(a) shows a plot of the total flux connecting the two magnetic elements as a function of time for parallel (black), anti-parallel (blue) and perpendicular (red) emergence with a 5 G overlying field. Both the non-linear force-free field (solid lines) and the corresponding potential field (dashed lines) are shown.
For the  perpendicular case, much more flux connects between the magnetic elements in the non-linear force-free simulations than in the potential field extrapolations. Due to the continuous nature of the magnetofrictional method, connections that are formed between the two magnetic elements as they first emerge are maintained as they move apart. For the perpendicular case, the flux still connects between the magnetic elements by the end of the non-linear force-free field simulation, but all connections have been completely severed in the potential field extrapolation. A comparison of the perpendicular simulation in Figure~\ref{fig:emerge1}(a) to the corresponding cancellation plot (Figure~\ref{fig:cancel1}(a), red line) indicates that, for this orientation, the emergence shows a much larger departure from the potential field. Less departure is found for the anti-parallel case, since much more reconnection occurs as the bipole emerges into the oppositely aligned overlying field.

   \begin{figure}
   \centerline{\hspace*{0.015\textwidth}
               \includegraphics[width=0.7\textwidth,clip=]{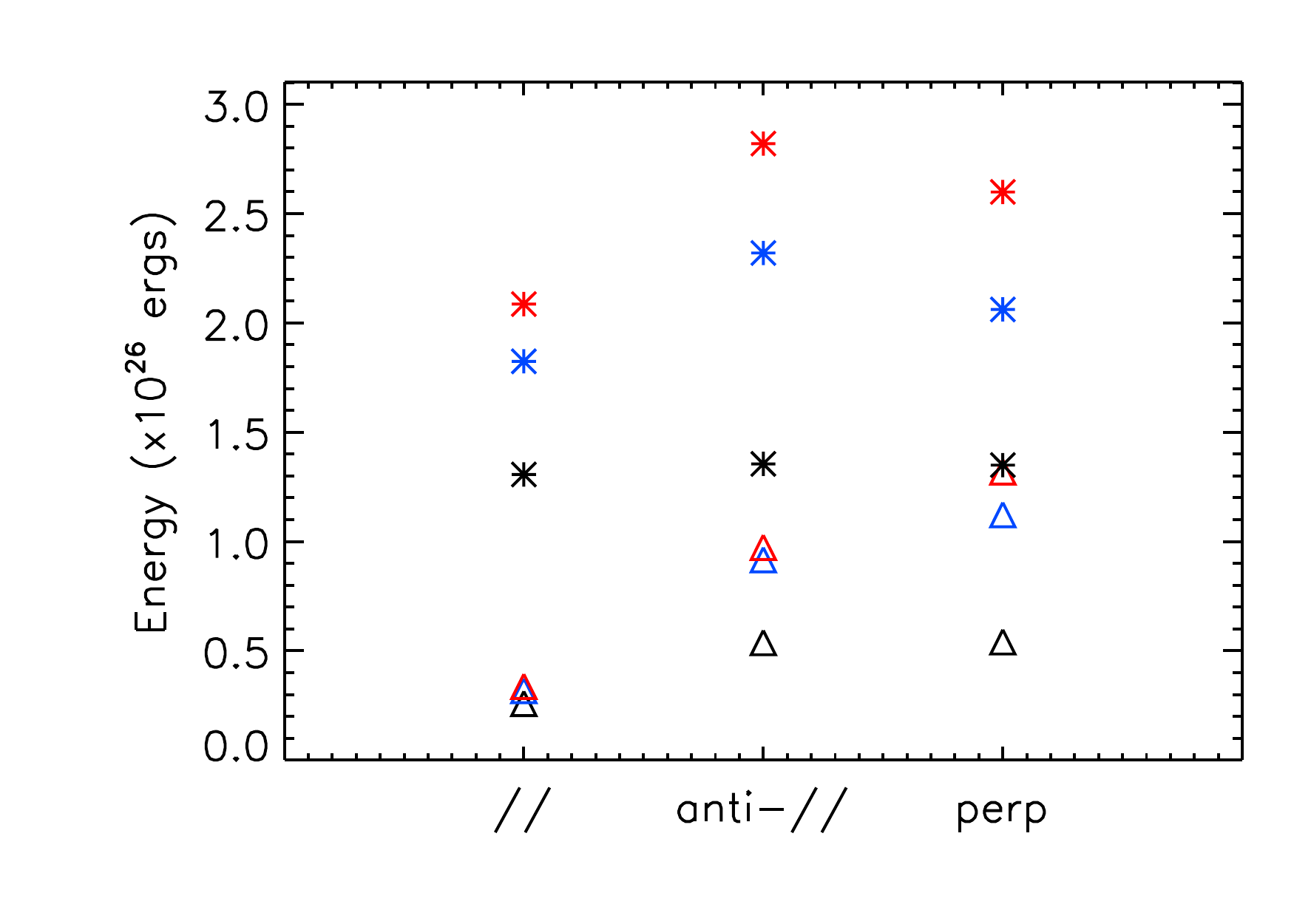}
              }

\caption{Free magnetic energy (triangles) and cumulative energy dissipated (stars) at the end of each emergence simulation, for a 1 G (black), 5 G (blue) and 10 G (red) overlying field.}\label{fig:allem}
   \end{figure}

Figure~\ref{fig:emerge1}(b) shows the free magnetic energy as a function of time, for each orientation of the bipole with a 5 G overlying field (lines are coloured as in Figure~\ref{fig:emerge1}(a)). The plot shows that perpendicular emergence results in the most free energy, and the parallel emergence in the least, where energy values range from $0.32-1.12\times 10^{26}$ ergs.
Figure~\ref{fig:emerge1}(c) shows a plot of the rate of energy dissipation, $Q$, integrated over the volume as a function of time. In all cases, the energy dissipation rate increases rapidly as the bipole first emerges, and the photospheric flux is increasing. Subsequently, the rate of energy dissipation continues to increase as the two magnetic elements move apart, but the increase is less rapid. As in the cancellation simulations, if we compare Figures~\ref{fig:emerge1}(b) and (c), it can be seen that the instantaneous energy dissipation is three orders of magnitude less than the free energy by the end of the simulation.
Figure~\ref{fig:emerge1}(d) shows the cumulative energy dissipated as a function of time (Equation~\ref{eqn:edt}), where values range from $1.82-2.32\times 10^{26}$ ergs. In contrast to Figure~\ref{fig:emerge1}(b) which shows that a perpendicular emergence results in the greatest build-up of free energy, Figures~\ref{fig:emerge1}(c) and (d) show that more energy is dissipated per unit time and cumulatively in an anti-parallel emergence. This happens because that in the anti-parallel case, larger gradients in $\alpha$ are produced and more reconnection takes place as the bipole emerges into the oppositely aligned overlying field.

Figure~\ref{fig:allem} compares the values of free energy (triangles) and total energy dissipated (stars) at the end of each simulation. The amount of free magnetic energy at the end of each emergence simulation follows the same trends as in the cancellation simulations: a stronger overlying field results in the build up of more free energy. The perpendicular emergence tends to result in the most free energy being built up, the parallel emergence in the least. We see that the most energy is dissipated in the anti-parallel case, and the least in the parallel case, as much more reconnection occurs in the anti-parallel case compared to the other two. As with the free energy, a stronger overlying field leads to more energy being dissipated.

To test the results of our simulations, it is possible to estimate the maximum amount of free magnetic energy that can be built up and stored for the anti-parallel emergence simulations. If we assume that no reconnection occurs, a current sheet (see e.g. \inlinecite{parnell2004}, \inlinecite{archontis2010}) would separate the bipole's field from the oppositely directed overlying field. The free energy is then given by:
\begin{equation}\label{eqn:emax}
E_{\textrm{\small max}}=\frac{B_0 \phi D}{2\pi},
\end{equation}
where $B_0$ is the strength of the overlying field, $\phi$ is the absolute flux of each magnetic element and $D$ is the photospheric separation of the magnetic elements. For the 5 G simulation, Equation~\ref{eqn:emax} gives $4.8\times 10^{26}$ ergs. This is the theoretical maximum value for the free energy built up. We note that within our simulations, the theoretical maximum cannot be obtained due to numerical diffusion and the fact that reconnection occurs as soon as the bipole emerges. We find that the occurrence of reconnection results in a smaller amount of free energy being stored. From Figure~\ref{fig:allem}, the amount of free energy stored at the end of the 5 G anti-parallel simulation is approximately 20\% of the theoretical maximum.
Although the free magnetic energy is one part of the energy calculated in the simulation, we also compute the energy dissipated due to relaxation processes and hyperdiffusion. For the 5 G case, a further $2.3\times 10^{26}$ ergs of magnetic energy is cumulatively dissipated (Figure~\ref{fig:emerge1}(d)). Summing the stored energy and the total energy dissipated, we obtain $3.2\times 10^{26}$ ergs, which is close to the theoretical maximum of $4.8\times 10^{26}$ ergs. Therefore, by taking into account not only the free energy, but also the energy dissipated when reconnection is allowed, we find a similar total amount energy to the theoretical maximum when reconnection is not allowed.

\subsection{Flyby}\label{fly}

\begin{figure}
 \begin{center}
  \includegraphics[width=0.7\textwidth]{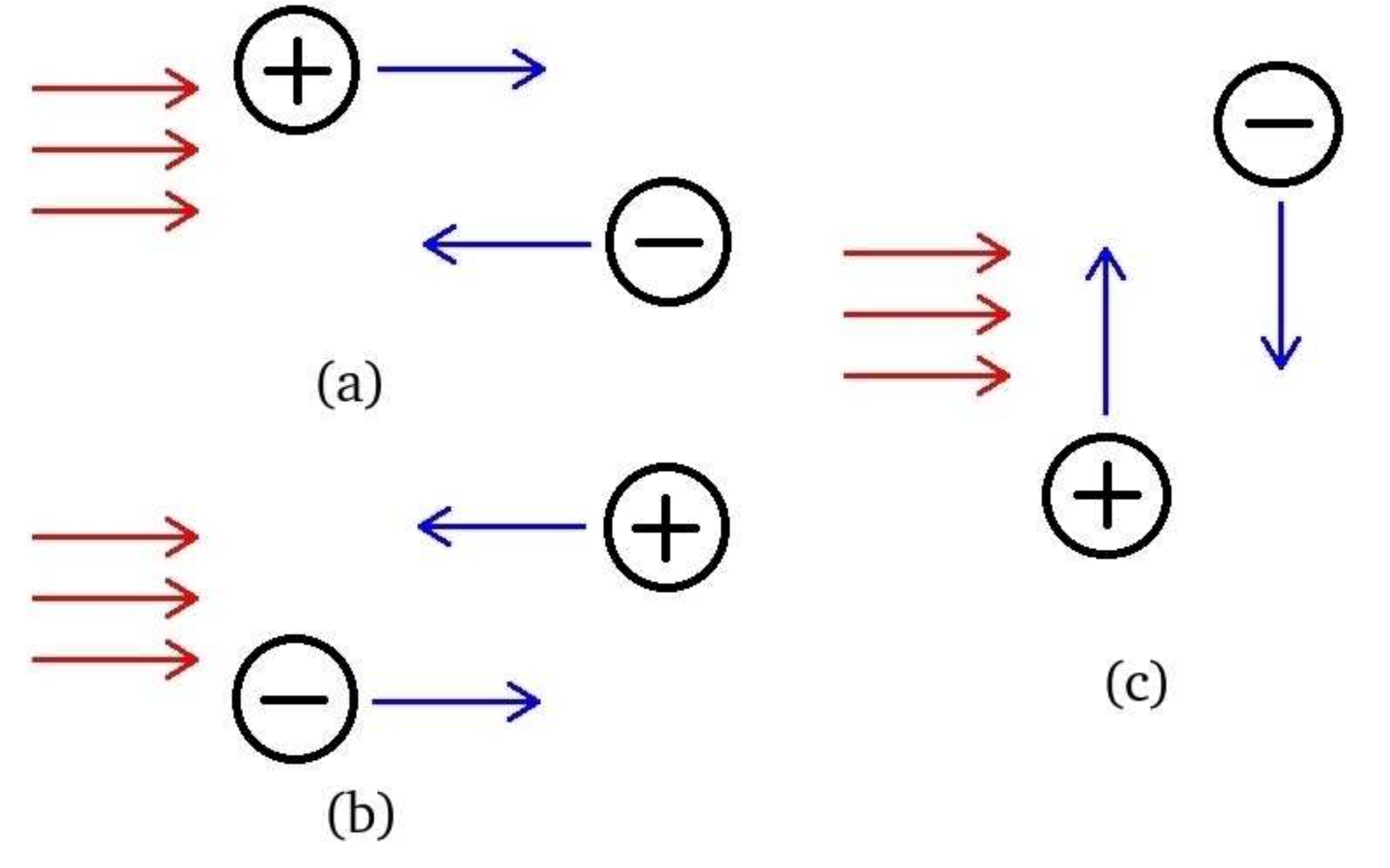}
 \end{center}
\caption{Flyby: The red arrows represent the direction of the overlying field, the blue arrows represent the direction of motion of the magnetic elements. The motion of the positive magnetic element is (a) parallel to,  (b) anti-parallel to, and (c) perpendicular to the overlying field.}\label{fig:flyby}
\end{figure}
The photospheric boundary distribution for the flyby simulations is slightly different from the previous simulations. The flyby simulations are run for 166.7 min instead of 100 min, and each magnetic element is advected a distance of 5 Mm rather than 3 Mm. The total absolute flux through the photosphere is constant throughout the simulation (Figure~\ref{fig:flux}, red line). The two magnetic elements are advected past one another under the presence of an overlying field, so that their final positions mirror their initial positions. An illustration of the initial set-up of each flyby simulation is shown in Figure~\ref{fig:flyby}. In two set-ups the magnetic elements are advected past one another in the $x$-direction. In the first case (Figure~\ref{fig:flyby}(a)) the motion of the positive magnetic element is parallel to the overlying field, in the second case (Figure~\ref{fig:flyby}(b)) it is anti-parallel to the overlying field. In the third set of simulations (Figure~\ref{fig:flyby}(c)) the magnetic elements are advected in the $y$-direction, so that their motion is perpendicular to the overlying field.

  \begin{figure}
   \centerline{\hspace*{-0.01\textwidth}
               \includegraphics[width=0.3\textwidth,clip=]{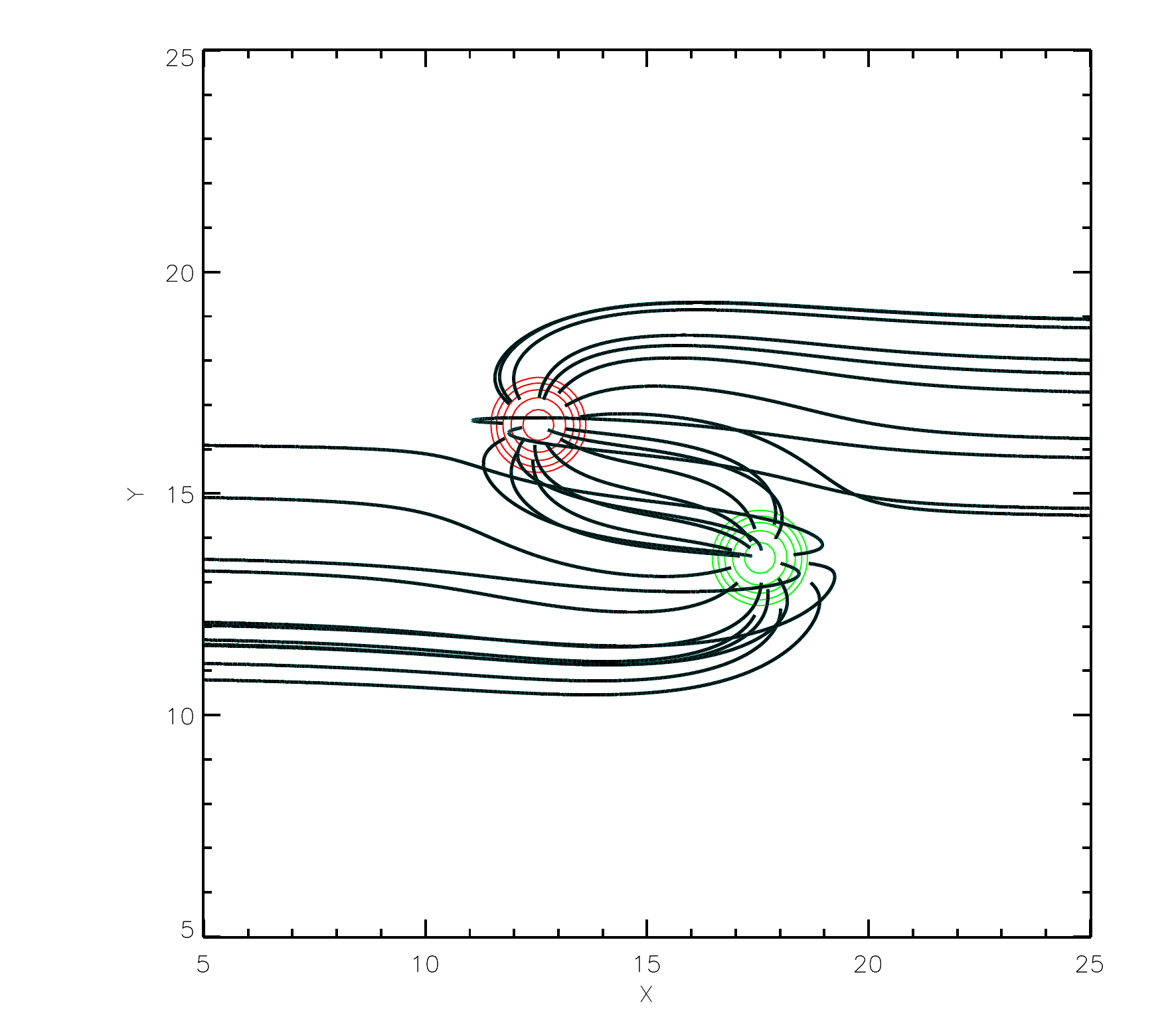}
               \hspace*{-0.0\textwidth}
               \includegraphics[width=0.3\textwidth,clip=]{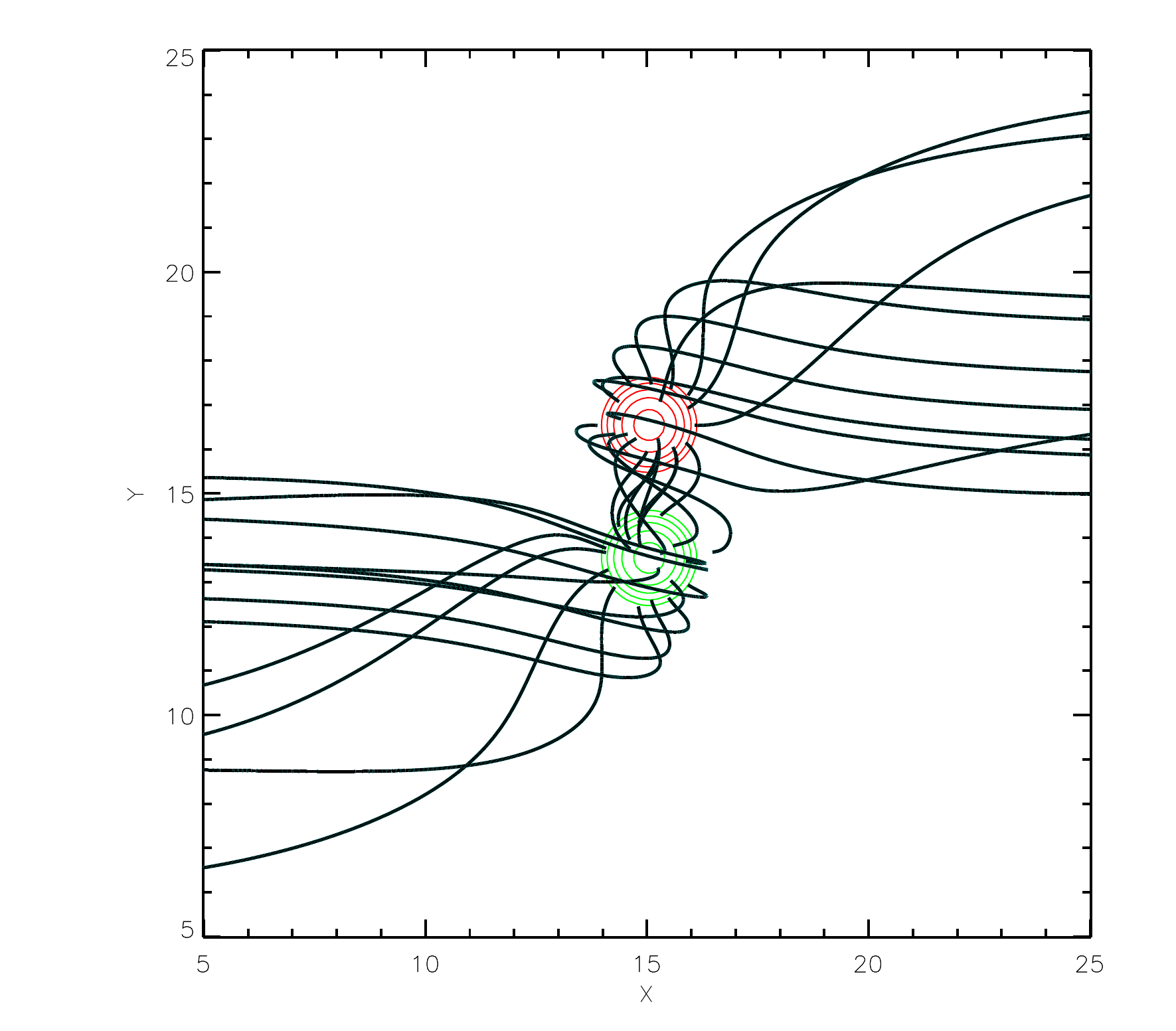}
               \hspace*{-0.0\textwidth}
               \includegraphics[width=0.3\textwidth,clip=]{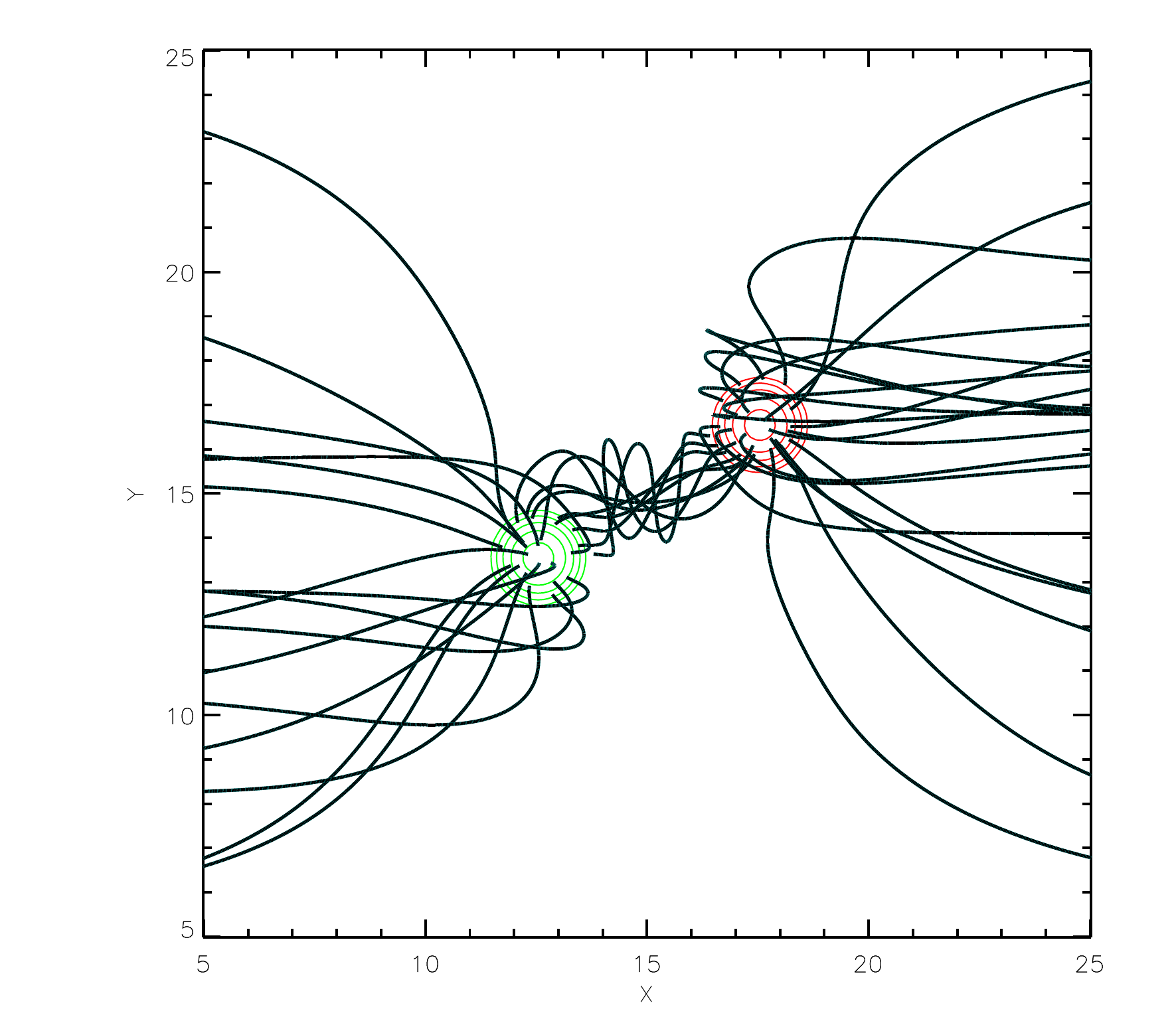}
              }
     \vspace{0.00\textwidth}   
     \centerline{ \bf     
      \hspace{0.15 \textwidth}  \color{black}{(i)}
      \hspace{0.253\textwidth}  \color{black}{(ii)}
      \hspace{0.253\textwidth}  \color{black}{(iii)}         \hfill}
     \vspace{0.02\textwidth}    

     \vspace{-0.23\textwidth}
     \centerline{ \bf     
      \hspace{-0.04 \textwidth}  \color{black}{\small{(a)}}
         \hfill}
     \vspace{0.19\textwidth}

   \centerline{\hspace*{-0.01\textwidth}
               \includegraphics[width=0.3\textwidth,clip=]{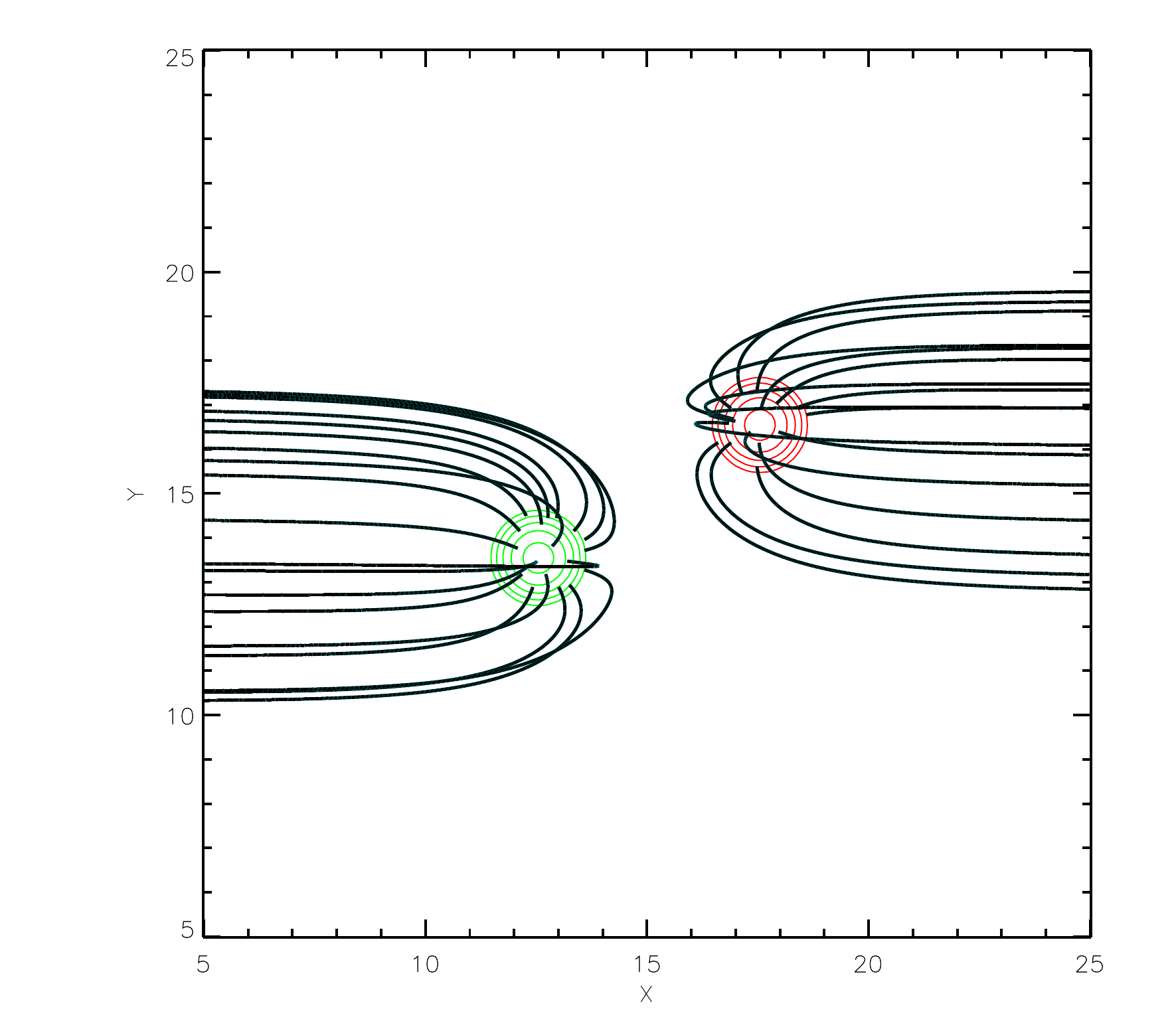}
               \hspace*{-0.0\textwidth}
               \includegraphics[width=0.3\textwidth,clip=]{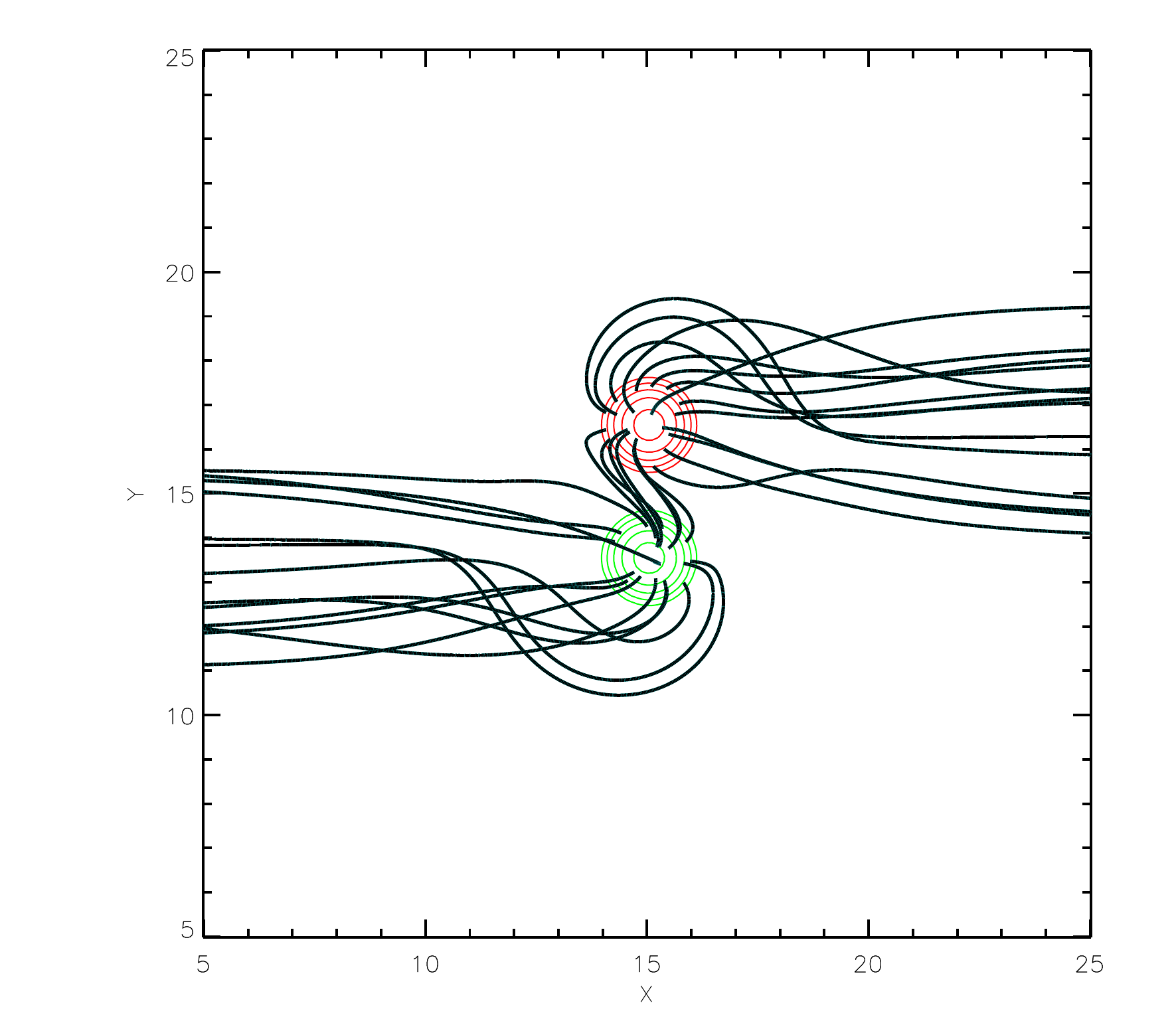}
               \hspace*{-0.0\textwidth}
               \includegraphics[width=0.3\textwidth,clip=]{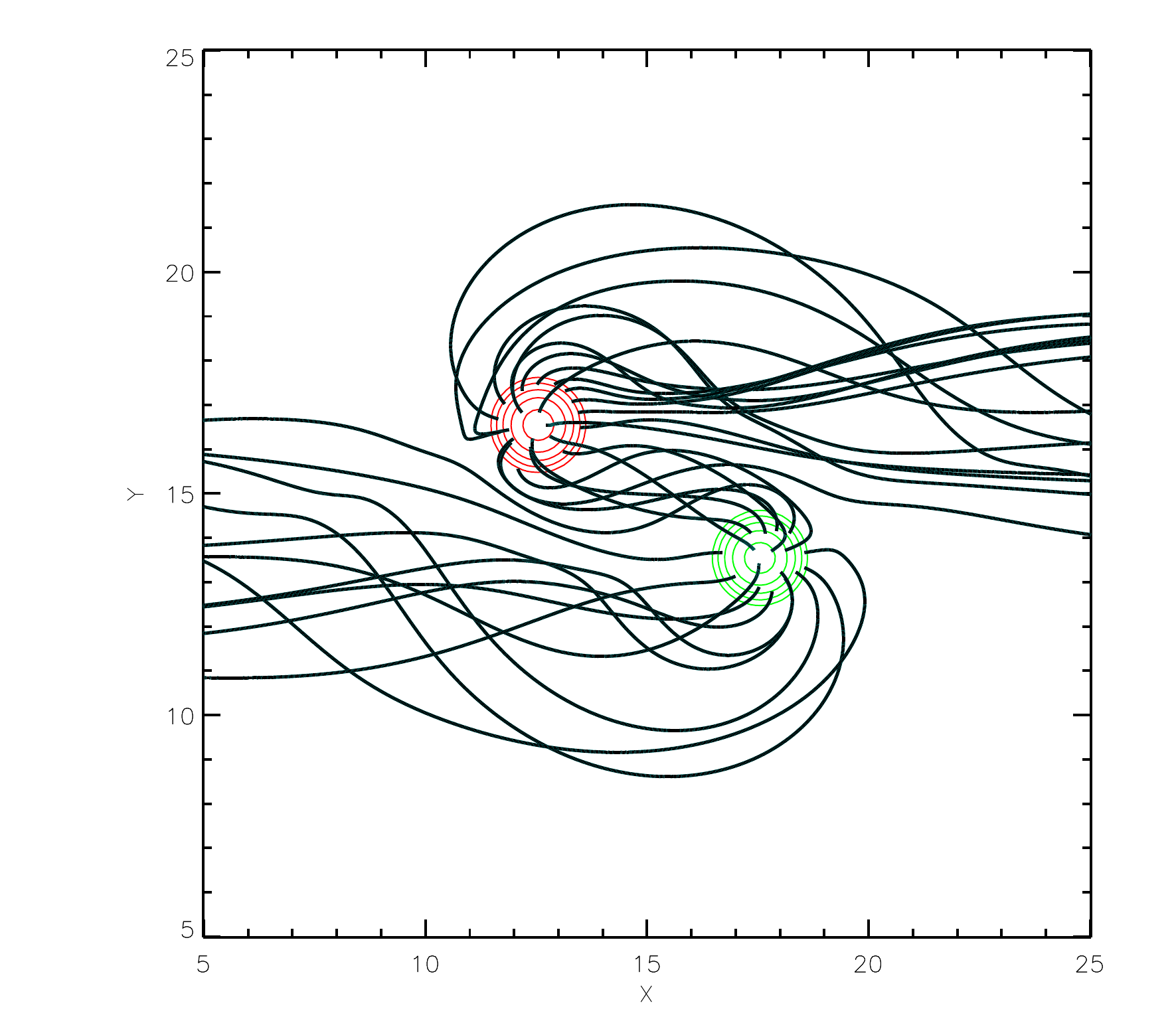}
              }
     \vspace{0.00\textwidth}   
     \centerline{ \bf     
      \hspace{0.15 \textwidth}  \color{black}{(i)}
      \hspace{0.253\textwidth}  \color{black}{(ii)}
      \hspace{0.253\textwidth}  \color{black}{(iii)}         \hfill}
     \vspace{0.02\textwidth}    

     \vspace{-0.23\textwidth}
     \centerline{ \bf     
      \hspace{-0.04 \textwidth}  \color{black}{\small{(b)}}
         \hfill}
     \vspace{0.19\textwidth}

  \centerline{\hspace*{-0.01\textwidth}
              \includegraphics[width=0.3\textwidth,clip=]{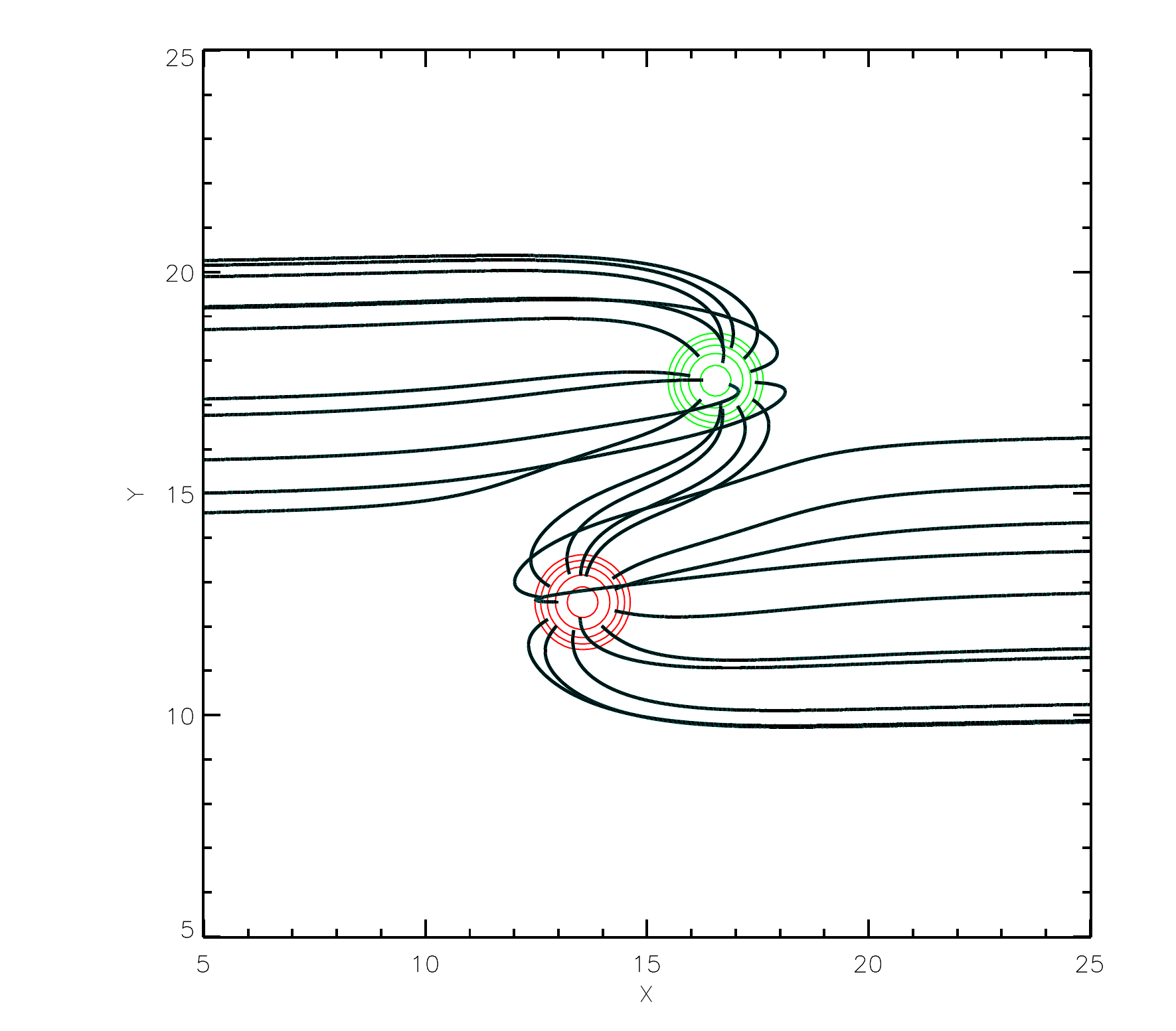}
              \hspace*{-0.0\textwidth}
              \includegraphics[width=0.3\textwidth,clip=]{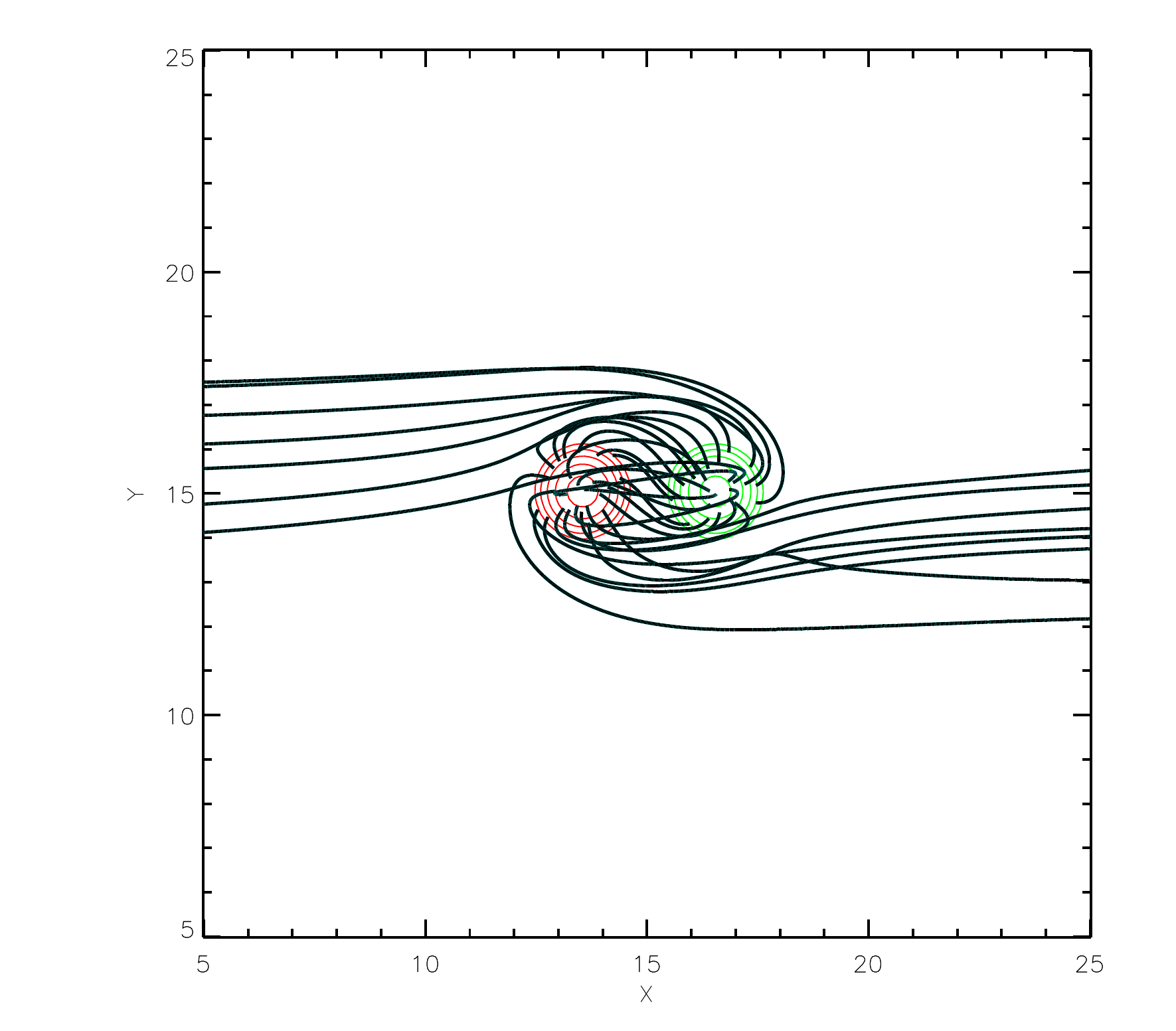}
              \hspace*{-0.0\textwidth}
              \includegraphics[width=0.3\textwidth,clip=]{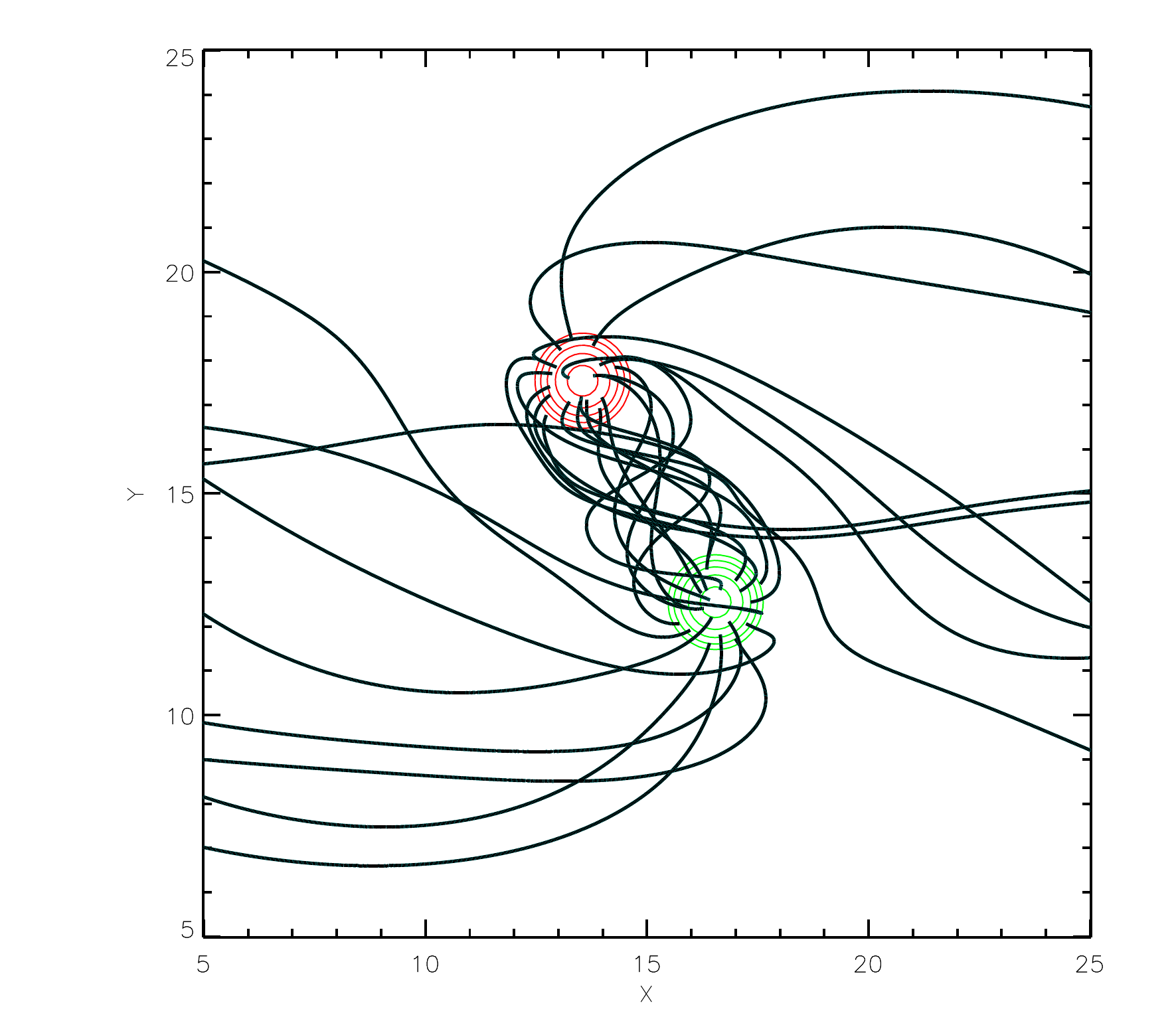}
             }
     \vspace{0.00\textwidth}   
     \centerline{ \bf     
      \hspace{0.15 \textwidth}  \color{black}{(i)}
      \hspace{0.253\textwidth}  \color{black}{(ii)}
      \hspace{0.253\textwidth}  \color{black}{(iii)}         \hfill}
     \vspace{0.02\textwidth}    

     \vspace{-0.23\textwidth}
     \centerline{ \bf     
      \hspace{-0.04 \textwidth}  \color{black}{\small{(c)}}
         \hfill}
     \vspace{0.19\textwidth}

\caption{A series of x-y plane images for each flyby case with a 5 G overlying field. The red and green contours represent positive and negative magnetic field. The images in each case are taken at (i) $t=0$ min, (ii) $t=83.3$ min and (iii) $t=166.7$ min, and a selection of field lines is plotted originating from the magnetic elements. The positive magnetic element is advected (a) parallel to, (b) anti-parallel to and (c) perpendicular to the overlying field.}\label{fig:flybyfield1}
   \end{figure}

\subsubsection{Field Lines}

Figure~\ref{fig:flybyfield1} shows a series of x-y plane images at $t=0$, 83.3 and 166.7 min for (a) parallel, (b) anti-parallel and (c) perpendicular flyby and a 5 G overlying field. Since the parallel flyby is the reverse of the anti-parallel flyby, the photospheric flux distribution at the start of each case is identical to that at the end of the other. This means that the initial potential fields in Figures~\ref{fig:flybyfield1}(a)(i) and (b)(i) may be compared with the final non-linear force-free fields in Figures~\ref{fig:flybyfield1}(b)(iii) and (a)(iii) respectively.
It is clear that the non-linear force-free field in each case is quite different from the corresponding potential field. In particular, significant differences can be seen for the parallel flyby. For the non-linear force-free field ((a)(iii)), strong connections exist between the magnetic elements. However, there are no such connections in the potential field ((b)(i)).
The field lines connecting the two elements are very twisted, and the bipole's magnetic field appears to occupy a much larger volume of the corona than in the corresponding potential field (Figure~\ref{fig:flybyfield1}(b)(i)).

It is also of interest to compare Figures~\ref{fig:flybyfield1}(a)(ii) and (b)(ii), as their photospheric distributions are also identical to one another. Even though both simulations have been running for the same amount of time ($83.3$ min), the shape of the bipole's field is very different. For the parallel simulation, the field lines that connect the magnetic elements are much more twisted. This illustrates that when the evolution of the coronal magnetic field is continuous, the properties of the field very much depend on its previous evolution and connectivity, not only on the photospheric boundary distribution.

For the perpendicular flyby (Figure~\ref{fig:flybyfield1}(c)), the photospheric boundary distributions in (c)(i) and (c)(iii) are symmetric to one another. At the midpoint, the bipole's axis becomes aligned fully parallel to the overlying field, and this results in two major occurrences of reconnection. The first occurs as the magnetic elements move towards one another, and the total flux connecting from one magnetic element to the other rapidly increases. The second occurs after the magnetic elements have passed one another at the midline, causing connections between the magnetic elements to break, and the total connecting flux decreases. More flux remains connected between the two magnetic elements in Figure~\ref{fig:flybyfield1}(c)(iii) than in the corresponding potential field in Figure~\ref{fig:flybyfield1}(c)(i). Again, this illustrates the effect of the memory of previous connectivity in our simulations.

\subsubsection{Flux Connectivity and Energetics}

   \begin{figure}
   \centerline{\hspace*{0.015\textwidth}
               \includegraphics[width=0.45\textwidth,clip=]{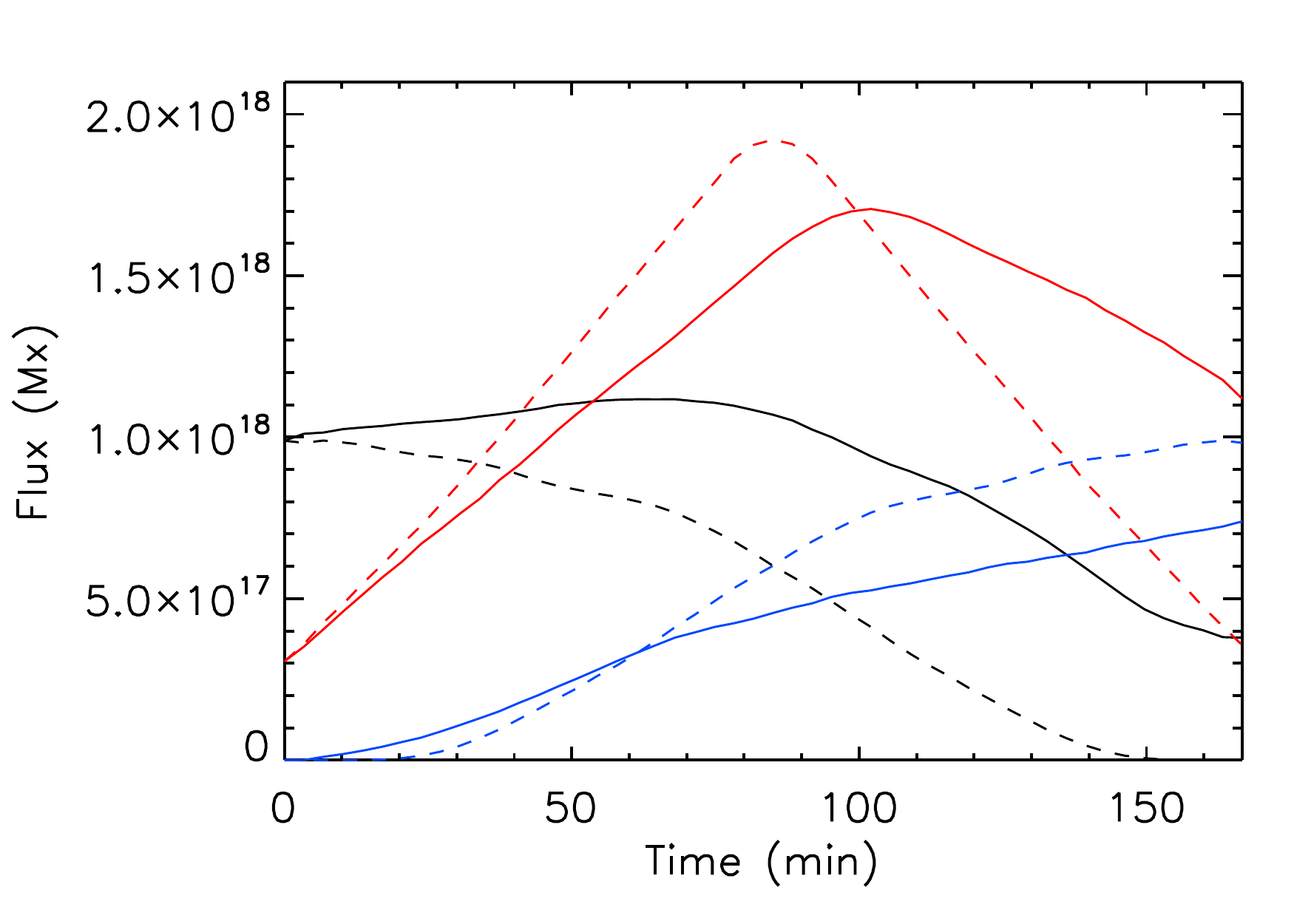}
               \hspace*{-0.0\textwidth}
               \includegraphics[width=0.45\textwidth,clip=]{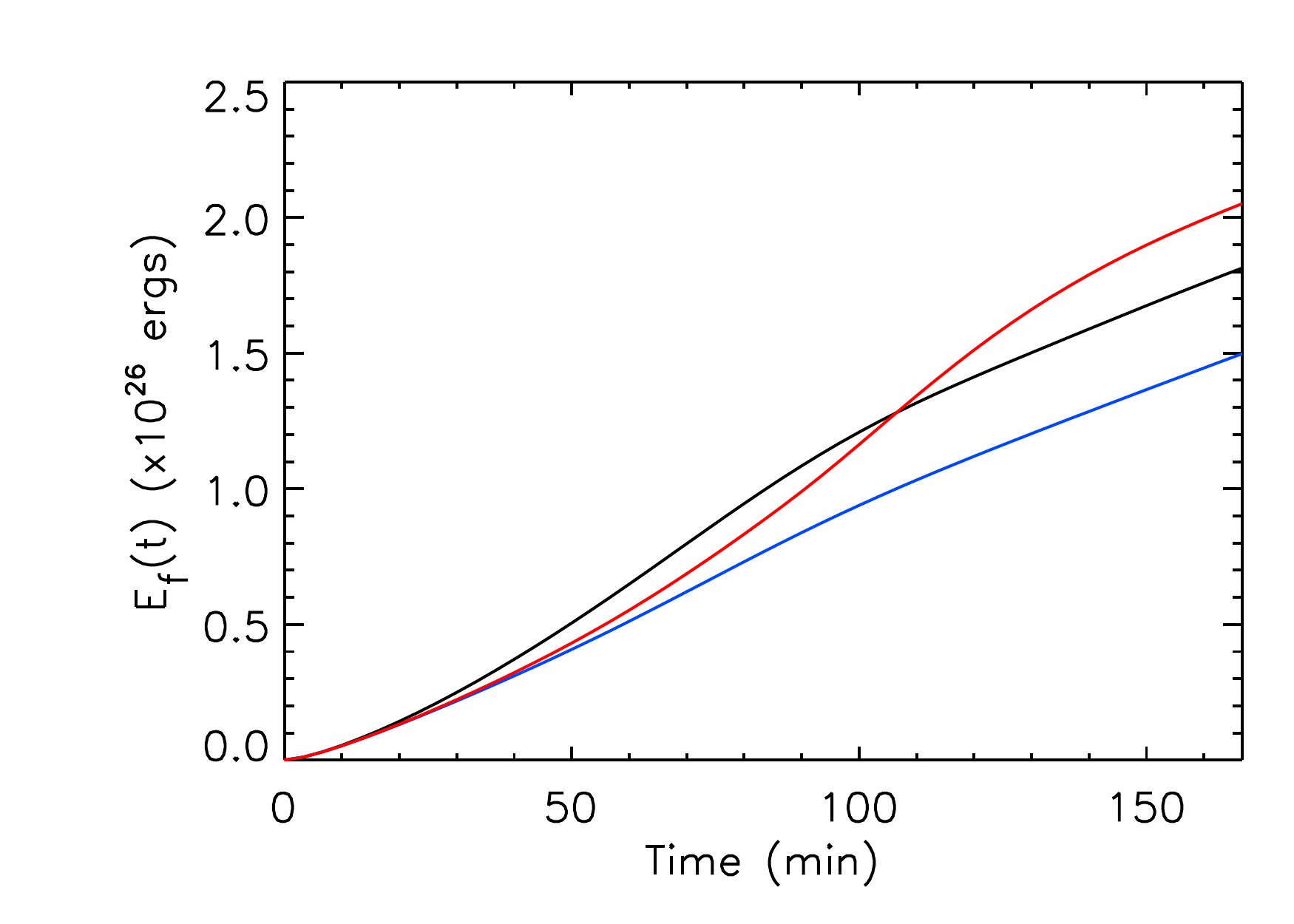}
              }
     \vspace{-0.33\textwidth}   
     \centerline{ \bf     
      \hspace{-0.0 \textwidth}  \color{black}{(a)}
      \hspace{0.42\textwidth}  \color{black}{(b)}
         \hfill}
     \vspace{0.3\textwidth}    
   \centerline{\hspace*{0.015\textwidth}
               \includegraphics[width=0.45\textwidth,clip=]{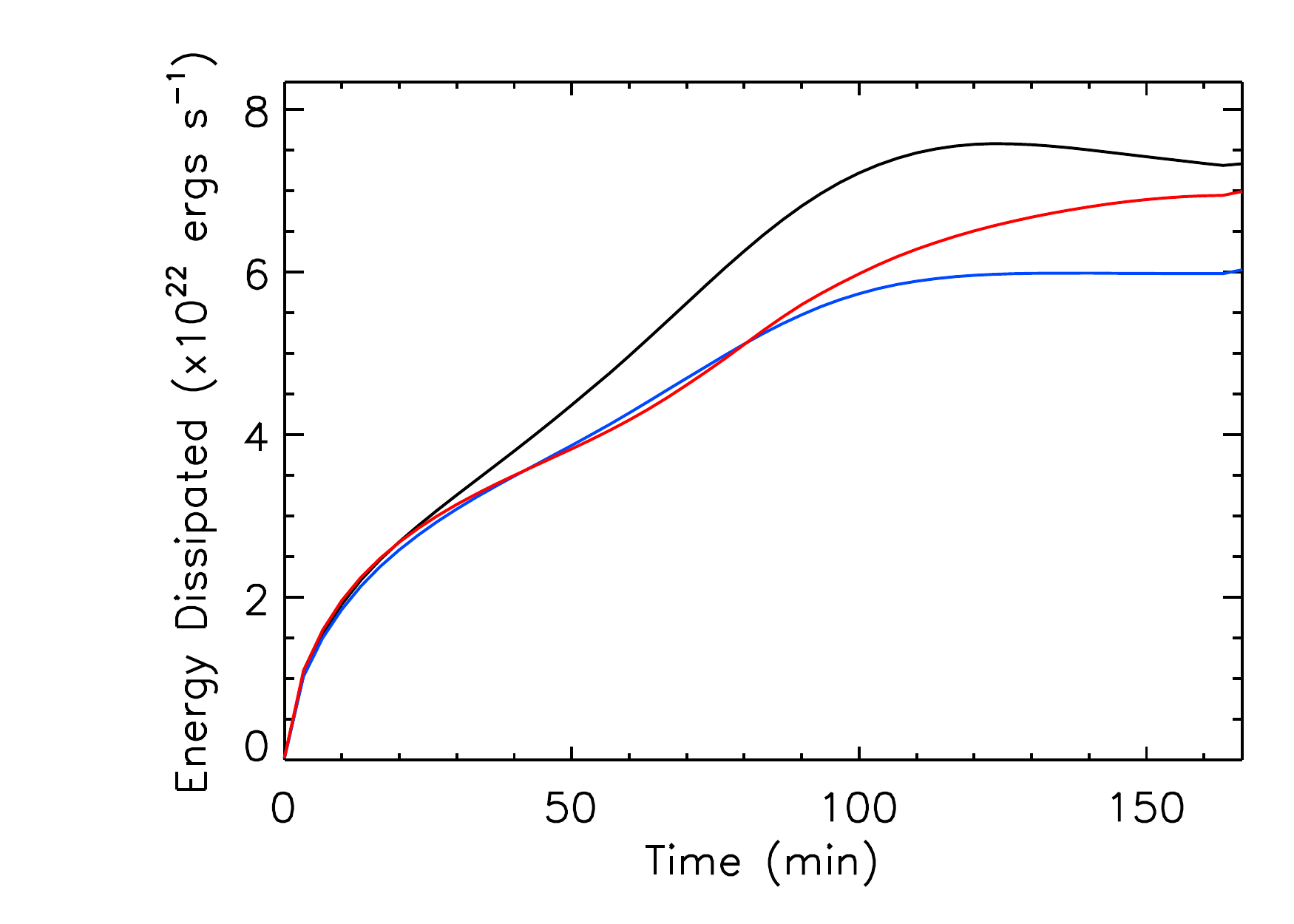}
               \hspace*{-0.0\textwidth}
               \includegraphics[width=0.45\textwidth,clip=]{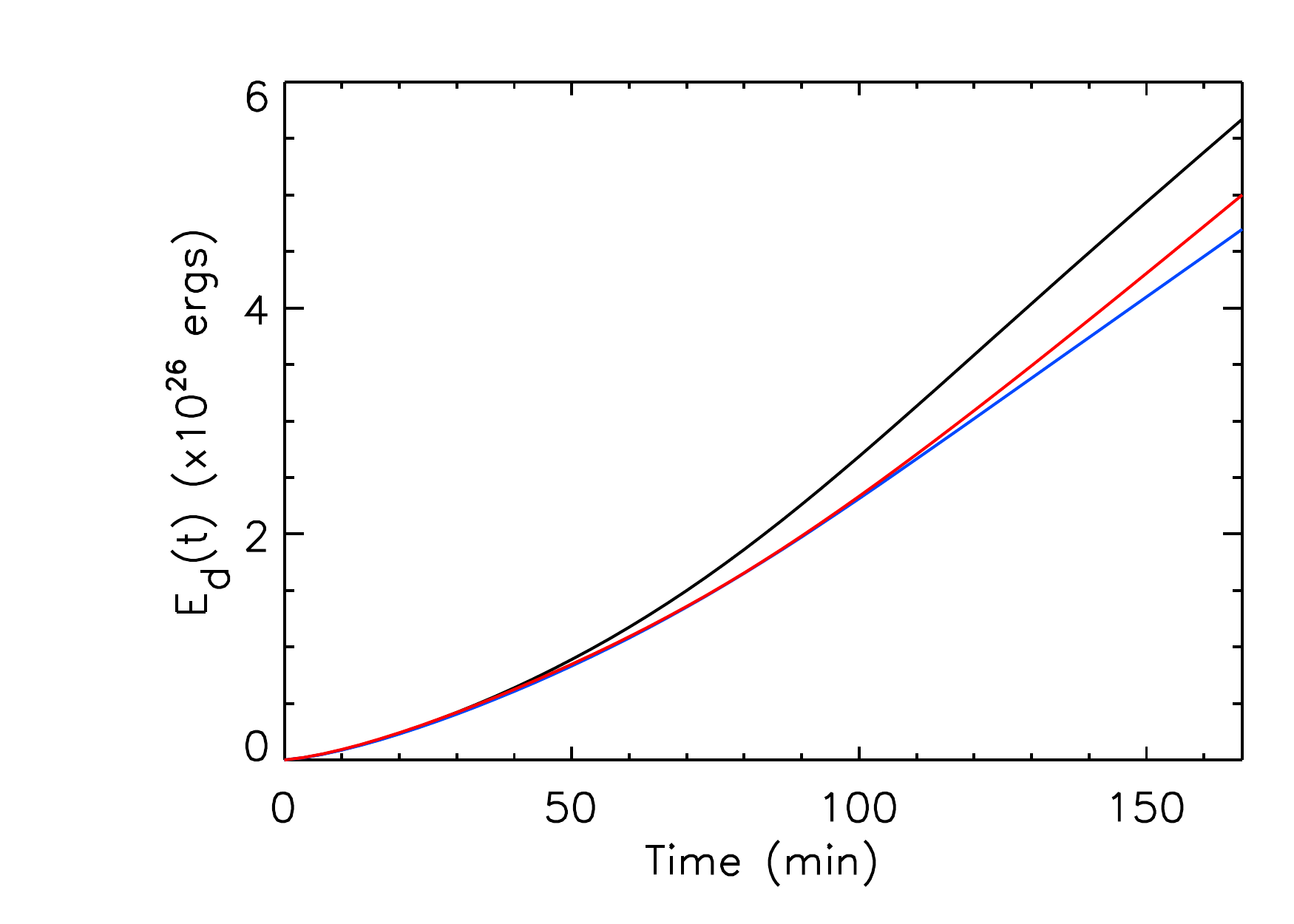}
              }
     \vspace{-0.34\textwidth}   
     \centerline{ \bf     
      \hspace{-0.0 \textwidth} \color{black}{(c)}
      \hspace{0.42\textwidth}  \color{black}{(d)}
         \hfill}
     \vspace{0.34\textwidth}    

\caption{Plots for a flyby, where the positive polarity is advected parallel to (black), anti-parallel to (blue) and perpendicular to (red) a 5 G overlying field, as a function of time: (a) total flux connecting the magnetic elements, (b) free magnetic energy, $E_\textrm{f}(t)$, (c) energy dissipated, $\int_V Q dV$, and (d) cumulative energy dissipated, $E_\textrm{d}(t)$.}\label{fig:flyby1}
   \end{figure}

Figure~\ref{fig:flyby1}(a) shows a plot of the total flux connecting the magnetic elements as a function of time, for a parallel (black), anti-parallel (blue) and perpendicular (red) flyby, with a 5 G overlying field. Both the non-linear force-free field (solid line) and corresponding potential field (dashed line) are shown.
For the parallel case, at all times, the total flux connecting the magnetic elements is greater for the non-linear force-free field than the corresponding potential field. In particular, flux still connects between the magnetic elements at the end of the simulation, but does not in the corresponding potential field.
For the anti-parallel case, initially, no flux connects between the magnetic elements. As the magnetic elements are advected past one another, the bipole's axis rotates to become aligned with the overlying field, and the total flux connecting the magnetic elements increases. Even though the connecting flux increases, the total flux connecting the elements is less than that of the potential field.

For the potential field case of the perpendicular flyby (red dashed line), the plot of the total flux connecting the magnetic elements is symmetric about the line $t=83.3$ min. In contrast, the connecting flux plot for the non-linear force-free field is non-symmetric. Before the polarities have aligned with the overlying field, the flux connecting the magnetic elements is lower than that of the potential field. However, after the elements pass one another at the midpoint, more flux connects between them in the non-linear force-free field than in the potential field case. This once again indicates significant differences between the potential field extrapolations and the non-linear force-free field simulations which retain a memory of flux connectivity.

   \begin{figure}
   \centerline{\hspace*{0.015\textwidth}
               \includegraphics[width=0.7\textwidth,clip=]{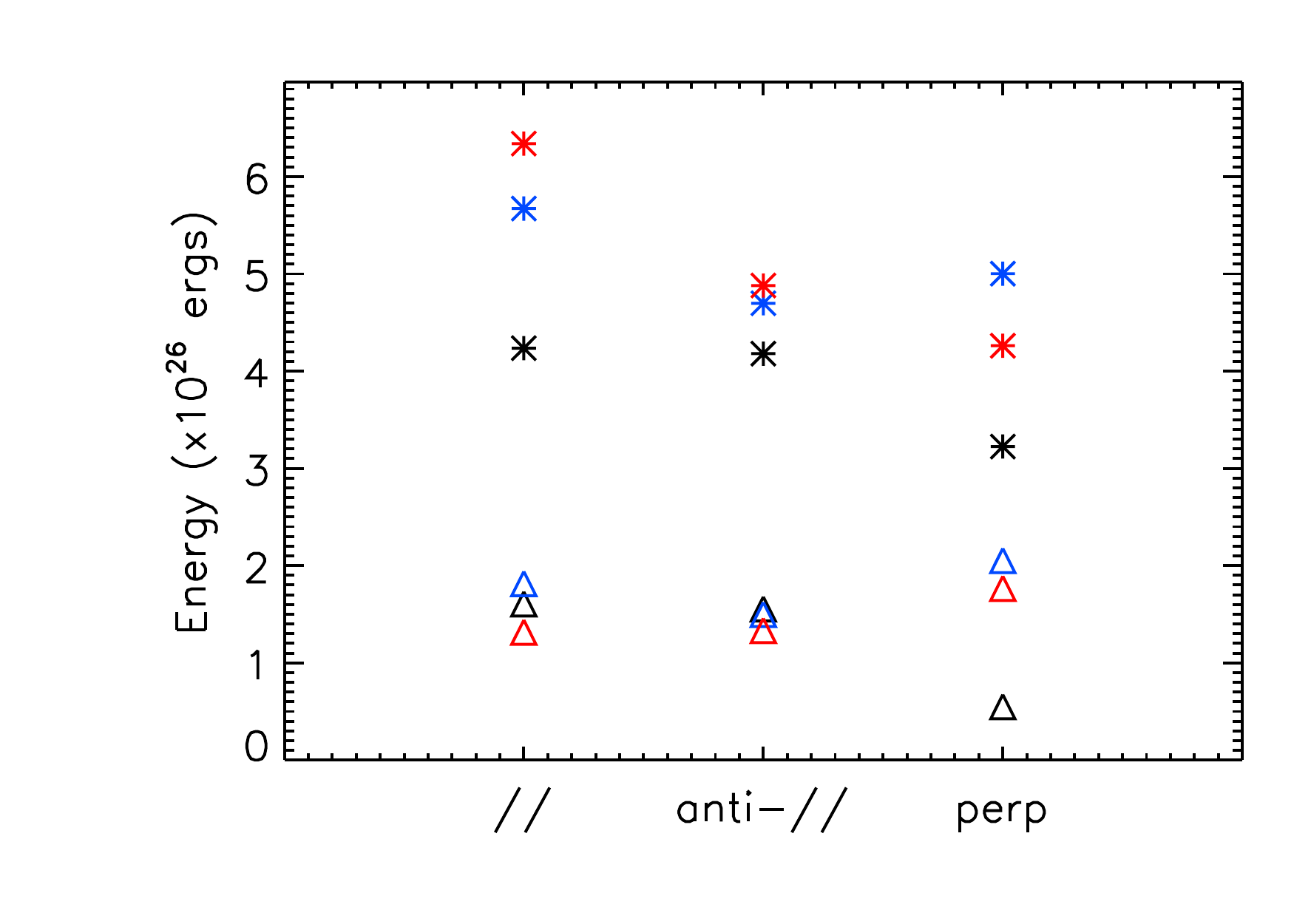}
              }

\caption{Free magnetic energy (triangles) and cumulative energy dissipated (stars) at the end of each flyby simulation, for a 1 G (black), 5 G (blue) and 10 G (red) overlying field.}\label{fig:allfly}
   \end{figure}

Figure~\ref{fig:flyby1}(b) shows a plot of the free magnetic energy as a function of time for the 5 G case of each orientation (lines are coloured as in Figure~\ref{fig:flyby1}(a)), where the final values range from $1.50-2.05\times 10^{26}$ ergs. The free energy stored by the end of the parallel flyby is greater than that of the anti-parallel flyby because, in the parallel case, more flux connects between the magnetic elements for a greater amount of time. However, the free energy is greatest for the perpendicular flyby, for two reasons. First, a greater volume of the magnetic field is disturbed by the magnetic elements, building up more free energy. Second, as the elements move past one another, flux from the positive magnetic element is forced to connect to the negative element. Numerous closed connections form and, due to the continuous nature of the magnetofrictional evolution, these flux connections are maintained as the magnetic elements move apart. Any free magnetic energy may be stored along them.

Figure~\ref{fig:flyby1}(c) shows a plot of the rate of energy dissipation, $Q$ integrated over the volume, as a function of time. In each case, the energy dissipated increases as the magnetic elements move towards one another, then levels off after they have passed one another at the midline. Initially, the curves for the anti-parallel and perpendicular cases are very similar. However, towards the end of the simulations, the rate of energy dissipation becomes greater in the perpendicular case. This happens because in the perpendicular case the amount of flux connecting the magnetic elements is still changing. This can be seen in Figure~\ref{fig:flyby1}(a), where towards the end of the simulation, the slope of the total flux connecting the magnetic elements is much steeper in the perpendicular than in the anti-parallel simulation. Figure~\ref{fig:flyby1}(d) shows the cumulative energy dissipated as a function of time, as calculated by Equation~\ref{eqn:edt}. One can see from Figures~\ref{fig:flyby1}(c) and (d) that the energy dissipation is greatest in the case of the parallel flyby ($5.67\times 10^{26}$ ergs in total) and least in the case of the anti-parallel flyby ($4.70\times 10^{26}$ ergs in total).

Figure~\ref{fig:allfly} compares the values of free magnetic energy (triangles) and total energy dissipated (stars) at the end of each flyby simulation. The most free energy tends to be stored in the 5 G case of each orientation, with the greatest amount of free energy resulting from the 5 G perpendicular flyby. Between $2-3$ times more energy is cumulatively dissipated by the end of each simulation than is stored as free energy. The greatest amount of energy is dissipated for the parallel flyby with a 10 G overlying field. The exact amount of free energy stored and energy dissipated by the end of each simulation depends on a balance between the strength and orientation of the overlying field, the volume of the coronal field that is disturbed, the amount of reconnection that occurs and, finally, the amount of connections that exist between the magnetic elements throughout their evolution.

In the next section, we compare the cancellation, emergence and flyby simulations to one another.

\subsection{Comparison of interactions} \label{compare}

For each of the three bipole interactions, three different orientations of the interaction with respect to an overlying magnetic field and three different strengths of overlying field have been considered. In this section, we compare the free energy and energy dissipated for all cases.
\begin{figure}
   \centerline{\hspace*{0.015\textwidth}
               \includegraphics[width=0.7\textwidth,clip=]{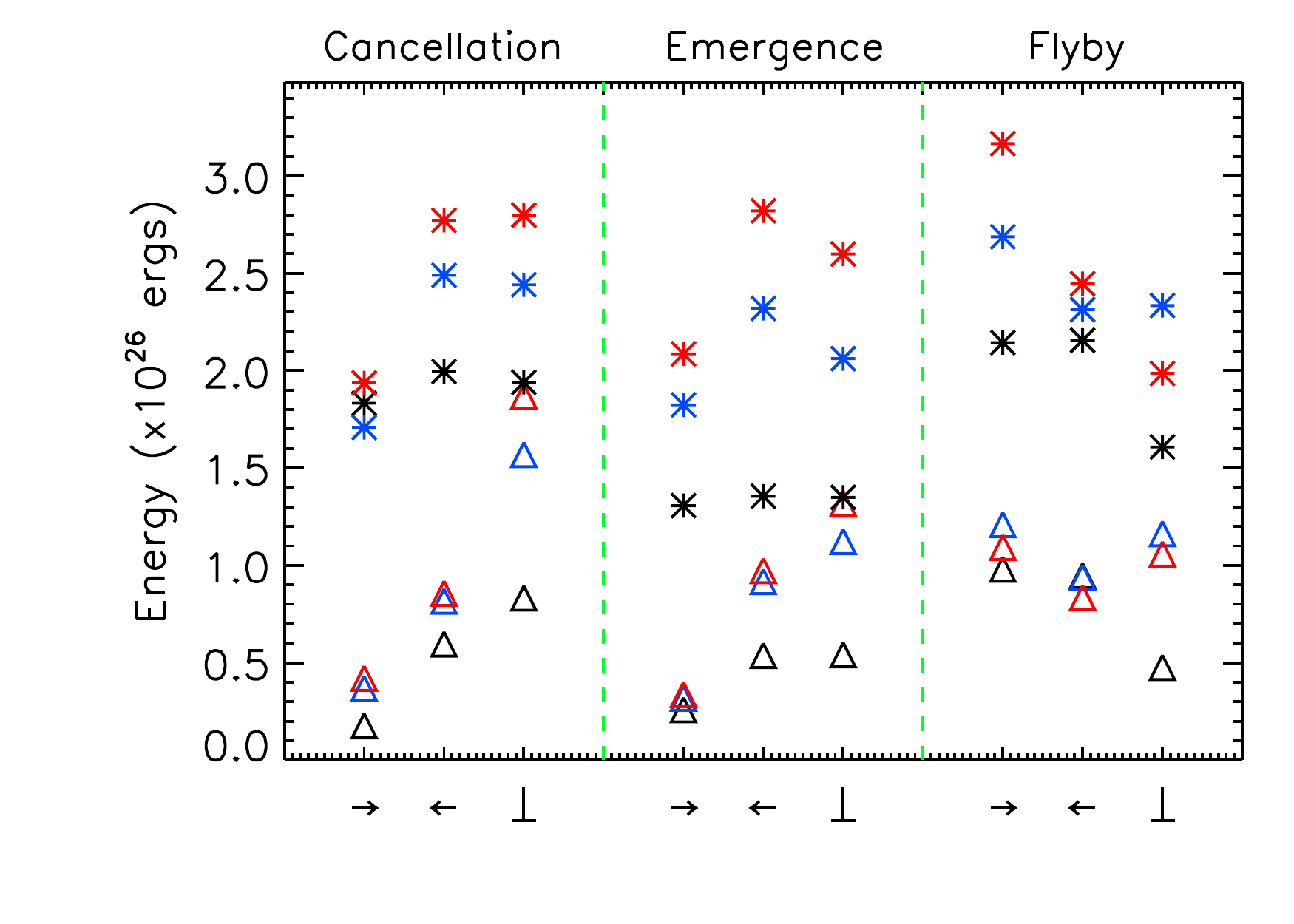}
              }

 \caption{Free magnetic energy, $E_f(t)$ (triangles), and cumulative energy dissipated, $E_\textrm{d}(t)$ (stars), at $t=100$ min in each simulation, for a 1 G (black), 5 G (blue) and 10 G (red) overlying field, in the parallel ($\rightarrow$), anti-parallel ($\leftarrow$) and perpendicular ($\bot$) cases. The green dashed lines separate the plot into three columns representing the cancellation (left), emergence (middle) and flyby (right) simulations.}\label{fig:all1}
\end{figure}
Figure~\ref{fig:all1} shows a plot comparing the free magnetic energy (triangles) and total energy dissipated (stars) at $t=100$ min for each simulation. The plot is split into three columns representing the cancellation (left), emergence (middle) and flyby (right) simulations. For each interaction, the parallel ($\rightarrow$), anti-parallel ($\leftarrow$) and perpendicular ($\bot$) cases are shown with a 1 G (black), 5 G (blue) and 10 G (red) overlying field.
Clearly, for the emergence and cancellation simulations, the smallest amount of free magnetic energy and energy dissipated arises in the parallel cases, while the anti-parallel and perpendicular cases lead to the largest energy stored and dissipated. In contrast, for the flyby, the parallel cases result in the most free energy and energy dissipated. However, for all cases the stronger the overlying field, the larger the energy values tend to be.

The amount of free energy stored in the magnetic field at $t=100$ min in each simulation varies between $0.2-1.9\times 10^{26}$ ergs, where the greatest value arises for the perpendicular cancellation with a 10 G overlying field. As discussed in Section~\ref{canc}, these values of free energy are small compared to the total energy within the volume (between $8.7\times 10^{26} - 6.3\times 10^{28}$ ergs depending on overlying field strength). However, the box is large compared to the size of the bipole and the bipole's area of interaction.
In reality on the Sun, many such magnetic elements would exist within such an area, with many of them continually interacting with one another. If we consider the power law distribution of \inlinecite{parnell2009}, we can determine the number of small magnetic elements that would exist in the simulation region. Taking
\begin{displaymath}
 N(\phi_1,\phi_2)=\int_{\phi_1}^{\phi_2} \frac{N_\textrm{f}}{\phi_0}\bigg(\frac{\phi}{\phi_0}\bigg)^{-1.85} d\phi,
\end{displaymath}
where $N_\textrm{f}=3.6\times 10^{-17}$ cm$^{-2}$ and $\phi_0=10^{16}$ Mx, the number of magnetic elements expected in a 30 $\times$ 30 Mm region with flux in the range $10^{17}-10^{18}$ Mx is 46, and the number with flux in the range $10^{16}-10^{18}$ Mx is 374.
Therefore in more complex simulations with a realistic number of magnetic elements, the free energy could be $1-2$ orders of magnitude higher.
Even though this is the case, for the simulations here, the amount of free energy built up in a single event is sufficient to explain sporadic events such as nanoflares, which release energy on the order of $10^{24}$ ergs \cite{golub1997}.

The total amount of energy dissipated by $t=100$ min varies between $1.3-3.2 \times 10^{26}$ ergs, where the simulation that results in the most energy dissipated is the parallel flyby.
We may compute the average rate of energy dissipation in ergs cm$^{-2}$ s$^{-1}$ for each simulation. We consider the central region in each case ($x=10-20$ Mm, $y=10-20$ Mm) to focus in on the interaction location.
We find that the average energy dissipation for each simulation occurs in the range $1.50-4.95\times 10^4$ ergs cm$^{-2}$ s$^{-1}$.
On comparing these values to the radiative losses of either the quiet Sun ($3\times 10^5$ ergs cm$^{-2}$ s$^{-1}$) or of a coronal hole ($8\times 10^5$ ergs cm$^{-2}$ s$^{-1}$, \inlinecite{withbroe1977}), we find that the rate of dissipation for a single event is too low. However, note that there would be many more magnetic elements on the Sun in a region of the size we have considered. Using the figures from above, $41$ magnetic elements of flux $10^{16}-10^{18}$ Mx would be expected, and this may increase the energy dissipation rate by an order of magnitude to observed levels.
In addition, we may compare the radiative losses to those of an XBP.
The radiative losses of an XBP have been measured as $5\times 10^4$ ergs cm$^{-2}$ s$^{-1}$ \cite{habbal1991}. This implies that the average energy dissipation in some of our simulations are of the correct size to explain the occurrence of such an event.

\section{Discussion and Conclusions} \label{conclusion}

The aim of this paper was to investigate the coronal consequences of three basic photospheric magnetic interactions: cancellation, emergence and flyby. Each interaction was simulated in the presence of an overlying uniform magnetic field, which was taken to be parallel, anti-parallel or perpendicular to the motion of the magnetic elements. The bipoles considered here were representative of small-scale photospheric magnetic features such as ephemeral regions or network features. Each bipole's physical extent was on the order of $3\,000-4\,000$ km and its total absolute flux was $2\times 10^{18}$ Mx.

In all cases, the 3D coronal magnetic field was initially in a potential state. A continuous evolution of the coronal field was then produced via a magnetofrictional relaxation technique that evolved the field through a series of quasi-static, non-linear force-free equilibria in response to applied photospheric boundary motions. Our treatment of the photospheric boundary evolution was discussed in Section~\ref{lower}. The continuous nature of the coronal evolution means that connectivity within the coronal volume was maintained from one step to the next. In many cases, this continuity allowed connections to exist longer than those found in potential field extrapolations. This allowed free energy to be built up and stored along closed field lines.

For each of the simulations, two forms of energy were studied. First there was the free magnetic energy which was stored in the non-potential magnetic field. This energy may be regarded as that available for sporadic coronal events such as XBPs or nanoflares. Second, energy was continually being dissipated, as described by Equation~\ref{eqn:q} ($Q$). This may be considered as energy that is available to be converted to heat or plasma motions, although for simplicity, here we did not follow the corresponding plasma processes. With the formalisation used in the present paper, the dissipated energy arose from the relaxation process employed, along with hyperdiffusion. We found that $Q$ mainly arose low down near the magnetic elements where the magnetic field departed most from a potential state, and at sites of changing magnetic topology.

The amount of free energy stored at $t=100$ min in each simulation ranges from $0.2-1.9\times10^{26}$ ergs. The cumulative energy dissipated in each simulation after the same amount of time is greater than the free energy stored; for each simulation, anywhere from $1.3-3.2\times 10^{26}$ ergs of energy has been dissipated after 100 min. The upper limits to both these values are higher if we consider the values at the end of the flyby simulations, which run for $166.7$ min ($2.1\times 10^{26}$ ergs for free energy, $6.3\times 10^{26}$ for dissipated energy). For cancellation and emergence, the amounts of free and dissipated energy are smallest when the motion of the magnetic elements is parallel to the overlying field, and largest when it is perpendicular. In contrast, for flyby, the amounts of free and dissipated energy are greatest in the parallel case. In all cases, a stronger overlying field tends to lead to greater energy storage and dissipation. The simulation that results in the most free energy is the perpendicular cancellation with a 10 G overlying field, while the simulation that results in the most energy cumulatively dissipated is the parallel flyby with a 10 G overlying field.
The exact amount of free energy stored and energy dissipated by the end of each simulation depends upon several factors: the strength and orientation of the overlying field, the volume of the overlying field that is disturbed, the amount of reconnection that occurs and, finally, the total flux connecting the magnetic elements.

The free magnetic energy built up in the present simulations is small compared to the total magnetic energy within the volume. However, in each case, the free energy is a significant fraction of the bipole's energy contribution ($8-86\%$), and is more than enough to account for small-scale, transient phenomena such as nanoflares or XBPs.
The rate of energy dissipation in each case is between $10^{22}-10^{23}$ ergs s$^{-1}$. This could provide a contribution towards the heating rate of an XBP of $3\times 10^{23}-10^{24}$ ergs s$^{-1}$, determined by \inlinecite{habbal1981}. We also find that for the inner $10\times 10$ Mm of each simulation, the energy dissipation rate is between $1.50-4.95\times 10^4$ ergs cm$^{-2}$ s$^{-1}$. This is equivalent to $5-17$\% of the energy required to heat the quiet Sun corona ($3\times 10^5$ ergs cm$^{-2}$ s$^{-1}$, \inlinecite{withbroe1977}).
Although it is at most $17\%$ of the coronal energy budget, on the Sun, many tens or hundreds of such small-scale magnetic elements would be found in a region of the size modelled here. The continual interaction of these magnetic elements with one another would result in a significantly larger build up of free magnetic energy and greater energy dissipation. With the expected number of magnetic elements in a region of this size, the free energy and energy dissipation rate may easily be $1-2$ orders of magnitude larger than those found for the simulations in this paper. This would bring it in line with coronal requirements. It is therefore of key importance to consider more complicated simulations of multiple magnetic elements.

From this study, the next step is to simulate the coronal evolution of the synthetic magnetograms constructed in Paper I. This would allow us to study the energetics of many events at the same time, as occur on the Sun. This will be the aim of the third paper in our series.
In these more complex simulations, we will study many aspects of the coronal evolution such as locations of coronal null points, electric currents, free magnetic energy and energy dissipation, and relate these to the dynamic processes occurring in the photospheric evolution. It will also be of interest to conduct a similar study by applying the same technique to real magnetogram data, such as those from SDO/HMI. Regions of interest in the resultant coronal magnetic field could then be compared with a wide variety of SDO coronal wavelengths.

\begin{acks}
We thank the anonymous referee, whose comments and suggestions have greatly improved the paper. In particular, we thank the referee for informing us of the calculation of the theoretical maximum free magnetic energy for an anti-parallel emergence in Section~\ref{emer}. KAM would like to thank ISSI (Bern) for their support of the team ``Solar small-scale transient phenomena and their role in coronal heating'', and the Harvard-Smithsonian Center for Astrophysics for their warm hospitality during the summers of 2010 and 2011. KAM and DHM acknowledge the financial support of the STFC. DHM would like to thank the Royal Society for their support through the research grant scheme, the EU-FP7 collaborative project SWIFF, and ISSI.
\end{acks}

\end{article} 


\begin{thebibliography}{}

\bibitem[\protect\citeauthoryear{Archontis et al.}{2004}]{archontis2004}
Archontis, V., Moreno-Insertis, F., Galsgaard, K., Hood, A., O'Shea, E.: 2004,
{\it Astron. Astrophys.}, {\bf 426}, 1047.


\bibitem[\protect\citeauthoryear{Archontis, Tsinganos and Gontikakis}{2010}]{archontis2010}
Archontis, V., Tsinganos, K., Gontikakis, C.: 2010,
{\it Astron. Astrophys.}, {\bf 512}, L2.


\bibitem[\protect\citeauthoryear{Bhattacharjee and Hameiri}{1986}]{bhattacharjee1986}
Bhattacharjee, A., Hameiri, E.: 1986,
{\it Phys. Rev. Lett.}, {\bf 57}, 206.


\bibitem[\protect\citeauthoryear{Boozer}{1986}]{boozer1986}
Boozer, A. H.: 1986,
{\it J. Plasma Phys.}, {\bf 35}, 133.


\bibitem[\protect\citeauthoryear{Browning et al.}{2004}]{browning2004}
Browning, P. K., van der Linden, R., Gerrard, C., Kevis, R., Hood, A.: 2004,
{\it ESASP}, {\bf 575}, 210.


\bibitem[\protect\citeauthoryear{Browning et al.}{2008}]{browning2008}
Browning, P. K., Gerrard, C., Hood, A. W., Kevis, R., van der Linden, R. A. M.: 2008,
{\it Astron. Astrophys.}, {\bf 485}, 837.


\bibitem[\protect\citeauthoryear{Cargill}{1993}]{cargill1993}
Cargill, P. J.: 1993,
{\it Solar Phys.}, {\bf 147}, 263.


\bibitem[\protect\citeauthoryear{Chae et al.}{2001}]{chae2001}
Chae, J., Martin, S. F., Yun, H. S., Kim, J., Lee, S., Goode, P. R., Spirock, T., Wang, H.: 2001, 
{\it Astrophys. J.}, {\bf 548}, 497.


\bibitem[\protect\citeauthoryear{Close et al.}{2003}]{close2003}
Close, R. M., Parnell, C. E., Mackay, D. H., Priest, E. R.: 2003, 
{\it Solar Phys.}, {\bf 212}, 251.


\bibitem[\protect\citeauthoryear{Close et al.}{2004}]{close2004}
Close, R. M., Parnell, C. E., Longcope, D. W., Priest, E. R.: 2004, 
{\it Astrophys. J.}, {\bf 612}, L81.


\bibitem[\protect\citeauthoryear{Cranmer and van Ballegooijen}{2010}]{cranmer2010}
Cranmer, S. R., van Ballegooijen, A. A.: 2010,
{\it Astrophys. J.}, {\bf 720} 824


\bibitem[\protect\citeauthoryear{de Wijn et al.}{2008}]{dewijn2008}
de Wijn, A. G., Lites, B. W., Berger, T. E., Frank, Z. A., Tarbell, T. D., Ishikawa, R.: 2008,
{\it Astrophys. J.}, {\bf 684}, 1469.


\bibitem[\protect\citeauthoryear{Galsgaard, Parnell and Blaizot}{2000}]{galsgaard2000}
Galsgaard, K., Parnell, C. E., Blaizot, J.: 2000,
{\it Astron. Astrophys.}, {\bf 362}, 395.


\bibitem[\protect\citeauthoryear{Galsgaard and Parnell}{2005}]{galsgaard2005}
Galsgaard, K., Parnell, C. E.: 2005,
{\it Astron. Astrophys.}, {\bf 439}, 335.


\bibitem[\protect\citeauthoryear{Golub and Pasachoff}{1997}]{golub1997}
Golub, L., Pasachoff, J. M.: 1997, {\it The Solar Corona}, Cambridge University Press.


\bibitem[\protect\citeauthoryear{Golub et al.}{1974}]{golub1974}
Golub, L., Krieger, A. S., Silk, J. K., Timothy, A. F., Vaiana, G. S. : 1974,
{\it Astrophys. J.}, {\bf 189}, L93.


\bibitem[\protect\citeauthoryear{Habbal and Withbroe}{1981}]{habbal1981}
Habbal, S. R., Withbroe, G. L.: 1981,
{\it Solar Phys.}, {\bf 69}, 77.


\bibitem[\protect\citeauthoryear{Habbal and Grace}{1991}]{habbal1991}
Habbal, S. R., Grace, E.: 1991,
{\it Astrophys. J.}, {\bf 382}, 667.


\bibitem[\protect\citeauthoryear{Hagenaar, Schrijver and Title}{1997}]{hagenaar1997}
Hagenaar, H. J., Schrijver, C. J., Title, A. M.: 1997,
{\it Astrophys. J.}, {\bf 481}, 988.


\bibitem[\protect\citeauthoryear{Hagenaar}{2001}]{hagenaar2001}
Hagenaar, H. J.: 2001, 
{\it Astrophys. J.}, {\bf 555}, 448.


\bibitem[\protect\citeauthoryear{Hagenaar, Schrijver and Title}{2003}]{hagenaar2003}
Hagenaar, H. J., Schrijver, C. J., Title, A. M.: 2003,
{\it Astrophys. J.}, {\bf 584}, 1107.


\bibitem[\protect\citeauthoryear{Hagenaar, DeRosa and Schrijver}{2008}]{hagenaar2008}
Hagenaar, H. J., DeRosa, M. L., Schrijver, C. J.: 2008,
{\it Astrophys. J.}, {\bf 678}, 541.


\bibitem[\protect\citeauthoryear{Harvey and Martin}{1973}]{harvey1973}
Harvey, K. L., Martin, S. F.: 1973,
{\it Solar Phys.}, {\bf 32}, 389.


\bibitem[\protect\citeauthoryear{Longcope}{1998}]{longcope1998}
Longcope, D. W.: 1998,
{\it Astrophys. J.}, {\bf 507}, 433.


\bibitem[\protect\citeauthoryear{Longcope and Kankelborg}{1999}]{longcope1999}
Longcope, D. W., Kankelborg, C. C.: 1999,
{\it Astrophys. J.}, {\bf 524}, 483.


\bibitem[\protect\citeauthoryear{Mackay and van Ballegooijen}{2009}]{mackay2009}
Mackay, D. H., van Ballegooijen, A. A.: 2009,
{\it Solar Phys.}, {\bf 260}, 321.


\bibitem[\protect\citeauthoryear{Mackay, Green and van Ballegooijen}{2011}]{mackay2011}
Mackay, D. H., Green, L. M., van Ballegooijen, A. A.: 2011,
{\it Astrophys. J.}, {\bf 729}, 97.


\bibitem[\protect\citeauthoryear{MacTaggart and Hood}{2009}]{mactaggart2009}
MacTaggart, D., Hood, A. W.: 2009, 
{\it Astron. Astrophys.}, {\bf 507}, 995.


\bibitem[\protect\citeauthoryear{Martin}{1988}]{martin1988}
Martin, S. F.: 1988, 
{\it Solar Phys.}, {\bf 117}, 243.


\bibitem[\protect\citeauthoryear{Martin}{1990}]{martin1990}
Martin, S. F.: 1990,
In: Stenflo, J. O. (ed.) {\it Solar Photosphere: Structure, Convection, and Magnetic Fields},
{\it IAU Symp.}, {\bf 138}, 129.


\bibitem[\protect\citeauthoryear{Meyer et al.}{2011}]{meyer2011}
Meyer, K. A., Mackay, D. H., van Ballegooijen, A. A., Parnell, C. E.: 2011,
{\it Solar Phys.}, {\bf 272}, 29.


\bibitem[\protect\citeauthoryear{Paniveni et al.}{2004}]{paniveni2004}
Paniveni, U., Krishan, V., Singh, J., Srikanth, R.: 2004,
{\it Mon. Not. R. Astron. Soc.}, {\bf 347}, 1279.


\bibitem[\protect\citeauthoryear{Parker}{1988}]{parker1988}
Parker, E. N.: 1988,
{\it Astrophys. J.}, {\bf 330}, 474.


\bibitem[\protect\citeauthoryear{Parnell}{2001}]{parnell2001}
Parnell, C. E.: 2001, 
{\it Solar Phys.}, {\bf 200}, 23.


\bibitem[\protect\citeauthoryear{Parnell}{2002}]{parnell2002}
Parnell, C. E.: 2002,
{\it Mon. Not. R. Astron. Soc.}, {\bf 335}, 389.


\bibitem[\protect\citeauthoryear{Parnell and Galsgaard}{2004}]{parnell2004}
Parnell, C. E., Galsgaard, K.: 2004,
{\it Astron. Astrophys.}, {\bf 428}, 595.


\bibitem[\protect\citeauthoryear{Parnell et al.}{2009}]{parnell2009}
Parnell, C. E., DeForest, C. E., Hagenaar, H. J., Johnston, B. A., Lamb, D. A., Welsch, B. T.: 2009,
{\it Astrophys. J.}, {\bf 698}, 75.


\bibitem[\protect\citeauthoryear{Parnell, Haynes and Galsgaard}{2010}]{parnell2010}
Parnell, C. E., Haynes, A. L., Galsgaard, K.: 2010,
{\it J. Geophys. Res.}, {\bf 115}, A02102.


\bibitem[\protect\citeauthoryear{Priest, Parnell and Martin}{1994}]{priest1994}
Priest, E. R., Parnell, C. E., Martin, S. F.: 1994,
{\it Astrophys. J.}, {\bf 427}, 459.


\bibitem[\protect\citeauthoryear{R{\'e}gnier}{2007}]{regnier2007}
R{\'e}gnier, S.: 2007,
{\it Memorie della Societa Astronomica Italiana}, {\bf 78}, 126.


\bibitem[\protect\citeauthoryear{R{\'e}gnier, Parnell and Haynes}{2008}]{regnier2008}
R{\'e}gnier, S., Parnell, C. E., Haynes, A. L.: 2008,
{\it Astron. Astrophys.}, {\bf 484}, L47.


\bibitem[\protect\citeauthoryear{Sakamoto, Tsuneta and Vekstein}{2009}]{sakamoto2009}
Sakamoto, Y., Tsuneta, S., Vekstein, G.: 2009,
{\it Astrophys. J.}, {\bf 703}, 2118.


\bibitem[\protect\citeauthoryear{Schrijver et al.}{1997}]{schrijver1997}
Schrijver, C. J., Title, A. M., van Ballegooijen, A. A., Hagenaar, H. J., Shine, R. A.: 1997,
{\it Astrophys. J.}, {\bf 487}, 424.


\bibitem[\protect\citeauthoryear{Schrijver and Title}{2002}]{schrijver2002}
Schrijver, C. J., Title, A. M.: 2002,
{\it Astrophys. J.}, {\bf 207}, 223.


\bibitem[\protect\citeauthoryear{Schrijver et al.}{2006}]{schrijver2006}
Schrijver, C. J., Derosa, M. L., Metcalf, T. R., Liu, Y., McTiernan, J., R{\'e}gnier, S., Valori, G.,
	Wheatland, M. S., Wiegelmann, T.: 2006,
{\it Solar Phys.}, {\bf 235}, 161.


\bibitem[\protect\citeauthoryear{Simon and Leighton}{1964}]{simon1964}
Simon, G. W., Leighton, R. B.: 1964,
{\it Astrophys. J.}, {\bf 140}, 1120.


\bibitem[\protect\citeauthoryear{Simon, Title and Weiss}{2001}]{simon2001}
Simon, G. W., Title, A. M., Weiss, N. O.: 2001,
{\it Astrophys. J.}, {\bf 561}, 427.


\bibitem[\protect\citeauthoryear{Strauss}{1988}]{strauss1988}
Strauss, H. R.: 1988,
{\it Astrophys. J.}, {\bf 326}, 412.


\bibitem[\protect\citeauthoryear{Title and Schrijver}{1998}]{title1998}
Title, A. M., Schrijver, A. M.: 1998,
{\it Cool Stars, Stellar Systems, and the Sun, ASPC}, {\bf 154}, 345.


\bibitem[\protect\citeauthoryear{Thornton and Parnell}{2011}]{thornton2011}
Thornton, L., Parnell, C. E.: 2011,
{\it Solar Phys.}, {\bf 269}, 13.


\bibitem[\protect\citeauthoryear{van Ballegooijen et al.}{1998}]{vanballegooijen1998}
van Ballegooijen, A. A., Nisenson, P., Noyes, R. W., L{\"o}fdahl, M. G., Stein, R. F., Nordlund, \AA., Krishnakumar, V.: 1998,
{\it Astrophys. J.}, {\bf 509}, 435.


\bibitem[\protect\citeauthoryear{van Ballegooijen, Priest and  Mackay}{2000}]{vanballegooijen2000}
van Ballegooijen, A. A., Priest, E. R., Mackay, D. H.: 2000,
{\it Astrophys. J.}, {\bf 539}, 983.


\bibitem[\protect\citeauthoryear{van Ballegooijen and  Cranmer}{2008}]{vanballegooijen2008}
van Ballegooijen, A. A., Cranmer, S. R.: 2008,
{\it Astrophys. J.}, {\bf 682}, 644.


\bibitem[\protect\citeauthoryear{von Rekowski, Parnell and Priest}{2006}]{vonrekowski2006}
von Rekowski, B., Parnell, C. E., Priest, E. R.: 2006,
{\it Mon. Not. R. Astron. Soc.}, {\bf 366}, 125.


\bibitem[\protect\citeauthoryear{Wang et al.}{1995}]{wang1995}
Wang, J., Wang, H., Tang, F., Lee, J. W., Zirin, H.: 1995,
{\it Solar Phys.}, {\bf 160}, 277.


\bibitem[\protect\citeauthoryear{Webb et al.}{1993}]{webb1993}
Webb, D. F., Martin, S. F., Moses, D., Harvey, J. W.: 1993,
{\it Solar Phys.}, {\bf 144}, 15.


\bibitem[\protect\citeauthoryear{Withbroe and Noyes}{1977}]{withbroe1977}
Withbroe, G. L., Noyes, R. W.: 1977,
{\it Annu. Rev. Astron. Astrophys.}, {\bf 15}, 363.


\bibitem[\protect\citeauthoryear{Yang, Sturrock and Antiochos}{1986}]{yang1986}
Yang, W. H, Sturrock, P. A., Antiochos, S. K.: 1986, 
{\it Astrophys. J.}, {\bf 309}, 383.


\bibitem[\protect\citeauthoryear{Yeates, Mackay and van Ballegooijen}{2008}]{yeates2008}
Yeates, A. R., Mackay, D. H., van Ballegooijen, A. A.: 2008,
{\it Solar Phys.}, {\bf 247}, 103.

\end{thebibliography}
\end{document}